\documentclass[a4paper,10pt,twoside]{article}
\usepackage[T1]{fontenc}
\usepackage[utf8]{inputenc}
\usepackage{a4wide}
\usepackage{fourier,bbm,bm}
\usepackage{amsmath,amsfonts,amsthm,amssymb}
\usepackage{color,xcolor,graphicx}
\usepackage{latexsym,mathtools,csquotes}
\usepackage{xargs,xspace,needspace,xparse,easyReview}
\usepackage[active]{srcltx}
\usepackage[colorlinks=true]{hyperref}
\usepackage{fancyhdr}
\usepackage[toc]{appendix}

\colorlet{darkblue}{blue!50!black}

\hypersetup{
    colorlinks,%
    citecolor=darkblue,%
    filecolor=red,%
    linkcolor=darkblue,%
    urlcolor=darkblue,%
    pdfnewwindow=true,%
    pdfstartview={FitH}
}
\usepackage[normalem]{ulem}
\usepackage[style=cap-alpha]{biblatex}
\newtheorem{theorem}{Theorem}[section]
\newtheorem{proposition}[theorem]{Proposition}
\newtheorem{lemma}[theorem]{Lemma}
\newtheorem{definition}[theorem]{Definition}
\newtheorem{corollary}[theorem]{Corollary}
\newtheorem{remark}[theorem]{Remark}

\newcommand{\bep}{\begin{proposition}}
\newcommand{\eep}{\end{proposition}}
\newcommand{\bel}{\begin{lemma}}
\newcommand{\eel}{\end{lemma}}
\newcommand{\bet}{\begin{theorem}}
\newcommand{\eet}{\end{theorem}}
\newcommand{\bed}{\begin{definition}}
\newcommand{\eed}{\end{definition}}
\newcommand{\bec}{\begin{corollary}}
\newcommand{\eec}{\end{corollary}}
\newcommand{\ber}{\begin{remark}}
\newcommand{\eer}{\end{remark}}
\newcommand{\beq}{\begin{equation}}
\newcommand{\eeq}{\end{equation}}
\newcommand{\bea}[1]{\begin{array}{#1}}
\newcommand{\eea}{\end{array}}

\def\remark{\noindent {\bf Remark.}\ \ }
\def\remarks{\noindent {\bf Remarks.}\ \ }

\newcommand{\ds}{\displaystyle}
\renewcommand{\i}{\mathrm{i}}

\newcommand{\e}{\mathrm{e}}
\renewcommand{\d}{\mathrm{d}}
\newcommand{\one}{\mathbbm{1}}
\renewcommand{\Im}{\operatorname{Im}}
\renewcommand{\Re}{\operatorname{Re}}
\newcommand{\slim}{\mathop{\mathrm{s-lim}}\limits}
\newcommand{\wlim}{\mathop{\mathrm{w-lim}}\limits}
\newcommand{\wstarlim}{\mathop{\mathrm{w}^\ast\mathrm{-lim}}\limits}
\renewcommand{\sp}{\operatorname{sp}}

\newcommand{\sign}{\operatorname{sign}}

\newcommand{\twol}[2]{\genfrac{}{}{0pt}{1}{#1}{#2}}

\newcommand{\Ran}{\operatorname{Ran}}
\newcommand{\Ker}{\operatorname{Ker}}
\newcommand{\Dom}{\operatorname{Dom}}
\newcommand{\Aut}{\operatorname{Aut}}

\newcommand{\tr}{\operatorname{tr}}
\newcommand\ie{\textsl{i.e.,\ }}
\newcommand{\argdot}{\,{\bm\cdot}\,}

\NewDocumentCommand{\ASS}{mm}{\expandafter\newcommand\csname #1\endcsname{{\hyperref[#1]{\bf (#2)}}}}
\NewDocumentCommand{\preASS}{mm}{\expandafter\newcommand\csname pre#1\endcsname{{\hyperref[#1]{\bf (#2)}}}}
\newcommand{\assuming}[3]{%arg1 = tag; arg2 = label = command name; arg3 = content
	\begin{quote}\label{#2}{\bf(#1) }%
		#3%
	\end{quote}%
	\ASS{#2}{#1}}
\newcommand{\cA}{\mathcal{A}}
\newcommand{\cB}{\mathcal{B}}

\newcommand{\cD}{\mathcal{D}}
\newcommand{\cE}{\mathcal{E}}
\newcommand{\cF}{\mathcal{F}}

\newcommand{\cH}{\mathcal{H}}

\newcommand{\cJ}{\mathcal{J}}
\newcommand{\cK}{\mathcal{K}}
\newcommand{\cL}{\mathcal{L}}
\newcommand{\cM}{\mathcal{M}}
\newcommand{\cN}{\mathcal{N}}
\newcommand{\cO}{\mathcal{O}}
\newcommand{\cP}{\mathcal{P}}

\newcommand{\cR}{\mathcal{R}}
\newcommand{\cS}{\mathcal{S}}

\newcommand{\cU}{\mathcal{U}}

\newcommand{\cX}{\mathcal{X}}

\newcommand{\cZ}{\mathcal{Z}}
\newcommand{\CC}{\mathbbm{C}}

\newcommand{\NN}{\mathbbm{N}}

\newcommand{\PP}{\mathbbm{P}}
\newcommand{\QQ}{\mathbbm{Q}}
\newcommand{\RR}{\mathbbm{R}}

\newcommand{\ZZ}{\mathbbm{Z}}
\newcommand{\fA}{\mathfrak{A}}

\newcommand{\fF}{\mathfrak{F}}
\newcommand{\fG}{\mathfrak{G}}

\newcommand{\fL}{\mathfrak{L}}
\newcommand{\fM}{\mathfrak{M}}

\newcommand{\fS}{\mathfrak{S}}

\newcommand{\fh}{\mathfrak{h}}

\newcommand{\fk}{\mathfrak{k}}

\newcommand{\fr}{\mathfrak{r}}

\newcommand{\bA}{\mathbf{A}}

\newcommand{\fin}{\mathrm{fin}}
\newcommand{\dec}{\mathrm{dec}}

\newcommand{\BH}{\cB(\cH)}

\newcommand{\CAR}{\operatorname{CAR}}

\newcommand{\ep}{\operatorname{ep}}

\newcommand{\Ent}{\operatorname{Ent}}
\newcommand{\Ep}{\operatorname{Ep}}

\newcommand{\wP}{\widehat{\PP}}

\newcommand{\cOs}{\cO_\mathrm{sa}}
\newcommand{\cOloc}{\cO_\mathrm{loc}}
\newcommand\fGfin{\fG_\mathrm{fin}}
\newcommand\cSI{\cS_\mathrm{I}}
\newcommand\cSeq{\cS_\mathrm{eq}}
\newcommand\ttm{\mathrm{ttm}}
\newcommand\TRI{{\rm TRI}}
\newcommand\BMV{\mathrm{BMV}}

\newcommand{\CCR}{\operatorname{CCR}}
\renewcommand{\Im}{\operatorname{Im}}
\renewcommand{\Re}{\operatorname{Re}}
\newcommand{\aFock}{\Gamma_{\mathrm a}}
\newcommand{\sFock}{\Gamma_{\mathrm s}}
\renewcommand{\l}{\mathrm{l}}
\renewcommand{\r}{\mathrm{r}}
\newcommand{\s}{\mathrm{s}}
\renewcommand{\a}{\mathrm{a}}
\renewcommand{\sp}{\mathrm{sp}}

\newcommand{\semi}{\mathrm{semi}}
\newcommand{\Id}{{\mathrm{Id}}}
\renewcommand{\i}{{\mathrm{i}}}
\renewcommand{\d}{{\mathrm{d}}}
\renewcommand{\ul}[1]{\textsl{#1}}
\def\bet{\begin{theorem}}
\def\eet{\end{theorem}}
\newcommand{\ben}{\begin{enumerate}}%[label=(\roman*)]}
\newcommand{\een}{\end{enumerate}}
\usepackage{marginnote}

\numberwithin{equation}{section}
\begin{document}
\title{Miniatures on  Open Quantum Systems}

\author{Jan Derezi\'nski${}^1$\ \ \ Vojkan Jak\v si\'c${^2}$\ \ \ Claude-Alain Pillet${}^3$
\\ \\
$^1$Department of Mathematical Methods in Physics, Faculty of Physics\\
 University of Warsaw, Pasteura 5, 02-093 Warszawa, Poland
\\ \\
$^2$Dipartimento di Matematica, Politecnico di Milano, \\
piazza Leonardo da Vinci, 32, 20133 Milano, Italy 
\\ \\
$^3$Universit\'e de Toulon, CNRS, CPT, UMR 7332, 83957 La Garde, France\\
Aix-Marseille Univ, CNRS, CPT, UMR 7332, Case 907, 13288 Marseille, France}
\maketitle
\thispagestyle{empty}

\bigskip\noindent{\small{\bf Abstract.}
We presents a unified and concise exposition of key topics in the
mathematical theory of open quantum systems, developed within the
framework of operator algebras. The manuscript consolidates and
extends a series of invited articles originally prepared for the
Modern Encyclopedia of Mathematical Physics, combining foundational
material with modern perspectives on non-equilibrium quantum
statistical mechanics. After introducing the C*- and W*-algebraic
formulation of quantum mechanics, the paper reviews quantum dynamical
systems, KMS states, and Tomita-Takesaki modular theory, as well as
CCR and CAR algebras for bosonic and fermionic systems. Particular
emphasis is placed on infinite systems, non-equilibrium steady states,
entropy production, and linear response theory. The later sections
develop a systematic treatment of small systems coupled to reservoirs,
open lattice quantum spin systems, culminating in a detailed
discussion of competing notions of quantum entropy production. The
presentation highlights structural insights, conceptual clarity, and
connections between equilibrium and non-equilibrium phenomena,
providing a self-contained reference for researchers and graduate
students in mathematical physics.}
%%%%%%%%%%%%%%%%%%%%%%%%%%%%%%%%%%%%%
\tableofcontents
%%%%%%%%%%%%%%%%%%%%%%%%%%%%%%%%%%%%%

\bigskip

\section{Introduction}
%%%%%
At the turn of the 21st century, there was a revival of C*-algebraic methods in
the study of equilibrium and non-equilibrium quantum statistical mechanics, a
development to which the authors contributed. This contribution was acknowledged
in 2006 through an invitation to take part in the ambitiously conceived
\textsl{Modern Encyclopedia of Mathematical Physics}.

\begin{figure}[h]
\centering
\includegraphics[width=0.25\textwidth]{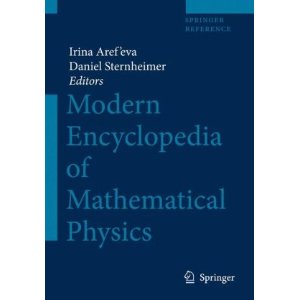}
\caption{The cover  of \textsl{Encyclopedia} from Springer 2011 web page.}
\end{figure}

The topics, titles, and length of invited articles were specified by the
Editors of the \textsl{Encyclopedia}. Our articles were completed in the summer
of 2007.

Unfortunately, the \textsl{Encyclopedia}’s publication was repeatedly delayed
and the project was eventually discontinued, so our articles never appeared in
print.

On the occasion of the thematic program \textsl{Quantum Mathematics @ Polimi}
held at the Politecnico di Milano in Spring 2025, in which the authors took
part, we have decided  to consolidate these contributions into a single, unified
publication. The articles in Parts I-IV appear largely in their original form,
as we have deliberately limited revisions and references to work published after
2007. Parts V–VIII extend and modernize our {\rm Encyclopedia article} on
\textsl{Quantum non-equilibrium statistical mechanics}.

The notation is not uniform across sections, and some themes are repeated, as in
the original articles. The  Parts I-IV correspond to individual articles and 
are self-contained. Parts V–VIII are interconnected and should be read together.

We hope these concise treatments of important topics in modern mathematical
physics will prove useful to readers.

\paragraph*{Acknowledgments.} The work of CAP and VJ was partly funded by the CY
Initiative grant Investisse\-ments d’Avenir, grant number ANR-16-IDEX-0008. VJ
acknowledges the support of the MUR grant "Dipartimento di Eccellenza 2023-2027"
of Dipartimento di Matematica, Politecnico di Milano. We also acknowledge the
support of the ANR project DYNACQUS, grant number ANR-24-CE40-5714.

%%%%%%%%%%%%%%%%%%%%%%%%%%%%%%%%%%%%%%
\part{Elements of  operator algebras}
%%%%%%%%%%%%%%%%%%%%%%%%%%%%%%%%%%%%%%

%%%%%%%%%%%%%%%%%%%%%%%%%%%%%%%%%%%%%%%%
\section{The $C^\ast$-algebra approach}
\label{sec:The C*-algebra approach}
%%%%%%%%%%%%%%%%%%%%%%%%%%%%%%%%%%%%%%%%

%%%%%%%%%%%%%%%%%%%%%%%%%%%%%%%%%%%%
\subsection{Why operator algebras?}
%%%%%%%%%%%%%%%%%%%%%%%%%%%%%%%%%%%%

 A quantum mechanical description of a system involving only a
finite number of particles or degrees of freedom is
given by a {Hilbert space} $\cH$ and a {Hamiltonian} $H$, a
{selfadjoint} operator on $\cH$.
 Such systems  will be 
   called  finite.
 States of those systems are
described by unit vectors $\psi\in\cH$ or more generally
statistical mixtures, \ie \ul{density matrices}. Physical
quantities, or observables, are described by all  selfadjoint operators on $\cH$.  
 Their time evolution is governed by its Hamiltonian.
In the {Schrödinger picture}, the state evolves according to the
Schrödinger equation $\i\partial_t\psi_t=H\psi_t$ while observables
are time-independent. In the equivalent {Heisenberg picture}, the state
is time-independent and observables evolve according to the Heisenberg
equation $\partial_tA_t=\i[H,A_t]$. As a consequence of the tight
relation between Hamiltonian and dynamics, the {spectrum} $\sp(H)$ of
$H$ contains a lot of information about the physics of the system.

From a mathematical perspective, constructing $\cH$ and
$H$ for a given physical system { can often be presented as} a problem in representation theory. 
In the case of a non-relativistic system of $N$ point-like
particles in Euclidean space $\RR^3$, the correspondence between the
{classical and  quantum description is based on} \ul{canonical quantization}.  This procedure provides a
representation of the positions $Q_1,\dots,Q_{3N}$ and conjugate
canonical momenta $P_1,\ldots,P_{3N}$ by selfadjoint operators
satisfying the \ul{canonical commutation relations} (CCR)
\begin{equation}
\i[P_i,Q_j]=\delta_{ij},\quad\i[P_i,P_j]=0,\quad\i[Q_i,Q_j]=0.\label{ccr}\end{equation}
If the particles have spin, then
$\cH$ has to carry $N$ representations of the Lie algebra of
$SU(2)$, the quantum mechanical rotation group. If the particles are
indistinguishable, then {Pauli's principle} imposes a definite
covariance (as prescribed by the {spin-statistics theorem}) with
respect to the natural action of the symmetric group $S_N$ on
$\cH$.

{It is a deep fact that for  finite systems representations of the  CCR
are } unique, up to unitary transformations and mostly
irrelevant multiplicities. This is the content of the celebrated
\ul{Stone-von Neumann} theorem (see~\cite{Rosenberg2004} for a review). 

When dealing with systems with an infinite number of particles or
degrees of freedom -- quantum fields or quantum statistical mechanics
in the thermodynamic limit -- we are faced with a radically different
situation. The breakdown of Stone-von Neumann theorem implies the
existence of a large number of unitarily inequivalent irreducible
representations of the CCR. This phenomenon is not a special feature
of CCR. The following example shows that it occurs also for
representations of the group $SU(2)$; see~\cite{Thirring2002} for 
a related discussion. 

Consider an infinite chain of quantum spins with $s=1/2$. To each site
$x\in\ZZ$ we associate observables $\sigma_x^{(1)}$, $\sigma_x^{(2)}$, 
$\sigma_x^{(3)}$, which satisfy the $SU(2)$ commutation relations
\beq
[\sigma_x^{(j)},\sigma_y^{(k)}]=2\i\delta_{xy}\epsilon_{jkl}\sigma_x^{(l)}.
\label{su2}
\eeq
The Hamiltonian is formally given by $H=J\sum_{x\in\ZZ}\sigma_x^{(3)}$,
so that
\beq
\i[H,\sigma_x^{(j)}]=-2J\epsilon_{3jk}\sigma_x^{(k)}.
\label{Heisen}
\eeq

Set $\fh_x=\CC^2$ for any $x\in\ZZ$. The naive candidate for the
Hilbert space of the system is the maximal tensor product of all the
spaces $\fh_x$, which we will denote $\otimes_{x\in\ZZ}\fh_x$. It is
defined as the completion of the pre-Hilbert space spanned by vectors
of the form $\otimes_{x\in\ZZ}\varphi_x$, where each $\varphi_x$ is a unit
vector in $\fh_x$.  The inner product between two such vectors is
defined by
\beq
(\otimes_{x\in\ZZ}\psi_x\,|\,\otimes_{x\in\ZZ}\varphi_x)
=\prod_{x\in\ZZ}(\psi_x\,|\,\varphi_x),
\label{innerprod}
\eeq
whenever the infinite product on the right-hand side  of~\eqref{innerprod} is absolutely
convergent. Otherwise, the inner product 
is set to be zero. The  space $\otimes_{x\in\ZZ}\fh_x$
was first considered by von Neumann in~\cite{vonNeumann1939}
(he called it the complete direct product of the family
$(\fh_x)_{x\in\ZZ}$). This space is much too big for most applications. In
particular, it is not separable, \ie it does not have a countable
orthonormal basis\footnote{Except for recent developments in quantum
gravity, most Hilbert spaces of quantum physics are separable.}.

Let us describe another candidate for the notion of the infinite tensor
product, which is more useful in quantum physics.
For all $x\in\ZZ$ fix an orthonormal basis $\{\chi_x^-,\chi_x^+\}$ of
$\fh_x$.  To each finite subset $X\subset\ZZ$ associate the vector
$$
e_X=\left(\bigotimes_{x\in X}\chi_x^+\right)\bigotimes
\left(\bigotimes_{x\in\ZZ\setminus X}\chi_x^-\right).
$$
According to~\eqref{innerprod} one has $(e_X|e_Y)=\delta_{XY}$. Thus,
finite linear combinations of the vectors $e_X$ form a pre-Hilbert
space. The Hilbert space $\cH$ obtained by completion is separable
since $\{e_X\,|\,X\subset\ZZ, |X|<\infty\}$ is a countable orthonormal
basis.

{ A pair $(\fh,\chi)$, where $\fh$ is a Hilbert space and
$\chi\in\fh$ a unit vector, is often} called a grounded Hilbert space. The
above construction is a special case of the tensor product of grounded
Hilbert spaces, namely $\cH=\otimes_{x\in\ZZ}(\fh_x,\chi_x^-)$. The
interested reader may consult~\cite{Baez1992} for the general construction.

Remark that the maximal tensor product $\otimes_{x\in\ZZ}\fh_x$
naturally splits into the direct sum of sectors, where each sector has
the form $\otimes_{x\in\ZZ}(\fh_x,\chi_x^-)$ for a certain sequence of
unit vectors $\chi_x^-\in\fh_x$.

If $J>0$, then the ground state of the chain has all spins pointing
down in the direction $3$. If we interpret $\chi_x^\pm$ as the eigenstates of
the spin at $x$ in direction $3$ with eigenvalue $\pm 1/2$, the
vector $e_\emptyset$ clearly describes this ground state. The
vector $e_X$ describes a local excitation of the chain, the spins at
$x\in X$ pointing up in  direction $3$. This immediately leads to the
following representation of the commutation relations~\eqref{su2} on
$\cH$:
$$
\sigma_{x}^{(1)+}e_X=e_{X\odot x},\quad
\sigma_{x}^{(2)+}e_X=\i s_X(x)e_{X\odot x},\quad
\sigma_{x}^{(3)+}e_X=s_X(x)e_{X},
$$
where
$$
X\odot x=\left\{\begin{array}{ll}
X\setminus\{x\}&\text{if}\ x\in X,\\
X\cup\{x\}&\text{if}\ x\not\in X,
\end{array}
\right.
\qquad
s_X(x)=\left\{\begin{array}{ll}
+1&\text{if}\ x\in X,\\
-1&\text{if}\ x\not\in X.
\end{array}
\right.
$$
We obtain a different representation of the commutation relations~\eqref{su2}
{on the same space $\cH$} if we think of $\chi_x^\pm$ as the eigenstates of the spin in
direction $3$ with eigenvalue $\mp 1/2$, namely
$$
\sigma_{x}^{(1)-}=\sigma_{x}^{(1)+},\quad
\sigma_{x}^{(2)-}=-\sigma_{x}^{(2)+},\quad
\sigma_{x}^{(3)-}=-\sigma_{x}^{(3)+}.
$$

Note that although  $\sigma_{x}^{(3)+}e_\emptyset=-e_\emptyset$ for all
$x\in\ZZ$,  one easily checks that there is no unit vector
$\Psi\in\cH$ such that $\sigma_{x}^{(3)-}\Psi=-\Psi$ for all
$x\in\ZZ$. Thus, in the second representation of the system the ground
state {\it does not belong to} $\cH$. In particular, there is
no unitary operator $U$ on $\cH$ such that
$U\sigma_{x}^{(3)-}U^\ast=\sigma_{x}^{(3)+}$ -- the two representations
are inequivalent. Since
$$
e_Y=\prod_{x\in X\Delta Y}\sigma^{(1)\pm}_{x}e_X,
$$
the two representations are also irreducible.

Expressing the formal Hamiltonian $H$ in the two representations leads
to
$$
He_X=J\sum_{x\in X}\pm e_X+J\sum_{x\not\in X}\mp e_X
=\left(\pm 2J|X|\mp J\sum_{x\in\ZZ}1\right)e_X.
$$
Discarding a constant -- the infinite energy of the state
$e_\emptyset$ -- we may thus set
$$
H_\pm e_X=\pm 2J|X|e_X,
$$
which defines two selfadjoint operators on $\cH$. The
physical meaning of $H_\pm$ is clear: it describes  the energy of the
system relative to the infinite energy of the state $e_\emptyset$.
One easily verifies that
$$
\i[H_\pm,\sigma_{x}^{(j)\pm}]=-2J\epsilon_{3jk}\sigma_{x}^{(k)\pm},
$$
\ie that the commutation relations~\eqref{Heisen} are
satisfied. Note, however, that the spectra of the two Hamiltonians 
are quite different,
$$
\sp(H_\pm)=\pm 2J\NN.
$$
This is not surprising,  since $H_\pm$ measures
the energy relative to two distinct reference states: one of them has all
spins up while the other has all spins down.

In conclusion, there { are many ways} to represent the algebraic
structure induced by commutation relations~\eqref{su2} and~\eqref{Heisen}
in a Hilbert space. To select such a representation one needs to
specify a reference state.  In equilibrium statistical mechanics this
fact does not lead to difficulties since it is always possible to
define thermodynamic quantities (free energy, pressure, ...)  as
limits of quantities related to finite systems. The situation is
different in non-equilibrium statistical mechanics where dynamics
plays a much more important role. To give a mathematically precise
sense to non-equilibrium steady states, for example, requires a
consideration of infinite systems from the outset;  see
Section~\ref{sec:NESS in quantum statistical mechanics}.

In the algebraic approach to quantum mechanics the central object is
the algebraic structure -- in the above example, Relations~\eqref{su2}
and~\eqref{Heisen}. Hilbert spaces and Hamiltonians only appear when 
this structure gets represented by linear operators. Such 
representations are induced by the states of the system, usually via 
the Gelfand-Naimark-Segal construction; see 
Section~\ref{sec:The C*-algebra approach}. States with different physical
properties (e.g. different particle or energy density) lead to inequivalent
representations and hence the
{Hamiltonians describing their dynamics may have} different spectral properties.

The mathematical framework of algebraic approach to quantum mechanics 
is the theory of \ul{ope\-rator al\-gebras}. \ul{$C^\ast$-algebras},
\ul{von Neumann algebras}, and \ul{$W^\ast$-algebras} are the most commonly 
used types of operator algebras in this context. There is a huge literature 
devoted to this subject. The reader may consult~\cite{Dixmier1977,Disxmier1981,
Kadison1983,Sakai1998,Stratila1979,Takesaki1979} for mathematical introductions and~\cite{Bratteli1987,Bratteli1981,Baez1992,Haag1996,Ruelle1969,Simon1993,
Derezinski2006a} for applications to quantum physics and statistical mechanics.

%%%%%%%%%%%%%%%%%%%%%%%%%%%%%%%%%%%%%%%%%%%
\subsection{Examples of $C^\ast$-algebras}
%%%%%%%%%%%%%%%%%%%%%%%%%%%%%%%%%%%%%%%%%%%

To illustrate the algebraic approach, we consider a few systems for
which $C^\ast$-algebras provide a natural framework. See also
Section~\ref{sec:Free Bose and Fermi gases -- the algebraic approach}.
We shall  be only concerned with operator algebras here. We refer to
Section~\ref{sec:Quantum dynamical systems} for examples of
dynamics on these algebras.

%%%%%%%%%
\subsubsection{Lattice spin systems}
%%%%%%%%%%

To describe a \ul{quantum spin system} on an infinite lattice $\Gamma$ (for
example, $\Gamma=\ZZ^d$ with $d\ge1$), let $\fh$ be the finite
dimensional Hilbert space of a single spin, and $\fh_x$  a copy of $\fh$
associated to  each $x \in \Gamma$.  For finite subsets $\Lambda \subset \Gamma$ we set
$\fh_{\Lambda} \equiv \otimes_{x \in \Lambda}\fh_x$, and define the local 
$C^{\ast}$-algebras as
\[
\fA_{\Lambda}\equiv\cB(\fh_{\Lambda}).
\]
If $\Lambda \subset \Lambda'$, the natural isometric injection $A \mapsto A
\otimes I_{\fh_{\Lambda' \setminus \Lambda}}$ allows one to
identify $\fA_{\Lambda}$ with a subalgebra of
$\fA_{\Lambda'}$. With this identification, we may define
$$
\|A\|=\|A\|_{\fA_{\Lambda}}\ \text{if}\ A\in\fA_\Lambda,
$$
unambiguously for all $A\in\fA_{\mathrm{loc}} \equiv
\cup_{\Lambda \subset \Gamma} \fA_{\Lambda}$, the union being
over finite subsets of $\Gamma$. This defines a $C^{\ast}$-norm on the
$\ast$-algebra $\fA_{\mathrm{loc}}$. Denote by $\fA=\fA(\Gamma,\fh)$
the $C^{\ast}$-algebra obtained as norm completion of
$\fA_{\mathrm{loc}}$. We can identify each local algebra
$\fA_{\Lambda}$ with the corresponding subalgebra
of $\fA$, hence
\beq
\fA=\left(\bigcup_{\Lambda\subset\Gamma}\fA_\Lambda\right)^\mathrm{
  cpl},
\label{SpinLocal}
\eeq
where each $\fA_\Lambda$ is a full matrix algebra,  and cpl
  denotes the completion.
$C^\ast$-algebras of this type are called \textsl{uniformly hyperfinite} (UHF)
or \textsl{Glimm algebras.}

\subsubsection{CAR algebras}
\label{ssec:CAR-algebra}

Let $\fh$ be the Hilbert space of a single \ul{fermion} (typically
$\fh=L^2(\RR^d)\otimes\CC^n$, but other geometries and additional
internal degrees of freedom may lead to different single particle Hilbert
spaces). A system of such fermions is described by the \ul{second
  quantization} formalism. Denote by $\aFock(\fh)$ the fermionic
(or anti-symmetric) \ul{Fock space} over $\fh$.  The fermion
\ul{creation/annihilation ope\-ra\-tors} $a^\ast(f)$, $a(f)$ are
bounded operators on $\aFock(\fh)$
satisfying the canonical anti-commutation relations (CAR)
\beq
[a(f),a^\ast(g)]_+=(f|g) \one,\quad[a(f),a(g)]_+=0,
\label{CAR}
\eeq
for all $f,g\in\fh$. A convenient choice of observables is given by 
polynomials in $a$, $a^\ast$, that is,    finite linear
combinations of monomials
$$
a^\#(f_1)\cdots a^\#(f_n),
$$
where each $a^\#$ stands for either $a$ or $a^\ast$ and the $f_j$ are
elements of $\fh$ or more generally of some subspace
$\fh_0\subset\fh$. This set is clearly a unital $\ast$-algebra.  Its
norm closure is a $C^\ast$-subalgebra of $\cB(\aFock(\fh))$.
It turns out that this algebra is completely characterized
by $\fh_0$ and the CAR~\eqref{CAR}.

\begin{theorem}Let $\fh_0$ be a pre-Hilbert space. Up to
$\ast$-isomorphisms, there exists a unique unital $C^\ast$-algebra
$\CAR(\fh_0)$ satisfying the  following two properties:
\begin{enumerate}%[(i)]
\item There exists an anti-linear map $a:\fh_0\rightarrow\CAR(\fh_0)$
such that the CAR~\eqref{CAR} hold for any $f,g\in\fh_0$.
\item The set of monomials 
$\{a^\#(f_1)\cdots a^\#(f_n)\,|\, f_1,\ldots,f_n\in\fh_0\}$ 
is total in the algebra $\CAR(\fh_0)$.
\end{enumerate}
\end{theorem}

\remarks\textbf{1.} For $f\not=0$, Equ.~\eqref{CAR} yields that
$a(f)\not=0$ and that $(a^\ast(f)a(f))^2=\|f\|^2a^\ast(f)a(f)$. Thus
\beq
\|a(f)\|=\|f\|
\label{5.1.2.4-CARcont}
\eeq
holds for any $f\in\fh_0$. In particular, the map $a$ is continuous,
extends to the completion $\fh$ of $\fh_0$, and
$\CAR(\fh)=\CAR(\fh_0)$.

\noindent\textbf{2.} By condition {\it (ii)}, the algebra $\CAR(\fh)$ is
separable if and only if the Hilbert space $\fh$ is.

\noindent\textbf{3.} An antilinear map $f\mapsto b(f)$ from $\fh$
to $\cB(\cH)$ satisfying~\eqref{CAR} extends to a faithful
representation of $\CAR(\fh)$ in the Hilbert space $\cH$.

\medskip
Set $\phi(f):=\frac12\big(a^*(f)+a(f)\big)$.
A direct consequence of this theorem is the following
\begin{corollary}
Let $\fh_1,\fh_2$ be two Hilbert spaces. Assume that
$R:\fh_1\to\fh_2$
  is  a bounded invertible real linear map such that
\begin{equation}
  \Re(f|g)=\Re(Rf|Rg).\end{equation}
Then there exists a unique $\ast$-isomorphism 
$\gamma:\CAR(\fh_1)\to\CAR(\fh_2)$ such that
$$
\gamma(\phi(f))=\phi(Rf).
$$
$\gamma$ is called the
  Bogoliubov transformation associated to $R$.

  Equivalently, if we decompose $R$ into its linear and antilinear
  part \footnote{Here, $\i$ denotes multiplication by the imaginary unit.}
  \begin{equation}
    R=P+Q,\quad P:=\frac12(R-\i R\i),\quad Q:=\frac12(R+\i
    R\i), \end{equation} then \eqref{boggo} can be written as
  \begin{eqnarray*}
P^\ast P+Q^\ast Q=\one_{\fh_1},&\quad& PP^\ast+QQ^\ast=\one_{\fh_2},\\
P^\ast Q+Q^\ast P=0_{\hphantom{\fh_1}},&\quad& PQ^\ast+QP^\ast=0.
\end{eqnarray*}
Moreover,
$$
\gamma(a(f))=a(Pf)+a^\ast(Qf).
$$
\end{corollary}

$\gamma$ is called the \textsl{Bogoliubov isomorphism} induced by $R$. In the special case of a complex linear $R$, when $R=U$ is 
  unitary, $\gamma$ is called the gauge-invariant or particle 
  preserving Bogoliubov transformation.

When dealing with compound Fermi systems the following result is
often useful.
\begin{theorem} [Exponential law for fermions] Let  $\fh_1,\fh_2$ be
two Hilbert spaces. There is a unique unitary operator
$U:\aFock(\fh_1\oplus\fh_2)\rightarrow\aFock(\fh_1)\otimes\aFock(\fh_2)$
such that
$$
U\Omega=\Omega \otimes \Omega,\quad
Ua(f_1\oplus f_2)U^\ast=a(f_1)\otimes \one+(-1)^N\otimes a(f_2),
$$
where $\Omega$ denotes the Fock vacuum vector and $N=\d\Gamma(I)$
the number operator.
\label{FermiExp}
\end{theorem}

Note that $N\in\CAR(\fh_1)$ if and only if $\fh_1$ is
finite dimensional. Thus, $\CAR(\fh_1\oplus\fh_2)$ is $\ast$-isomorphic
to the $C^\ast$-tensor product $\CAR(\fh_1)\otimes\CAR(\fh_2)$ if
and only if at least one of the spaces $\fh_1$, $\fh_2$ is finite
dimensional.

%%%%%%%%%%%%%%%%%%%%%%%%%%%%%
\subsubsection{CCR algebras}
\label{ssec:CCR-algebra}
%%%%%%%%%%%%%%%%%%%%%%%%%%%%%

Systems of \ul{bosons} are described in an analogous way by
creation/annihilation ope\-ra\-tors $a^\ast(f)$ and $a(f)$ on the
bosonic (or symmetric) Fock space $\sFock(\fh)$. These operators
satisfy the canonical commutation relations
\beq
 [a(f),a^\ast(g)]=(f|g),\quad [a(f),a(g)]=0,
\label{CCR}
\eeq
for $f,g\in\fh$.  However, dealing with bosonic systems is more
delicate since the operators $a^\ast(f)$ and $a(f)$ are unbounded.
This follows readily from the algebraic structure described by the
CCR. Indeed, suppose that $a(f)$ is bounded. Then, since
$a^\ast(f)a(f)$ is positive, it follows from the CCR that
$\|a(f)a^\ast(f)\|=\|a^\ast(f)a(f)\|+\|f\|^2$, which contradicts the
fact that $\|a(f)a^\ast(f)\|=\|a^\ast(f)a(f)\|=\|a(f)\|^2$.

Thus it is not a priori clear how the CRR should   interpreted without
referring to some domain $\cD\subset\sFock(\fh)$ on which they
are supposed to hold. The operator
$$
\frac{1}{\sqrt2}\left(a^\ast(f)+a(f)\right)
$$
is essentially selfadjoint on the dense subspace
$\Gamma_\mathrm{s,fin}(\fh)$ of finite particle vectors of
$\sFock(\fh)$. Its selfadjoint closure is called \textsl{Segal field operator}
and denoted by $\phi(f)$. Segal field operators satisfy the
commutation relations $[\phi(f),\phi(g)]=\i\,\Im(f|g)$ which are
formally equivalent to~\eqref{CCR}. The unitary operators
$$
 W(f)=\e^{\i{\phi(f)}}
$$
are called \textsl{Weyl operators.} They satisfy the \textsl{Weyl relations}
\beq
 W(f)W(g)=\e^{-\i\,\Im(f,g)/2}W(f+g).
\label{Weyl}
\eeq
Finite linear combinations of Weyl operators build a $\ast$-algebra.
Its closure is a $C^\ast$-algebra which is completely characterized by
the Weyl relations~\eqref{Weyl}.

\begin{theorem}\label{CCRUniq}
Let $\fh_0$ be a pre-Hilbert space. Up to $\ast$-isomorphisms,
there exists a unique unital $C^\ast$-algebra $\CCR(\fh_0)$
with the following properties:
 \begin{enumerate}%[(i)]
 \item There is a map $f\mapsto W(f)$ from $\fh_0$ to $\CCR(\fh_0)$ such that
 $$
 W(-f)=W(f)^\ast,\quad W(0)\not=0,
 $$
 and the Weyl relations~\eqref{Weyl} are satisfied for all $f,g\in\fh_0$.
 \item The set $\{W(f)\,|\,f\in\fh_0\}$ is total in $\CCR(\fh_0)$.
 \end{enumerate}
 \end{theorem}

\remarks\textbf{1.} It follows from~\eqref{Weyl} and conditions 
\textit{(i)-(ii)} that $W(0)=\one$ and that $W(f)^\ast=W(f)^{-1}$, \ie that 
$W(f)$ is unitary. Moreover, if $f\not=g$, then $\|W(f)-W(g)\|=2$.

\noindent\textbf{2.} Unlike in the CAR-case, if $\fh_0\not=\fh_1$, then
 $\CCR(\fh_0)\not=\CCR(\fh_1)$.  Moreover, $\CCR(\fh_0)$ is not
 separable if $\fh_0\not= \{0\}$.

\noindent{\bf 3.} A map $f\mapsto W_\pi(f)$ from $\fh_0$ to the unitary
 operators on $\cH$ satisfying the Weyl relations~\eqref{Weyl} extends
 to a representation $(\cH,\pi)$ of $\CCR(\fh_0)$.

\medskip
Bogoliubov isomorphisms between CCR algebras are defined in a
similar way as in the CAR case.

\begin{corollary} Let $\fh_1$, $\fh_2$ be two pre-Hilbert spaces and
$ R:\fh_1\to\fh_2$ an invertible real-linear map such that
\beq
\mathrm{Im}( Rf|  Rg)=\mathrm{Im}(f|g)\label{boggo}\eeq
Then there is a unique
$\ast$-isomorphism $\gamma:\CCR(\fh_1)\to\CCR(\fh_2)$ such
that $\gamma(W(f))=W( Rf)$. $\gamma$ is called the
  Bogoliubov transformation associated to $R$.

  Equivalently, if we decompose $R$ into the linear and antilinear
  part
  \begin{equation}
    R=P+Q,\quad P:=\frac12(R-\i R\i),\quad Q:=\frac12(R+\i
    R\i), \end{equation} then \eqref{boggo} can be written as
  \begin{eqnarray*}
P^\ast P-Q^\ast Q=\one_{\fh_1},&\quad& PP^\ast-QQ^\ast=\one_{\fh_2},\\
P^\ast Q-Q^\ast P=0_{\hphantom{\fh_1}},&\quad& PQ^\ast-QP^\ast=0.
\end{eqnarray*}
Moreover, (in a representation where we can define creation and
annihilation operators, so that we can extend the Bogoliubov isomorphism
to affiliated unbounded operators)
$$
\gamma(a(f))=a(Pf)+a^\ast(Qf).
$$
\end{corollary}

In the special case of a complex linear $R$, when $R=U$ is
  unitary, $\gamma$ is called the gauge-invariant or particle
  preserving Bogoliubov transformation.

An exponential law similar to Theorem~\ref{FermiExp} holds for bosons.

\begin{theorem}[Exponential law for bosons]
Let $\fh_1$, $\fh_2$ be two Hilbert spaces. There is a unique
unitary operator
$U:\sFock(\fh_1\oplus\fh_2)\rightarrow\sFock(\fh_1)\otimes\sFock(\fh_2)$
such that
$$
U \Omega=\Omega\otimes \Omega,\quad
UW(f_1\oplus f_2)U^\ast= W(f_1)\otimes  W(f_2),
$$
where ${\Omega}$ denotes the Fock vacuum vector.
Moreover, $\CCR(\fh_1\oplus\fh_2)$ is $\ast$-isomorphic to the
\ul{$C^\ast$-tensor product} $\CCR(\fh_1)\otimes\CCR(\fh_2)$. \footnote{Since CCR algebras are nuclear $C^\ast$-algebras, their $C^\ast$-tensor product is unique.}
\end{theorem}
In practice the $C^\ast$-algebra $\CCR(\fh_0)$ of Theorem~\ref{CCRUniq}
is not very convenient and one often prefers to work with von Neumann
algebras when dealing with bosons.
Let $(\cH,\pi)$ be a representation of $\CCR(\fh_0)$.
The von Neumann algebra on $\cH$ generated
by $\{\pi(A)\,|\,A\in\CCR(\fh_0)\}$ is given by the \ul{bicommutant}
$$
\fM_\pi(\fh_0)=\pi(\CCR(\fh_0))''.
$$
It is the enveloping von Neumann algebra of the representation $\pi$,
see Section~\ref{sec:Free Bose and Fermi gases -- the algebraic approach}
for an example.

%%%%%%%%%%%%%%%%%%%%%%%%%%%%%%%%%%%%%%%%%%%%%%%%%%%%%%%%%%%%%%%%%%
\subsubsection{Quasi-local structure of $\fA(\Gamma,\fh)$
 and of $\CAR(\fh)$}
%%%%%%%%%%%%%%%%%%%%%%%%%%%%%%%%%%%%%%%%%%%%%%%%%%%%%%%%%%%%%%%%%%

As already mentioned, in most physical applications the single-fermion Hilbert
space is $\fh=L^2(\RR^d)\otimes\CC^n$ or some straightforward variant of it. To each bounded open subset $\Lambda\subset\RR^d$ a local Hilbert
space $\fh_\Lambda=L^2(\Lambda)\otimes\CC^n$. The canonical isometric injections
$\fh_\Lambda\hookrightarrow\fh$ yield injections
$\CAR(\fh_\Lambda)\hookrightarrow\CAR(\fh)$, allowing us to  identify  the
local algebra $\CAR(\fh_\Lambda)$ with a $C^\ast$-subalgebra of $\CAR(\fh)$. It
follows immediately from remark~1 that
$$
\CAR(\fh)=\left(\bigcup_{\Lambda\subset\RR^d}
\CAR(\fh_\Lambda)\right)^\mathrm{cl},
$$
to be compared with Equ.~\eqref{SpinLocal}. Note, however, that while
$\Lambda\cap\Lambda'=\emptyset$ implies $[\fA_\Lambda,\fA_{\Lambda'}]=\{0\}$, 
the CAR algebras of disjoint subsets do not commute. This
is of course due to the fact that $a(f)$ and $a(g)$  anticommute. Let $\theta$
be the $\ast$-automorphism defined by $\theta(a(f))=-a(f)$ and denote by
$$
\CAR_\pm(\fh)=\{A\in\CAR(\fh)\,|\,\theta(A)=\pm A\},
$$
the even and odd parts of $\CAR(\fh)$ with respect to $\theta$. Alternatively,
these sets are the closed linear spans of monomials of even and odd degree in the
$a^\#$. Then one has
$\CAR(\fh_\Lambda)=\CAR_+(\fh_\Lambda)\oplus\CAR_-(\fh_\Lambda)$, and one easily
checks that
$$
[\CAR_\pm(\fh_\Lambda),\CAR_\pm(\fh_{\Lambda'})]=\{0\},\quad
[\CAR_\pm(\fh_\Lambda),\CAR_\mp(\fh_{\Lambda'})]_+=\{0\}.
$$
From a physical point of view,  observables localized in disjoint regions of
space should be simultaneously measurable. Hence physical observables of a
fermionic system should be elements of the even subalgebra $\CAR_+(\fh)$. In
fact, the stronger requirement of gauge-invariance further reduces the
observable algebra to the subalgebra of $\CAR_+(\fh)$ generated by monomials in
the $a^\#$ containing the same number of $a$ and $a^\ast$; see 
Section~\ref{sec:Araki-Wyss representation} for a discussion of this point. In 
both the UHF-algebra $\fA$ and the CAR-algebra $\CAR(\fh)$, the family of  local 
subalgebras define the   so-called quasi-local structure. We refer the reader 
to~\cite[Section~2.6]{Bratteli1987} for a general discussion.

%%%%%%%%%%%%%%%%%%%%%%%%%%%%%%%%%%%%%%%%%%%%%
\subsection{States and the GNS construction}
\label{ssec:States and the GNS construction}
%%%%%%%%%%%%%%%%%%%%%%%%%%%%%%%%%%%%%%%%%%%%%

Let $\cO$ be a $C^\ast$-algebra. A linear functional $\varphi$ on $\cO$ is
positive if $\varphi(A^\ast A)\ge0$ for all $A\in\cO$. A positive linear
functional is automatically bounded, \ie is an element of the dual
$\cO^{\scriptscriptstyle\#}$. It is called a state if $\|\varphi\|=1$. If $\cO$
has a unit $\one$ then $\|\varphi\|=\varphi(\one)$ for any positive linear
functional $\varphi$.

A representation of $\cO$ on a Hilbert space $\cH$ is a $\ast$-morphism
$\pi:\cO\to\cB(\cH)$. Given such a representation and a unit vector
$\Omega\in\cH$, the formula $\varphi(A)=(\Omega|\pi(A)\Omega)$ defines a state
on $\cO$. The GNS construction shows that any state on $\cO$ is of this form.

\begin{theorem} Let $\omega$ be a state on the $C^\ast$-algebra
$\cO$. Then there exist a Hilbert space $\cH_\omega$, a representation
$\pi_\omega$ of $\cO$ in $\cH_\omega$ and a unit vector
$\Omega_\omega\in\cH_\omega$ such that
\begin{enumerate}%[(i)]
\item $\omega(A)=(\Omega_\omega|\pi_\omega(A)\Omega_\omega)$ for all
  $A\in\cO$.
\item $\pi_\omega(\cO)\Omega_\omega$ is dense in $\cH_\omega$.
\end{enumerate}
The triple $(\cH_\omega,\pi_\omega,\Omega_\omega)$ is unique
up to unitary equivalence. It is called the GNS representation or
the cyclic representation of $\cO$ induced by the state $\omega$.
\end{theorem}

An important object associated with the GNS representation is the
\textsl{enveloping von~Neumann algebra:} The $\sigma$-weak closure $\cO_\omega$
of $\pi_\omega(\cO)$ in $\cB(\cH_\omega)$. By von Neumann's
\ul{bicommutant theorem}, it is given by the bicommutant
$$
\cO_\omega=\pi_\omega(\cO)''.
$$
We note that if $\cO$ is itself a von Neumann or a $W^\ast$-algebra and $\omega$
is $\sigma$-weakly continuous (\ie a \ul{normal state}) then $\pi_\omega$ is
$\sigma$-weakly continuous and $\cO_\omega=\pi_\omega(\cO)$.

Two states $\omega$, $\nu$ on $\cO$ are quasi-equivalent if there exists a
$\ast$-isomorphism $\phi:\cO_\omega\to\cO_\nu$ such that
$\pi_\nu=\phi\circ\pi_\omega$.

The folium of a state $\omega$ is the set $\cN_\omega$ of all states of
the form $\nu(A)=\tr (\rho\pi_\omega(A))$ for some density matrix $\rho$ on
$\cH_\omega$. A state $\nu\in\cN_\omega$ is said to be $\omega$-normal.
Thus, $\omega$-normal states on $\cO$ are characterized by the fact that
they extend to normal states on the enveloping von Neumann algebra $\cO_\omega$.

\begin{theorem} The following propositions are equivalent.
\begin{enumerate}%[(i)]
\item $\mu$ and $\nu$ are quasi-equivalent.
\item $\mu$ and $\nu$ have the same folium.
\item There exist Hilbert spaces $\cK_\mu$ and $\cK_\nu$ 
and a unitary map $V:\cH_\mu\otimes\cK_\mu\to\cH_\nu\otimes\cK_\nu$ such that
$$
\pi_\nu\otimes \one=V(\pi_\mu\otimes \one)V^\ast.
$$
\end{enumerate}
\end{theorem}
The reader should consult~\cite{Bratteli1987} for a more detailed discussion.

%%%%%%%%%%%%%%%%%%%%%%%%%%%%%%%%%%%%%%%
\section{Quantum dynamical systems}
\label{sec:Quantum dynamical systems}
%%%%%%%%%%%%%%%%%%%%%%%%%%%%%%%%%%%%%%%

In the most common approach to quantum physics,  observables are described by 
bounded operators on a  Hilbert space. This formalism is usually sufficient at zero temperature. To describe quantum systems in the thermodynamic limit
at positive temperatures, or more generally, at positive densities, it is
convenient  to use a more sophisticated formalism where observables are
described by elements of some operator algebra, see Sections~\ref{sec:The
C*-algebra approach} and~\ref{sec:Free Bose and Fermi gases -- the algebraic
approach}. In this algebraic approach, quantum dynamics in the Heisenberg
picture is given by a one-parameter group of automorphisms of the algebra of
observables. In analogy with the theory of classical dynamical systems, such a group defines a quantum dynamical system. There are two main versions of
the algebraic approach, which  differ in the topological properties of the algebra
and of the dynamical group: the $C^\ast$- and the $W^\ast$-approach.

%%%%%%%%%%%%%%%%%%%%%%%%%%%%%%%%%%%%%%%%%
\subsection{$C^\ast$-dynamical systems}
%%%%%%%%%%%%%%%%%%%%%%%%%%%%%%%%%%%%%%%%%

\begin{definition}
A $C^\ast$-dynamics on a \ul{$C^\ast$-algebra} $\cO$ is a strongly
continuous one-parameter group $\RR\ni t\to\tau^t$ of \ul{$\ast$-automorphisms} of $\cO$. A $C^\ast$-dynamical system is a pair $(\cO,\tau)$,
where $\cO$ is a $C^\ast$-algebra and $\tau$ a $C^\ast$-dynamics on
$\cO$.
\end{definition}

The strong continuity of $\tau$ means that the map $t\mapsto\tau^t(A)$ is
norm-continuous for any $A\in\cO$. From the general theory of
\ul{strongly continuous groups} on a Banach space, a $C^\ast$-dynamics $\tau$
admits  a densely defined, closed \ul{infinitesimal generator} $\delta$ such that
\beq
\delta(A)=\lim_{t\to0}\frac{\tau^t(A)-A}{t}
\label{deltadef}
\eeq
for $A\in\Dom(\delta)$. In particular, if $\cO$ has a unit $\one$, then
$\one\in\Dom(\delta)$ and $\delta(\one)=0$. It follows immediately that $\delta$ is a
$\ast$-derivation, \ie that
\begin{enumerate}%[(i)]
\item $\Dom(\delta)$ is a $\ast$-algebra.
\item $\delta(AB)=\delta(A)B+A\delta(B)$ for all $A,B\in\Dom(\delta)$.
\item $\delta(A^\ast)=\delta(A)^\ast$ for all $A\in\Dom(\delta)$.
\end{enumerate}
Generators of $C^\ast$-dynamics are characterized by the following
simple adaptation of the \ul{Hille-Yosida Theorem},
see~\cite[Section~3.2]{Bratteli1987}).

\begin{theorem} Let $\cO$ be a $C^\ast$-algebra.
A norm densely defined and closed operator $\delta$ on $\cO$
generates a $C^\ast$-dynamics if and only if
\begin{enumerate}%[(i)]
\item $\delta$ is a $\ast$-derivation.
\item $\Ran(\Id+\lambda\delta)=\cO$ for
all $\lambda\in\RR$.
\item $\|A+\lambda\delta(A)\|\ge\|A\|$ for all $\lambda\in\RR$ and
$A\in\Dom(\delta)$.
\end{enumerate}
\label{Cstargen}
\end{theorem}

\noindent\textit{Example 1.} Let $\cH$ be a Hilbert space and $H$ a bounded
selfadjoint operator on $\cH$. Then $\tau^t(A)=\e^{\i tH}A\e^{-\i tH}$ is a
$C^\ast$-dynamics on $\cB(\cH)$. Its generator $\delta(A)=\i[H,A]$ is bounded.
Note that boundedness of $H$ is required for  the strong continuity of $t\to\tau^t$.

\noindent\textit{Example 2.} The following is based on the material of
Section~\ref{ssec:CAR-algebra}. Let $\fh$ be a Hilbert space and $h$ a
selfadjoint operator on $\fh$. The group of gauge invariant Bogoliubov
automorphisms of the $C^\ast$-algebra $\CAR(\fh)$ defined by
$\tau^t(a(f))=a(\e^{\i th}f)$ is a $C^\ast$-dynamics. This is a consequence of
the strong continuity of the unitary group $t\mapsto\e^{\i th}$ and of the norm
continuity of the map $f\mapsto a(f)$ from $\fh$ to $\CAR(\fh)$. The
$\ast$-subalgebra generated by $\{a(f)\,|\,f\in\Dom(h)\}$ is contained in the
domain of the generator $\delta$ of $\tau$ and $\delta(a(f))=a(\i hf)$.

%%%%%%%%%%%%%%%%%%%%%%%%%%%%%%%%%%%%%%%%
\subsection{$W^\ast$-dynamical systems}
\label{ssec:W*-dynamical systems}
%%%%%%%%%%%%%%%%%%%%%%%%%%%%%%%%%%%%%%%%

In some cases, such as  systems of bosons, the $C^\ast$-approach is
not adequate and one has to use the $W^\ast$ setting.

\begin{definition}
Let $\fM$ be a \ul{von Neumann algebra} or a \ul{$W^\ast$-algebra}. A
$W^\ast$-dynamics on $\fM$ is a $\sigma$-weakly continuous group $\RR\ni
t\mapsto\tau^t$ of $\ast$-automorphisms of $\fM$. A $W^\ast$-dynamical system is
a pair $(\fM,\tau)$, where $\fM$ is a von Neumann algebra and $\tau$ a
$W^\ast$-dynamics on $\fM$.
\end{definition}

The continuity condition on the group $\tau$ means that, for any $A\in\fM$, the
map $t\mapsto\tau^t(A)$ is continuous in the \ul{$\sigma$-weak topology} of
$\fM$. The generator $\delta$ of a $W^\ast$-dynamics $\tau$ on a von Neumann
algebra $\fM$ is defined by Equ.~\eqref{deltadef}, as in the $C^\ast$-case,
except that the limit is now understood in the $\sigma$-weak topology. It is a
$\sigma$-weakly densely defined and closed $\ast$-derivation on $\fM$ such that
$\one\in\Dom(\delta)$ and $\delta(\one)=0$. Generators of $W^\ast$-dynamics are
characterized by the following analog of Theorem~\ref{Cstargen},
see~\cite[Section~5.2]{Bratteli1987}.

\begin{theorem} Let $\fM$ be a von Neumann algebra.
A $\sigma$-weakly densely defined and closed operator $\delta$ on
 $\fM$ generates a $W^\ast$-dynamics if and only if
\begin{enumerate}%[(i)]
\item $\delta$ is a $\ast$-derivation and $\one\in\Dom(\delta)$.
\item $\Ran(\Id+\lambda\delta)=\fM$ for
all $\lambda\in\RR$.
\item $\|A+\lambda\delta(A)\|\ge\|A\|$ for all $\lambda\in\RR$ and
$A\in\Dom(\delta)$.
\end{enumerate}
\end{theorem}

\noindent\textit{Example 3.} Let $\cH$ be a Hilbert space and $H$ a selfadjoint
operator on $\cH$. Then $\tau^t(A)=\e^{\i tH}A\e^{-\i tH}$ is a
$W^\ast$-dynamics on $\cB(\cH)$.

\noindent{\it Example 4.} The following is based on the material of
Section~\ref{ssec:CCR-algebra}. Let $\fh$ be a Hilbert space and $h$ a
selfadjoint operator on $\fh$. Denote by $\fh_0$ a subspace of $\fh$ invariant
under the unitary group $\e^{\i th}$. The following formula
  defines a group of gauge-invariant automorphisms of the $C^\ast$-algebra $\CCR(\fh_0)$:
$$
\tau^t(W(f))=W(\e^{\i th}f). 
$$
Except in trivial cases, this group fails to be  a $C^\ast$-dynamics because $\|\tau^t(W(f))-W(f)\|=2$ 
for all $t\in\RR$ and $f\in\fh_0$ such that $\e^{\i th}f\not=f$.

Denote by $\fM$ the von Neumann algebra acting on the bosonic Fock space
$\sFock(\fh)$ and generated by the \ul{Weyl operators} $\{W(f)\,|\,f\in\fh_0\}$.
Then $\tau$ has an extension to $\fM$ given by
$$
\tau^t(A)=\e^{\i t\d\Gamma(h)}A\e^{-\i t\d\Gamma(h)}. 
$$
It defines a $W^\ast$-dynamics on $\fM$.

%%%%%%%%%%%%%%%%%%%%%%%%%%%%%%%%%%%%%%%%%%%%%%%
\subsection{Invariant states and Liouvilleans}
\label{ssec:Liouvilleans}
%%%%%%%%%%%%%%%%%%%%%%%%%%%%%%%%%%%%%%%%%%%%%%%

In this and the following sections we use freely the notation introduced 
in  Section~\ref{ssec:States and the GNS construction}. We shall say
that $(\cO,\tau)$ is a \textsl{quantum dynamical system} if it is either
a $C^\ast$- or a $W^\ast$-dynamical system.

\begin{definition}
Let $(\cO,\tau)$ be a quantum dynamical system. A \ul{state}
$\omega$ on $\cO$ is $\tau$-invariant if $\omega\circ\tau^t=\omega$
holds for all $t\in\RR$.
\end{definition}

As in the theory of classical dynamical systems, invariant states (and more
specifically normal invariant states in the $W^\ast$-case) play an important
role in the analysis of quantum dynamical systems. As an illustration, we
explore the \ul{GNS representation} induced by an invariant state.

Let $\omega$ be an invariant state of the quantum dynamical system $(\cO,\tau)$
and $(\cH_\omega,\pi_\omega,\Omega_\omega)$ the GNS representation of $\cO$
induced by $\omega$. The uniqueness of the GNS representation implies that there
exists a unique one-parameter group $t\mapsto U_\omega(t)$ of unitary operators
on $\cH_\omega$ such that
$$
\pi_\omega(\tau^t(A))=U_\omega(t)\pi_\omega(A)U_\omega(t)^\ast,\quad
U_\omega(t)\Omega_\omega=\Omega_\omega,
$$
for any $t\in\RR$ and $A\in\cO$. Assuming $\omega$ to be normal in the
$W^\ast$-case, it is easy to show that the group $U_\omega$ is
strongly continuous. Hence, by Stone's theorem, there exists a
unique selfadjoint operator $L_\omega$ such that
$$
\e^{\i tL_\omega}\pi_\omega(A)\e^{-\i tL_\omega}
=\pi_\omega(\tau^t(A)),
\quad L_\omega\Omega_\omega=0.
$$
The operator $L_\omega$ is sometimes called the $\omega$-Liouvillean
of $(\cO,\tau)$. Important information about  the dynamics of the system
can be deduced from its spectral properties, see Sections~\ref{sec:Tomita--Takesaki theory},
\ref{sec:Quantum Koopmanism}
and~\ref{sec:Spectral analysis of small quantum systems interacting with a reservoir}.

Next we note that
$$
\tilde\omega(A)=(\Omega_\omega|A\Omega_\omega),\quad{ A\in\cO_\omega,}
$$
defines a normal extension of the state $\omega$ to the \ul{enveloping von Neumann 
algebra} $\cO_\omega$: $\omega=\tilde\omega\circ\pi_\omega$. Similarly,
$$
\tilde\tau^t(A)=\e^{\i tL_\omega}A\e^{-\i tL_\omega}
$$
defines a $W^\ast$-dynamics on $\cO_\omega$ such that
$\tilde\tau^t\circ\pi_\omega=\pi_\omega\circ\tau^t$. Thus,  the GNS
construction maps a $C^\ast$-dynamical system $(\cO,\tau)$ with invariant state
$\omega$ into a $W^\ast$-dynamical system $(\cO_\omega,\tilde\tau)$ with normal
invariant state $\tilde\omega$.

The above construction can be performed under weaker continuity conditions than
the strong/$\sigma$-weak continuity used here; see~\cite{Pillet2006} for a more
general definition of quantum dynamical systems.

%%%%%%%%%%%%%%%%%%%%%%%%%%%%%%%%%%%%%
\subsection{Perturbation theory}
\label{ssec:Perturbation theory}
%%%%%%%%%%%%%%%%%%%%%%%%%%%%%%%%%%%%%

Let $(\cO,\tau)$ be a $C^\ast$-dynamical system and $V\in\cO$ a selfadjoint
element. If $\delta$ denotes the generator of $\tau$, then
$$
\delta_V=\delta+\i[V,\argdot]
$$
is well defined on the domain $\Dom(\delta)$ and generates
a perturbed dynamics $\tau_V^t=\e^{t\delta_V}$
on $\cO$. One says that
$\tau_V$ is a local perturbation of $\tau$.
Such local perturbations play an important role
in the theory of $C^\ast$-dynamical systems.

Iterating the integral equation (Duhamel formula)
$$
\tau_V^t(A)=\tau^t(A)+\int_0^t\tau^{t-s}(\i[V,\tau_V^s(A)])\,\d s
$$
leads to the Araki-Dyson expansion
$$
\tau_V^t(A)=\tau^t(A)+\sum_{n=1}^\infty
\int_0^t\!\d t_1\int_0^{t_1}\!\d t_2\cdots\int_0^{t_{n-1}}\!\d t_n\,
\i[\tau^{t_n}(V),\i[\cdots,\i[\tau^{t_1}(V),\tau^t(A)]\cdots]],
$$
which is norm convergent for any $t\in\RR$ and $A\in\cO$.
Another useful representation of the locally perturbed dynamics
is given by the interaction picture
$\tau_V^t(A)=\Gamma_V^t\tau^t(A)\Gamma_V^{t\ast}$.
The operator $\Gamma_V^t$ is the solution of the Cauchy
problem
$$
\partial_t\Gamma_V^t=\i\Gamma_V^t\tau^t(V),\quad \Gamma_V^0=\one.
$$
It follows that $\Gamma_V^t\in\cO$ is unitary and has the norm
convergent Dyson expansion
$$
\Gamma_V^t=\one+\sum_{n=1}^\infty\i^n
\int_0^t\!\d t_1\int_0^{t_1}\!\d t_2\cdots\int_0^{t_{n-1}}\!\d t_n\,
\tau^{t_n}(V)\cdots\tau^{t_1}(V),
$$
which satisfies the cocycle relation
$$
\Gamma_V^{t+s}=\Gamma_V^t\tau^t(\Gamma_V^s)=
\tau_V^t(\Gamma_V^s)\Gamma_V^t.
$$
Local perturbations of $W^\ast$-dynamical systems can be
handled in a similar way, replacing the norm topology with
the $\sigma$-weak topology and interpreting all integrals
in the weak-$\ast$ sense.

If $\omega$ is an {invariant state} for the unperturbed dynamical
system $(\cO,\tau)$ (supposed to be normal in the $W^\ast$-case)
then, in the induced GNS representation and with the
notation of the previous subsection,
the perturbed dynamics is implemented by the unitary group generated by
$L_\omega+Q$ where $Q=\pi_\omega(V)$,
\begin{equation*}
\pi_\omega(\tau_V^t(A))=\e^{\i t(L_{\omega}+Q)}
\pi_\omega(A)\e^{-\i t(L_{\omega}+Q)}.
\end{equation*}
Note that the perturbed unitary group is related to the
cocycle $\Gamma_V$ through the interaction picture formula
$$
\e^{\i t(L_{\omega}+Q)}=\widetilde\Gamma_V^t\,\e^{\i tL_{\omega}},\quad
\widetilde\Gamma_V^t=\pi_\omega(\Gamma_V^t)=
\one+\sum_{n=1}^\infty\i^n
\int_0^t\!\d t_1\int_0^{t_1}\!\d t_2\cdots\int_0^{t_{n-1}}\!\d t_n\,
\tilde\tau^{t_n}(Q)\cdots\tilde\tau^{t_1}(Q).
$$
Consequently, the perturbed dynamics extends to a $W^\ast$-dynamics
$$
\tilde\tau_V^t(A)=\e^{\i t(L_{\omega}+Q)}
A\e^{-\i t(L_{\omega}+Q)}=
\widetilde\Gamma_V^t\tilde\tau^t(A)\widetilde\Gamma_V^{t\ast},
$$
on the enveloping von Neumann algebra $\cO_\omega$. In the $W^\ast$-case
this formula is the starting point for an extension of
perturbation theory to unbounded perturbations.
If $Q$ is a selfadjoint operator on $\cH_\omega$ affiliated
to $\cO_\omega=\pi_\omega(\cO)$ and such that $L_\omega+Q$ is essentially
selfadjoint on $\Dom(L_\omega)\cap\Dom(Q)$,
then the unitary group $\e^{\i t(L_\omega+Q)}$ defines
a $W^\ast$-dynamics on $\cO_\omega$. This extension
of perturbation theory has been developed in~\cite{Derezinski2003a}.

Except for its important role in Araki's perturbation theory of
KMS-states discussed in Section~\ref{sec:KMS--States}, the operator
$L_{\omega}+Q$ is of little value in the study of dynamical properties
of $\tau_V$.  This is due to the fact that it is not adapted to the
structure of the enveloping von Neumann algebra $\cO_\omega$.  The
standard Liouvillean, introduced in Section~\ref{sec:Tomita--Takesaki theory},
corrects this problem.

If $t\mapsto V(t)=V(t)^\ast\in\cO$ is continuous, then equation
$\partial_t\tau_V^{s\to t}(A)=\tau_V^{s\to t}(\delta_{V(t)}(A))$
together with the condition $\tau_V^{s\to s}=\Id$ defines a
two-parameter family of $\ast$-automorphisms of $\cO$ such that
$\tau_V^{s\to t}\circ\tau_V^{t\to r} =\tau_V^{s\to r}$. Perturbation
theory can be developed as in the time-independent case starting from
the integral equation
$$
\tau_V^{s\to t}(A)=\tau^{t-s}(A)+\int_s^t\tau_V^{s\to u}\left(
\i[V(u),\tau^{t-u}(A)]\right)\,\d u.
$$

%%%%%%%%%%%%%%%%%%%%%%%%
\section{KMS--States}
\label{sec:KMS--States}
%%%%%%%%%%%%%%%%%%%%%%%%

The KMS condition plays a fundamental role in quantum statistical mechanics
where it provides a general abstract definition of equilibrium state. It is also
deeply rooted in the mathematical structure of von Neumann algebras, see
Section~\ref{sec:Tomita--Takesaki theory}. Consequently, there is an enormous literature on the
subject and the following section only provides a crude and  condensed
introduction. We refer the reader to~\cite[Chapters 5.3+5.4]{Bratteli1981}
and~\cite[Chapter~5]{Haag1996} for a more elaborate introduction. We also
recommend reading the pioneering article~\cite{Haag1967}.

%%%%%%%%%%%%%%%%%%%%%%%%%%%%%%%%%%%%%%%%
\subsection{Motivation and definition}
%%%%%%%%%%%%%%%%%%%%%%%%%%%%%%%%%%%%%%%%

Consider a quantum system with finite dimensional Hilbert space $\cH$ (e.g. an
$N$-level atom). Such a system is described by a \ul{$C^\ast$-dynamical system}
$(\cB(\cH),\tau)$ where
$$
\tau^t(A)=\e^{\i tH}A\e^{-\i tH},
$$
and $H=H^\ast$ denotes the Hamiltonian. For any $\beta\in\RR$ this system has a
unique thermal equilibrium state $\omega_\beta$ at inverse temperature $\beta$
given by the Gibbs-Boltzmann prescription
$$
\omega_\beta(A)=\frac{\tr(\e^{-\beta H}A)}{\tr(\e^{-\beta H})}.
$$
Note that the equilibrium correlation function
\beq
F_\beta(A,B;t)=\omega_\beta(A\tau^t(B))
\label{KMSone}
\eeq
is an entire function of $t$.
The cyclicity of the trace yields the identity
$$
\tr(\e^{-\beta H}A\tau^t(B))=
\tr(\e^{-\beta H}A\e^{\i tH}B\e^{-\i tH})=
\tr(\e^{-\i(t-\i\beta) H}A\e^{\i tH}B).
$$
Analytic continuation from $t\in\RR$ to $t\in\RR+\i\beta$ further gives
$$
\tr(\e^{-\beta H}A\tau^{t+\i\beta}(B))=
\tr(\e^{-\i t H}A\e^{\i (t+\i\beta)H}B)=
\tr(\e^{-\beta H}\tau^t(B)A),
$$
from which we conclude that
\beq
F_\beta(A,B;t+\i\beta)=\omega_\beta(\tau^t(B)A).
\label{KMStwo}
\eeq
Relations~\eqref{KMSone} and~\eqref{KMStwo} relate the
values of the analytic function $F_\beta(A,B;z)$ on the boundary
of the strip
$$
S_\beta=\{z\in\CC\,|\,0<\Im(z\,\mathrm{sign}\beta)<|\beta|\}
$$
to the state $\omega_\beta$.
They are called Kubo-Martin-Schwinger (KMS) boundary conditions.
It is a simple exercise in linear algebra to show that the Gibbs state
$\omega_\beta$ is the only state on $\cB(\cH)$ satisfying
the KMS boundary conditions~\eqref{KMSone} and~\eqref{KMStwo}
for all $A,B\in\cB(\cH)$. This fact motivates the following
general definition.

\begin{definition}
Let $(\cO,\tau)$ be a $C^\ast$- or $W^\ast$-dynamical
system. A state $\omega$ on $\cO$, assumed  to be
normal in the $W^\ast$-case, is $(\tau,\beta)$-KMS for some
$\beta\in\RR$ if the following holds: for any $A,B\in\cO$
there exists a function $z\mapsto F_\beta(A,B;z)$ analytic in the strip
$S_\beta$, continuous on its closure, and satisfying the
Kubo-Martin-Schwinger conditions~\eqref{KMSone}--\eqref{KMStwo}
on its boundary.
\label{KMSdef}
\end{definition}

\remarks\textbf{1.} KMS states for negative temperatures  rarely 
have a physical meaning. However, for historical reasons, they are
widely used in the mathematical literature. For example,
any modular state on a von Neumann algebra is a KMS
state at inverse temperature $\beta=-1$ for its modular group,
see Section~\ref{sec:Tomita--Takesaki theory}.

\noindent\textbf{2.} In the special case $\beta=0$ the KMS conditions degenerate
to $\omega(AB)=\omega(BA)$. In mathematics such states are called {\it tracial.}
Physicists sometimes refer to these infinite-temperature equilibrium states as
{\it stochastic}.

\noindent\textbf{3.} If $\omega$ is $(\tau,\beta)$-KMS, then it is also
$(\tau^{\gamma t},\beta/\gamma)$-KMS. Note however that there is no simple
connection between KMS states at different temperatures for the same dynamics
$\tau^t$.

\noindent\textbf{4.} If $\omega$ is a $\beta$-KMS state for the $C^\ast$-dynamical system $(\cO,\tau)$, then its normal extension $\tilde\omega$
to the enveloping von Neumann algebra $\cO_\omega$ is a $\beta$-KMS state for
the induced $W^\ast$-dynamical system $(\cO_\omega,\tilde\tau)$ defined in
Section~\ref{ssec:W*-dynamical systems}.

%%%%%%%%%%%%%%%%%%%%%%%%%%%%%%%%
\subsection{Characterizations}
%%%%%%%%%%%%%%%%%%%%%%%%%%%%%%%%

Let $(\cO,\tau)$ be a $C^\ast$- or $W^\ast$-dynamical system. An element
$A\in\cO$ is called $\tau$-analytic if the function $t\mapsto\tau^t(A)$ extends
to an entire function on $\CC$. The set $\cO_\tau$ of $\tau$-analytic
elements is a dense $\ast$-subalgebra of $\cO$ in the appropriate topology
(uniform topology in the $C^\ast$-case, $\sigma$-weak topology in the $W^\ast$-case). If $\omega$
is a $(\tau,\beta)$-KMS state and $A,B\in\cO_\tau$, then the function
$F_\beta(A,B;\argdot)$ of Definition~\ref{KMSdef} is given by
$F_\beta(A,B;z)=\omega(A\tau^z(B))$. In particular one has
$\omega(A\tau^{\i\beta}(B))=\omega(BA)$. The following theorem  shows that this
property characterizes  KMS states.

\begin{theorem}\label{thm:KMScharone}
Let $(\cO,\tau)$ be a $C^\ast$- or $W^\ast$-dynamical system. A state $\omega$
on $\cO$, assumed  to be normal in the $W^\ast$-case, is a $(\tau,\beta)$-KMS state 
for some $\beta\in\RR$ if and only if the following holds. There exists a dense,
$\tau$-invariant $\ast$-subalgebra $\cM\subset\cO$ of $\tau$-analytic elements
such that $\omega(A\tau^{\i\beta}(B))=\omega(BA)$ for all $A,B\in\cM$.
\end{theorem}

A fundamental  property of KMS states -- namely the invariance under
time evolution -- is a simple corollary of Theorem~\ref{thm:KMScharone}.
\begin{theorem}
If $\omega$ is a $(\tau,\beta)$-KMS state then it is
$\tau$-invariant, \ie $\omega\circ\tau^t=\omega$
holds for all $t\in\RR$.
\end{theorem}

Remarkably, Definition~\ref{KMSdef} which involves
global properties of the dynamics $\tau$ can be rephrased
in terms of its infinitesimal generator. To formulate
this result let us set
$$
s(x,y)=\left\{\begin{array}{ll}
x(\log x-\log y)&x,y>0,\\
0&x=0,y\ge0,\\
+\infty&x>0,y=0.
\end{array}
\right.
$$

\begin{theorem}[Araki~\cite{Araki1978b}]\label{thm:ArakiPerturb}
Let $(\cO,\tau)$ be a $C^\ast$-dynamical system and
denote by $\delta$ the infinitesimal generator of $\tau$.
A state $\omega$ is $(\tau,\beta)$-KMS if and only if
it is $\tau$-invariant and satisfies the so-called
differential $\beta$-KMS condition
\beq
-\i\beta\omega(A^\ast\delta(A))\ge
s(\omega(A^\ast A),\omega(AA^\ast))
\label{dKMS}
\eeq
for all $A\in\Dom(\delta)$.
\end{theorem}

KMS states satisfy various correlation inequalities of the
type~\eqref{dKMS}, and some of them completely characterize them,
see~\cite{Bratteli1981,Fannes1977}. They play an important
role in proving other characterizations of KMS states:
Araki's {\it Gibbs condition} (a quantum substitute for the
DLR equation) and the variational principle for lattice spin
systems~\cite[Chapter~6.2]{Bratteli1981} and lattice
fermions~\cite{Araki2003}; see also Section~\ref{sec-lattice-eq}.

%%%%%%%%%%%%%%%%%%%%%%%%%%%%%%%%%%
\subsection{Perturbation theory}
%%%%%%%%%%%%%%%%%%%%%%%%%%%%%%%%%%

As thermodynamic equilibrium states, KMS states enjoy several stability
properties. Here we discuss only one of them: structural stability with respect
to local perturbations of the dynamics. For additional information, the reader
may consult~\cite[Section~5.4]{Bratteli1981}.

Let $(\cO,\tau)$ be a $C^\ast$- or $W^\ast$-dynamical system and $\omega$
a $(\tau,\beta)$-KMS state. Consider the local perturbation $\tau_V$ of $\tau$
by a selfadjoint element $V\in\cO$.
In the GNS representation $(\cH_\omega,\pi_\omega,\Omega_\omega)$ the perturbed
dynamics is unitarily implemented by
$$
\pi_\omega(\tau_V^t(A))=\e^{\i t(L_\omega+Q)}\pi_\omega(A)
\e^{-\i t(L_\omega+Q)},
$$
where $L_\omega$ is the $\omega$-Liouvillean and $Q=\pi_\omega(V)$.

\begin{theorem}[Araki~\cite{Araki1973/74}]
The cyclic vector $\Omega_\omega$ belongs to the domain of
$\e^{-\beta(L_\omega+Q)/2}$ and
$$
\omega^V(A)=(\Psi_V|\pi_\omega(A)\Psi_V),\qquad
\Psi_V=\frac{\e^{-\beta(L_\omega+Q)/2}\Omega_\omega}
{\|\e^{-\beta(L_\omega+Q)/2}\Omega_\omega\|},
$$
is a $(\tau_V,\beta)$-KMS state. The GNS representation of $\cO$ induced by
$\omega^V$ is $(\cH_\omega,\pi_\omega,\Psi_V)$. Moreover, the map $\omega\mapsto\omega^V$
is a bijection between the set of $(\tau,\beta)$-KMS states and the set
of $(\tau_V,\beta)$-KMS states.
\end{theorem}

In the $W^\ast$-case, Araki's theorem extends to perturbed dynamics generated by
unbounded perturbations $Q$ affiliated to $\pi_\omega(\cO)$,
see~\cite{Derezinski2003a}.

%%%%%%%%%%%%%%%%%%%%%%%%%%%%%%%%%%%%%
\section{Tomita--Takesaki theory}
\label{sec:Tomita--Takesaki theory}
%%%%%%%%%%%%%%%%%%%%%%%%%%%%%%%%%%%%%

%%%%%%%%%%%%%%%%%%%%%%%%%%%%
\subsection{Modular states}
%%%%%%%%%%%%%%%%%%%%%%%%%%%%

Let $\fM$ be a \ul{von Neumann algebra} acting on a
Hilbert space $\cH$. Its \ul{commutant}
$$
\fM'=\{A\in\cB(\cH)\,|\,AB=BA\text{ for all }B\in\fM\}
$$
is also a von Neumann algebra.  Von Neumann's
\ul{bicommutant theorem} states that $\fM''=\fM$.

\begin{definition}A vector $\Psi\in\cH$ is cyclic for $\fM$ if the subspace
$\fM\Psi$ is dense in $\cH$. It is separating for $\fM$ if $A\Psi=0$ for some
$A\in\fM$ implies $A=0$. It is modular if it is both cyclic and separating for
$\fM$.
\end{definition}

A vector $\Psi\in\cH$ is separating for $\fM$ if and
only if the corresponding normal state $\omega_\Psi(A)=
(\Psi|A\Psi)$ is faithful.

The support $s_\omega$ of a \ul{normal state} $\omega$ on $\fM$ is the
smallest orthogonal projection $P\in\fM$ such that $\omega(P)=1$. It
follows that $\omega(A^\ast A)=0$ if and only if $As_\omega=0$.  In
particular, $\omega$ is faithful if and only if $s_\omega=\one$.

\begin{lemma}
The support of the vector
state $\omega_\Psi$ is the orthogonal projection
on the closure of $\,\fM'\Psi$. Consequently,
a vector $\Psi\in\cH$ is separating for $\fM$
if and only if it is cyclic for $\fM'$.
\end{lemma}

\remark Since $\fM''=\fM$, it follows that
$\Psi$ is cyclic for $\fM$ if and only if it is separating
for $\fM'$.

\medskip
Let $\cO$ be a $C^\ast$-algebra and $\omega$ a state
on $\cO$. Denote by
 $(\cH_\omega,\pi_\omega,\Omega_\omega)$
the GNS representation of $\cO$ induced
by $\omega$.

\begin{definition}
The state $\omega$ is modular if the vector $\Omega_\omega$
is modular for the enveloping von Neumann algebra
$\cO_\omega=\pi_\omega(\cO)''$.
\end{definition}

We warn the reader that this definition of a modular state is 
more general than the one we will adopt in Section~\ref{sec-eqsm-gen}.

Note that the state $\omega$ is modular if and only if
the vector state induced by $\Omega_\omega$ is
faithful on $\cO_\omega$. This does not imply, nor
is it implied by the faithfulness of $\omega$.
However, if $\cO$ is a von Neumann algebra, then
a faithful normal state $\omega$ is modular.

The following result links modular theory with
the theory of KMS states. It is often useful
in applications to statistical mechanics.

\begin{theorem}
Let $(\cO,\tau)$ be a $C^\ast$- or $W^\ast$-dynamical system.
Then every  $(\tau,\beta)$-KMS state for  $\beta\in\RR$ is modular.
\label{KMSmod}
\end{theorem}

%%%%%%%%%%%%%%%%%%%%%%%%%%%%%%%
\subsection{Modular structure}
%%%%%%%%%%%%%%%%%%%%%%%%%%%%%%%

Let $\Psi$ be a modular vector for $\fM$.
Since $\Psi$ is separating for $\fM$, the map 
$$
A\Psi\mapsto A^\ast\Psi,
$$
defines an anti-linear involution $S_0$ of $\fM\Psi$.
Inspection of the graph of $S_0$ and the fact that
$\Psi$ is cyclic for $\fM'$ show that $S_0$ is closable and that
its closure $S$ is involutive. Since
$\Psi$ is cyclic for $\fM$, $S$ is densely defined and hence
has a densely defined adjoint $S^\ast=S_0^\ast$.
Define the selfadjoint operator $\Delta=S^\ast S$ and write
the polar decomposition of $S$  as $S=J\Delta^{1/2}$.
Since $S$ is injective and has dense range $J$ is anti-unitary.
From $\one=S^2=J\Delta^{1/2}J\Delta^{1/2}$ we conclude
that $J\Delta^{1/2}=\Delta^{-1/2}J^\ast$. It follows that
$J^2\Delta^{1/2}=J\Delta^{-1/2}J^\ast$ and the uniqueness 
of the polar decomposition yields $J^2=\one$, \ie $J=J^\ast$.

\begin{definition} The positive selfadjoint operator
$\Delta$ is the modular operator and the anti-unitary involution
$J$ the modular conjugation of the pair $(\fM,\Psi)$.
\end{definition}

The fundamental  algebraic properties of the modular operator and conjugation
are the content of Tomita-Takesaki's theorem:

\begin{theorem}Let $\Psi$ be a modular vector for a von Neumann
algebra $\fM$. If $\Delta$ and $J$ are the corresponding modular
operator and modular conjugation, then the following hold:
\begin{enumerate}%[(i)]
\item $J\fM J=\fM'$.
\item For any $t\in\RR$ one has $\Delta^{\i t}\fM\Delta^{-\i t}=\fM$.
\end{enumerate}
\end{theorem}

Since $S$ is an unbounded operator, the proof of this theorem is
technically involved. It was first published in~\cite{Takesaki1970}; a more
concise  exposition can be found  in~\cite{Bratteli1987}.
For a technically simpler proof see~\cite{Rieffel1977}.

\begin{definition}The group of $\ast$-automorphisms of $\fM$
defined by $\sigma^t(A)=\Delta^{\i t}A\Delta^{-\i t}$
is the modular group of the pair $(\fM,\Psi)$.

More generally,
if $\omega$ is a faithful normal state on the von Neumann
algebra $\fM$ and $\Delta$ the modular operator
of $(\fM,\Omega_\omega)$
then
$\sigma_\omega^t(A)=
\pi_\omega^{-1}(\Delta^{\i t}\pi_\omega(A)\Delta^{-\i t})$
is the modular group of $\omega$.
\end{definition}

The main property of the modular group is the following result
due to Takesaki which can be seen as a reverse of Theorem~\ref{KMSmod}.

\begin{theorem}Let $\omega$ be a faithful normal state on
the von Neumann algebra $\fM$. Then $\omega$
is a KMS state for the modular group $\sigma_\omega$
at inverse temperature $\beta=-1$. Moreover, the modular group
is the only dynamics on $\fM$
for which $\omega$ has this property.
\end{theorem}

The modular conjugation allows us to construct another central
object of modular theory.

\begin{definition} The natural cone associated to the pair
$(\fM,\Psi)$ is the closed subset of $\cH$ defined
by
$$
\cH_+=\{AJAJ\Psi\,|\,A\in\fM\}^\mathrm{cl}.
$$
\end{definition}

The most important properties of the natural cone are the
following.

\begin{theorem} The natural cone $\cH_+$ is self-dual, \ie
$$
\cH_+=
\widehat{\cH}_+ :=
\{\Omega\in\cH\,|\,(\Phi\,|\,\Omega)\ge0\ \text{for all}\ \Phi\in\cH_+\}.
$$
In particular, $\cH_+$ is convex. Moreover, the following hold:
\begin{enumerate}%[(i)]
\item $J\Phi=\Phi$ for all $\Phi\in\cH_+$.
\item $AJA\cH_+\subset\cH_+$ for all $A\in\fM$.
\item $JAJ=A^\ast$ for all $A\in\fM\cap\fM'$.
\end{enumerate}
\label{coneess}
\end{theorem}

%%%%%%%%%%%%%%%%%%%%%%%%%%%%%%%%%%%%%
\subsection{Standard representation}
\label{sec-standard-rep}
%%%%%%%%%%%%%%%%%%%%%%%%%%%%%%%%%%%%%

\begin{definition}A quadruple $(\cH,\pi,J,\cH_+)$ is a standard representation
of the $W^\ast$-algebra $\fM$ if $\pi:\fM\to\cB(\cH)$ is a
representation of $\,\fM$, $J$ an antiunitary involution on $\cH$ and
$\cH_+$ a self-dual cone in $\cH$ satisfying the following conditions:
\begin{enumerate}%[(i)]
\item $J\pi(\fM)J=\pi(\fM)'$;
\item $J\pi(A)J=\pi(A)^\ast$ for all $A\in\fM\cap\fM'$;
\item $J\Psi=\Psi$ for all $\Psi\in\cH_+$;
\item $\pi(A)J\pi(A)\cH_+\subset\cH_+$ for all $A\in\fM$.
\end{enumerate}
\end{definition}

One of the key results in the theory of $W^\ast$-algebras
is the following.

\begin{theorem}Any $W^\ast$-algebra $\fM$ has a faithful
standard representation. Moreover, this representation is unique, up
to unitary equivalence.
\end{theorem}

If the von Neumann algebra $\fM$ is such that any family of mutually orthogonal projections is at most
countably infinite, then  it has a faithful normal state $\omega$, and it
follows from Theorem~\ref{coneess} that the corresponding GNS representation is
standard. This applies, in particular, to many von Neumann algebras arising in
physical applications. See~\cite[Proposition~2.5.6]{Bratteli1987}, and~\cite{Stratila1979} for the general case.

The standard representation has two
properties which are of crucial importance in the study
of quantum dynamical systems. The first one
concerns  normal states.

\begin{theorem}\label{thm:NormalStates}
Let $(\cH,\pi,J,\cH_+)$ be a standard representation of
$\fM$. Any normal state $\omega$ on
$\fM$ has a unique vector representative
$\Phi_\omega\in\cH_+$
such that $\omega(A)=(\Phi_\omega|\pi(A)\Phi_\omega)$. Moreover,
$$
\|\Phi_\omega-\Phi_\nu\|\le\|\omega-\nu\|\le
\|\Phi_\omega-\Phi_\nu\|\,\|\Phi_\omega+\Phi_\nu\|
$$
holds for all normal states $\omega,\nu$. Thus, there
is a homeomorphic correspondence between normal
states and unit vectors of $\cH_+$. Finally,
$\left(\pi(\fM)\Phi_\omega\right)^\mathrm{cl}
=J\left(\pi(\fM)'\Phi_\omega\right)^\mathrm{cl}$,
and in particular
$$
\omega\ \text{is faithful}\Leftrightarrow
\Phi_\omega\ \text{is separating for}\ \pi(\fM)
\Leftrightarrow
\Phi_\omega\ \text{is cyclic for}\ \pi(\fM).
$$
\end{theorem}
The second property concerns the unitary implementation
of $\ast$-automorphisms of $\fM$ in a standard representation
$(\cH,\pi,J,\cH_+)$. Denote by $\Aut(\fM)$
the topological group of $\ast$-automorphisms of $\fM$
with the topology of pointwise $\sigma$-weak convergence.
Let  $\cU$ be the set of unitaries of $\cH$
satisfying  $U\pi(\fM)U^\ast=\pi(\fM)$ and $U\cH_+\subset\cH_+$.
Equipped with the strong operator topology, $\cU$ is
a topological group and $\tau_U(A)=\pi^{-1}(U\pi(A)U^\ast)$ defines
a continuous morphism $\cU\to\mathrm{Aut}(\fM)$.

\begin{theorem}The map $U\mapsto\tau_U$ is a topological
isomorphism. Moreover, for any $U\in\cU$
and any normal state $\omega$ on $\fM$,
one has
\begin{enumerate}%[(i)]
\item $JUJ=U$.
\item $U\pi(\fM)'U^\ast=\pi(\fM)'$.
\item $U^\ast\Phi_\omega=\Phi_{\omega\circ\tau_U}$.
\end{enumerate}
\end{theorem}

In particular, if $(\fM,\tau)$ is a $W^\ast$-dynamical system,
then there exists a unique selfadjoint
operator $L$ on $\cH$ such that $\pi(\tau^t(A))=\e^{\i tL}\pi(A)\e^{-\i tL}$
and $\e^{\i tL}\cH_+\subset\cH_+$.

\begin{definition}
The generator $L$ is called  the standard Liouvillean of the dynamical system $(\fM,\tau)$.
\end{definition}

The standard Liouvillean is uniquely defined up to unitary equivalence.
If $\omega$ is a modular $\tau$-invariant state, then the induced GNS representation is
standard, and the $\omega$-Liouvillean of Section~\ref{ssec:Liouvilleans}
coincides with the standard Liouvillean. This is in particular the case if $\omega$
is a KMS state for $\tau$.

Spectral properties of the standard Liouvillean are intimately related to those of the corresponding dynamical system. As an illustration, the following
result is a direct consequence of Theorem~\ref{thm:NormalStates};  
see Section~\ref{sec:Quantum Koopmanism} for additional information.
\begin{theorem} Let $L$ be the standard Liouvillean of a $W^\ast$-dynamical
system $(\fM,\tau)$.
\begin{enumerate}%[(i)]
\item $L$ has no eigenvalues if and only if there is no normal $\tau$-invariant state on $\fM$.
\item $\Ker(L)$ is one-dimensional if and only if there is a unique
normal $\tau$-invariant state $\omega$ on $\fM$. In this case $\Phi_\omega$
 is the unique unit vector in $\Ker(L)\cap\cH_+$.
\end{enumerate}
\end{theorem}

%%%%%%%%%%%%%%%%%%%%%%%%%%%%%%%%%%%%%%%%%
\subsection{The finite dimensional case}
%%%%%%%%%%%%%%%%%%%%%%%%%%%%%%%%%%%%%%%%%

It is instructive to work out the standard representation of a finite
dimensional von Neumann algebra $\fM\subset\cB(\CC^N)$. This case is
particularly simple since $\cB(\CC^N)$ is itself a Hilbert space for the inner
product $(X|Y)=\tr(X^\ast Y)$. One has $\cB(\CC^N)=\fM\oplus\fM^\perp$ and the
predual $\fM_\ast$ can be identified with $\fM$. Since any $A\in\fM$ can be
written as a linear combination of $4$ non-negative elements of $\fM$, it is easy
to see that there exists a basis $\rho_1,\ldots,\rho_n$ of $\fM$ such that
$\rho_j\ge0$ and $\tr\,\rho_j=1$. It follows that
$$
\omega=\frac1n\sum_{j=1}^n\rho_j
$$
defines a faithful state on $\fM$. Consider $\cH=\fM$ as a Hilbert space (a
subspace of $\cB(\CC^N)$). Then $\Omega=\omega^{1/2}$ is a unit vector in $\cH$.
Moreover, the map $\pi:\fM\to\cB(\cH)$ defined by $\pi(A)X=AX$ is a
$\ast$-morphism such that $\omega(A)=(\Omega|\pi(A)\Omega)$. Denote by $P$ the
orthogonal projection on $(\Ker\,\omega)^\perp$. Clearly $P\in\fM$ and, since
$\rho_j\ge0$,  one has $\Ker\,\omega\subset\cap_j\Ker\,\rho_j$ and 
$\Ker\,\omega\subset\Ker\,A$ for all $A\in\fM$. It follows that
$\Ran\,A\subset\Ran\,P$ and hence $A=PA$ for all $A\in\fM$. Since the last
identity is equivalent to $A^\ast=A^\ast P$, we conclude that $A=AP=PA=PAP$ for
all $A\in\fM$, \ie that $P$ is the unit of $\fM$. Since there exists $T\in\fM$
such that $T\omega^{1/2}=P$ we can write $\pi(XT)\Omega=X$ and conclude that
$\pi(\fM)\Omega=\cH$. We have shown that $(\cH,\pi,\Omega)$ is the GNS
representation of $\fM$ induced by $\omega$. Note that since $\omega$ is a
normal faithful state, this representation is itself faithful.

The formulas $JX=X^\ast$ and $\Delta^{1/2}X=\omega^{1/2}XT$ define an
anti-unitary involution and a positive selfadjoint operator on $\cH$ such that
$$
J\Delta^{1/2}\pi(A)\Omega=
\pi(A)^\ast\Omega.
$$
Thus, $J$ and $\Delta$ are the modular conjugation and the modular
operator of the pair $(\pi(\fM),\Omega)$.

Elements of the natural cone are given
by $\pi(A)J\pi(A)\Omega=
A\omega^{1/2}A^\ast$, from which we can conclude that
$$
\cH_+=\{A\in\fM\,|\,A\ge0\},
$$
and one easily checks the validity of Theorem~\ref{coneess}.

Let $C$ be an element of $\pi(\fM)'$. For all $A\in\fM$ and
$X\in\cH$ one has
$$
C(AX)=C(\pi(A)X)=\pi(A)(C(X))=AC(X).
$$
Setting $X=P$, the unit of $\fM$, and $B=C(P)\in\fM$, we get
$C(A)=AB$. We conclude that $\pi(\fM)'$ consists of the linear maps
$X\mapsto XB$ with $B\in\fM$. Thus, $J\pi(\fM)J=\pi(\fM)'$ and we have
obtained the standard representation of the finite dimensional von
Neumann algebra $\fM$.

Any normal state $\nu$ on $\fM$ is given by
$\nu(A)=\tr(\rho A)$ for a density matrix
$\rho\in\fM$. It follows that
$\nu\mapsto\Phi_\nu=\rho^{1/2}\in\cH_+$
is the homeomorphism described in Theorem~\ref{thm:NormalStates}.

%%%%%%%%%%%%%%%%%%%%%%%%%%%%
\part{CCR and CAR algebras}
%%%%%%%%%%%%%%%%%%%%%%%%%%%%

%%%%%%%%%%%%%%%%%%%%%%%%%%%%%%%%%%%%%%%%%%%%%%%%%%%%%%%%%%%%%%%%%
\section{Free Bose and Fermi gases -- the algebraic approach}
\label{sec:Free Bose and Fermi gases -- the algebraic approach}
%%%%%%%%%%%%%%%%%%%%%%%%%%%%%%%%%%%%%%%%%%%%%%%%%%%%%%%%%%%%%%%%%

We will describe the  algebraic approach to free Bose and Fermi gases, which
allows us to discuss these systems in the thermodynamic limit at a positive
temperature, or more generally, at a positive density. We will use the
terminology of Section~\ref{sec:The C*-algebra approach}.

First we will follow the $W^*$-approach, working in a concrete Hilbert space.
More precisely, we will describe Araki-Woods representations of the CCR and
Araki-Wyss representations of the CAR, often used to describe free  Bose and
Fermi gases at a positive density. Then we will describe the $C^*$-algebraic
approach to Bose and Fermi gases, which has a number of conceptual advantages.
Unfortunately, as we shall see,  the $C^*$-algebraic
approach is somewhat problematic in the bosonic case.

For an in-depth treatment of the material of this chapter, see~\cite{Derezinski2013}.

%%%%%%%%%%%%%%%%%%%%%%%%%%%%%%%%%%%%%%%%%%%%%%%%%%%%
\subsection{Araki-Woods and Araki-Wyss operators}
%%%%%%%%%%%%%%%%%%%%%%%%%%%%%%%%%%%%%%%%%%%%%%%%%%%%%
We fix a nonnegative function $\xi\mapsto\gamma(\xi)$ and define
\beq
\rho(\xi) :=(\gamma(\xi)^{-1}\mp1)^{-1},
\label{rho}
\eeq
where the minus sign corresponds to the bosonic case and the plus sign to the
fermionic case. The function $\gamma$, or equivalently, $\rho$ will parametrize
a certain class of representations of the \ul{canonical commutation
relations} (CCR) and \ul{canonical anticommutation relations} (CAR). 
As we shall see, $\rho(\xi)$ has the physical interpretation of the density of the mode~$\xi$.

We consider the bosonic/fermionic Fock space $\Gamma_{\s/\a}\left(L^2(\Xi)\oplus
L^2(\Xi)\right)$, in which the one-particle space has been ``doubled'' to
account for  both ``excitations'', corresponding to the ``left space'', as well
as ``holes'', corresponding to the ``right space''. The creation/annihilation
operators corresponding to the left/right space are distinguished by the
subscript $\l/\r$. The subscript $\s/\a$ indicates that we consider
the bosonic/fermionic Fock space.

Define
\beq
\begin{split}
a_{\gamma,\l}^\ast(\xi)
&:=(1\pm\rho(\xi))^{{\frac12}}a_\l^*(\xi)+\rho(\xi)^{\frac12}a_\r(\xi),\\[4pt]
a_{\gamma,\l}(\xi)
&:=(1\pm\rho(\xi))^{{\frac12}}a_\l(\xi)+\rho(\xi)^{\frac12}a_\r^*(\xi),
\end{split}
\label{eq:arakiw}
\eeq
where the plus corresponds to the bosonic case and minus to the fermionic case.
In the bosonic case, \eqref{eq:arakiw} are called the (left) \ul{Araki-Woods
creation/annihilation operators}. They satisfy the usual CCR. In the fermionic
case, \eqref{eq:arakiw} are called the (left) \ul{Araki-Wyss creation/annihilation
operators}. They satisfy the usual CAR.

Let $\Omega$ denote the vacuum vector for
$\Gamma_{\s/\a}\left(L^2(\Xi)\oplus L^2(\Xi)\right)$.
Then $(\Omega\mid\argdot\,\Omega)$ is an example
of a \ul{quasifree state}. Its 2-point function  is
$$
(\Omega\mid a_{\gamma,\l}^\ast(\xi)a_{\gamma,\l}(\xi')\Omega)
=\delta(\xi,\xi')\rho(\xi),
$$
where $\delta(\xi,\xi')$ denotes the delta function. This justifies  the name ``the density of the mode $\xi$'' for $\rho(\xi)$;  
see~\cite{Araki1963,Derezinski2006a,Bratteli1981}.

%%%%%%%%%%%%%%%%%%%%%%%%%%%%%%%%%%%%%%%%%%%%%%%%%%%%%%%%%%%%%%%%
\subsection{The commutant of the algebra of the Bose/Fermi gas}
%%%%%%%%%%%%%%%%%%%%%%%%%%%%%%%%%%%%%%%%%%%%%%%%%%%%%%%%%%%%%%%%

Let $\fM_{\gamma,\l}^{\rm AW}$ be the $W^*$-algebra in
$\cB(\Gamma_{\s/\a}(L^2(\Xi)\oplus L^2(\Xi)))$  generated by smeared-out
bosonic/fermionic operators introduced in the previous section (in the bosonic case in addition we need to take
bounded functions of these operators). Note that $\fM_{\gamma,\l}^{\rm AW}$ can
be a rather nontrivial $W^*$-algebra -- typically it is of \ul{type III}.

For nonzero $\rho$ the state $(\Omega\mid\argdot\,\Omega)$ on the algebra
$\fM_{\gamma,\l}^{\rm AW}$ is faithful. Therefore, we can apply
\ul{Tomita-Takesaki theory}, see Section~\ref{sec:Tomita--Takesaki theory},
obtaining in particular the so-called modular conjugation $J_{\s/\a}$, which
is an antiunitary operator on $\Gamma_{\s/\a}(L^2(\Xi\oplus\Xi))$
satisfying
\[
J_{\s/\a}^2=\one,\qquad J_{\s/\a}\Omega=\Omega,\qquad
J_{\s/\a}\fM_{\gamma,\l}^{\rm AW}J_{\s/\a}=\left(\fM_{\gamma,\l}^{\rm AW}\right)',
\]
where $\left(\fM_{\gamma,\l}^{\rm AW}\right)'$ denotes the commutant of
$\fM_{\gamma,\l}^{\rm AW}$. Let $\epsilon$ be the antiunitary operator on
$L^2(\Xi)\oplus L^2(\Xi)$ defined by
\[
\epsilon(f_1,f_2) :=(\bar f_2,\bar f_1).
\]
Then, in the bosonic case, $J_\s=\Gamma(\epsilon)$. In the fermionic case,
$J_\a=\Gamma(\epsilon)(-1)^{N(N-1)/2}$. Here $\Gamma(\epsilon)$ denotes the
second quantization of the operator $\epsilon$ and $N$ is the number
operator.

Later on we will need the Araki-Woods/Araki-Wyss creation and annihilation
operators transformed by $J_{\s/\a}$ (which will be called the ``right'' operators,
as opposed to the ``left'' operators introduced above). The right Araki-Woods
creation/annihilation operators are given by
\begin{align*}
a_{\gamma,\r}^\ast(\xi)& := J_\s a_{\gamma,\l}^\ast(\xi)J_\s
=\rho(\xi)^{{\frac12}}a_\l^*(\xi)+(1+\rho(\xi))^{\frac12}a_\r(\xi),\\[4pt]
a_{\gamma,\r}(\xi)&:= J_\s a_{\gamma,\l}(\xi)J_\s
=\rho(\xi)^{{\frac12}}a_\l(\xi)+(1+\rho(\xi))^{\frac12}a_\r^*(\xi).
\end{align*}
The right Araki-Wyss creation/annihilation operators are given by
\begin{align*}
a_{\gamma, \r}^\ast(\xi)& := J_\a a_{\gamma,\l}^\ast(\xi)J_\a
= \rho(\xi)^{{\frac12}}a_\l^*(\xi)(-1)^N-(1-\rho(\xi))^{\frac12}a_\r(\xi)(-1)^N,\\[4pt]
a_{\gamma, \r}(\xi)& := J_\a a_{\gamma, \l}(\xi)J_\a
=-\rho(\xi)^{{\frac12}}a_\l(\xi)(-1)^N+(1-\rho(\xi))^{\frac12}a_\r^*(\xi)(-1)^N.
\end{align*}
Note that the left and the right operators commute with one another.

%%%%%%%%%%%%%%%%%%%%%%%%%%%%%%%%%
\subsection{Quasi-free dynamics}
%%%%%%%%%%%%%%%%%%%%%%%%%%%%%%%%%

Consider  Bose or Fermi gas with the 1-particle space $L^2(\Xi)$ and the
1-particle excitation spectrum $\omega(\xi)$. One usually assumes that $\omega$
is positive.

At zero temperature,  the Hilbert space of the system is 
$\Gamma_{\s/\a}(L^2(\Xi))$,  and its Hamiltonian is
$$
H=\int\omega(\xi)a^*(\xi)a(\xi)\d\xi.
$$
Although an  advanced algebraic language in this case is
not necessary,  it is often convenient to introduce  the $W^*$-dynamical system
$\left(\cB(\Gamma_{\s/\a}(L^2(\Xi))),\e^{\i tH}\argdot\e^{-\i tH}\right)$. Clearly,
\beq
\e^{\i tH}a^*(\xi)\e^{-\i tH}=\e^{\i t\omega(\xi)} a^*(\xi),\qquad
\e^{\i tH}a(\xi)\e^{-\i tH}=\e^{-\i t\omega(\xi)}a(\xi).
\label{dyna}
\eeq

We now consider the Bose/Fermi gas at a
positive density described by  $\gamma$. Its observables are elements 
of  $\fM_{\gamma,\l}^{\rm AW}$.  Its dynamics is given by
$\tau_{t}(A)=\e^{\i tL}A\e^{-\i tL}$, where the  \ul{Liouvillean} $L$ is  given by
\[
L :=  \int \omega(\xi)a_\l^*(\xi)a_\l(\xi)\d\xi-\int \omega(\xi)a_\r^*(\xi)a_\r(\xi)\d\xi,
\]
see Section~\ref{sec:The C*-algebra approach}. Note that
$$
\tau_{t}(a_{\gamma,\l}^*(\xi))=\e^{\i t\omega(\xi)} a_{\gamma,\l}^*(\xi),\qquad
\tau_{t}(a_{\gamma,\l}(\xi))=\e^{-\i t\omega(\xi)} a_{\gamma,\l}(\xi),
$$
and so  the dynamics $\tau^t$ acts on left creation/annihilation operators in the
same way as the zero-temperature dynamics in~\eqref{dyna}.
However
$$
\tau_{t}(a_{\gamma,\r}^*(\xi))=\e^{-\i t\omega(\xi)} a_{\gamma,\r}^*(\xi),\qquad
\tau_{t}(a_{\gamma,\r}(\xi))=\e^{\i t\omega(\xi)} a_{\gamma,\r}(\xi),
$$
indicates that the ``particles'' created by $a_{\gamma,\r}^*(\xi)$ behave like a ``hole''.
We emphasize  that the $W^*$-dynamical systems $\left(\fM_{\gamma,\l}^{\rm AW},\tau\right)$ for
distinct $\gamma$ are often non-equivalent.

The state $(\Omega\mid\argdot\,\Omega)$ is $\beta$-KMS for the dynamics $\tau$ iff
$\gamma(\xi)=\e^{-\beta\omega(\xi)}$. In the bosonic case the density is then given
by the \ul{Planck law}: $\rho(\xi)=(\e^{\beta\omega(\xi)}-1)^{-1}$. In the fermionic
case we obtain the \ul{Fermi-Dirac law}: $\rho(\xi)=(\e^{\beta\omega(\xi)}+1)^{-1}$.

%%%%%%%%%%%%%%%%%%%%%%%%%%%%%%%%%%%%%%%%%%%%%%%%
\subsection{$C^*$-algebraic approach to the CAR}
%%%%%%%%%%%%%%%%%%%%%%%%%%%%%%%%%%%%%%%%%%%%%%%%

The CAR algebra over $L^2(\Xi)$, sometimes denoted $\CAR(L^2(\Xi))$, is defined
as the $C^*$-algebra generated by the elements $\{a(f)\ : f\in L^2(\Xi)\}$
satisfying
$$
[a(f_1),a(f_2)]_+=0,\qquad
[a(f_1),a^*(f_2)]_+=2\Re(f_1|f_2)\one.
$$
One can show that for an infinite dimensional separable space
$L^2(\Xi)$ this algebra is isomorphic to the uniformly hyperfinite algebra
${\rm UHF}(2^\infty)$. In the case of a finite dimension, this algebra
coincides with the complex Clifford algebra. $a(f)$ and $a^*(f)$
are called (abstract) annihilation and creation operators of $f$.

If $\e^{\i t\omega }$ is a unitary group on $L^2(\Xi)$, then
$\tau_t(a(f)) := a(\e^{t \i\omega}f)$ extends to a unique $C^*$-dynamics on
$\CAR(L^2(\Xi))$. Such dynamics are often called 
  gauge-invariant quasi-free dynamics.

Let $\rho,\gamma$ be as in~\eqref{rho}. Then there exists a unique state
$\psi_\gamma$ on $\CAR(L^2(\Xi))$ such that
$$
\psi_\gamma\left(a^*(f_1)\cdots a^*(f_n)a(g_m)\cdots a(g_1)\right)
=\begin{cases}
\ds\sum_{\sigma\in S_n}\prod_{j=1}^n\sign(\sigma)
\int\bar f_j(\xi)\rho(\xi)g_{\sigma( j)}(\xi)\d\xi,&\text{if }n=m;\\[16pt]
0&\text{otherwise,}\\[4pt]
\end{cases}
$$
where $S_n$ denotes the set of permutations of $\{1,\dots,n\}$ and
$\sign(\sigma)$ is the signature of $\sigma\in S_n$.

Let $a_{\gamma,\l}^*(\xi)$, $a_{\gamma,\l}(\xi)$ be the Araki-Wyss
representation of the CAR. Then
\[
\pi_{\gamma}(a(f)):= \int a_{\gamma,\l}(\xi)\bar{f}(\xi)\d\xi
\]
extends uniquely to a representation of $\CAR(L^2(\Xi))$ in
$\Gamma_\a(L^2(\Xi)\oplus L^2(\Xi))$, which
is a GNS representation generated by the state $\psi_\gamma$.

%%%%%%%%%%%%%%%%%%%%%%%%%%%%%%%%%%%%%%%%%%%%%%%%%
\subsection{$C^*$-algebraic approach to the CCR}
%%%%%%%%%%%%%%%%%%%%%%%%%%%%%%%%%%%%%%%%%%%%%%%%%

In the bosonic case, several  algebras could be referred to as 
$\CCR(L^2(\Xi))$.  The choice made by most authors is the $C^*$-algebra
generated by $\{W(f)\mid f\in L^2(\Xi)\}$  satisfying the relations
$$
W(f)^*=W(-f),\qquad W(f_1)W(f_2)=\e^{-\frac{\i}{2}\Im(f_1|f_2)}W(f_1+f_2).
$$
We will denote it $\CCR(L^2(\Xi))$. $W(f)$'s are called Weyl operators, and this  algebra is
sometimes called the Weyl CCR algebra. An important result about this algebra
was obtained by J.~Slawny~\cite{Bratteli1981}. This is the reason for another
name for $\CCR(L^2(\Xi))$, the Slawny CCR algebra.

If $\e^{\i t\omega}$ is a unitary group on $L^2(\Xi)$, then
$\tau_t(W(f)) := W(\e^{\i t\omega }f)$ extends to a unique 1-parameter
group of $*$-automorphisms of $\CCR(L^2(\Xi))$, often called a gauge-invariant
quasi-free dynamics. Note that $t\mapsto \tau_t$ is not continuous, hence it is
not a $C^*$-dynamics. This is a signature of the difficulties the
$C^*$-algebraic approach encounters in the CCR case.

Let $\rho,\gamma$ be as in~\eqref{rho}. Then
\[
\psi_\gamma(W(f)):= \exp\left(-\frac{1}{4}\int|f(\xi)|^2\rho(\xi)\d\xi\right)
\]
extends by linearity to a state on $\CCR(L^2(\Xi))$.

Let $a_{\gamma,\l}^*(\xi),a_{\gamma,\l}(\xi)$ be the Araki-Woods
creation/annihilation operators. Then
\[
\pi_\gamma(W(f))
:= \exp\left(\frac{\i}{\sqrt2}\int\left(f(\xi)
a_{\gamma,\l}^*(\xi)+\bar{f}(\xi)a_{\gamma,\l}(\xi)
\right)\d\xi\right)
\]
extends to a representation of $\CCR(L^2(\Xi))$ in
$\Gamma_\s(L^2(\Xi)\oplus L^2(\Xi))$, which is a
GNS representation induced by the state $\psi_\gamma$.

%%%%%%%%%%%%%%%%%%%%%%%%%%%%%%%%%%%%%%%%
\section{Araki-Wyss representation}
\label{sec:Araki-Wyss representation}
%%%%%%%%%%%%%%%%%%%%%%%%%%%%%%%%%%%%%%%%

%%%%%%%%%%%%%%%%%%%%%%%%%%%%%%%%%%%%%%%%
\subsection{Non-interacting fermions}
%%%%%%%%%%%%%%%%%%%%%%%%%%%%%%%%%%%%%%%%

A system of non-interacting fermions is specified by a one-particle Hilbert space $\fh$ and a one-particle Hamiltonian $h$,
a selfadjoint operator on $\fh$. Within the \ul{$C^\ast$-algebraic
approach}, observables of this system are elements of the
$C^\ast$-algebra of \ul{Canonical Anti\-commu\-tation Relations} $\CAR(\fh)$.
Their time evolution is given by the group of \ul{Bogoliubov automorphisms}
$$
\tau^t(a(f))=a(\e^{\i th}f)
$$
associated to $h$. Thus, the dynamics of the system is described by
the $C^\ast$-dynamical system $(\CAR(\fh),\tau)$. Taking gauge-invariance
into account, we should in fact restrict the algebra to its gauge-invariant
part $\CAR_0(\fh)$, see Section~\ref{ssec:Gauge invariant states}.
It is often more convenient to retain  the full CAR algebra and consider
only gauge-invariant states.

%%%%%%%%%%%%%%%%%%%%%%%%%%%%%%%%%%%%%%%%%%%%%%%%
\subsection{Gauge-invariant quasi-free states}
%%%%%%%%%%%%%%%%%%%%%%%%%%%%%%%%%%%%%%%%%%%%%%%%

In the \ul{Fock representation} the dynamical group $\tau$ as well
as the gauge group $\vartheta$ are unitarily implemented by
the second quantized Hamiltonian $H=\d\Gamma(h)$ and the
number operator $N=\d\Gamma(\one)$,
$$
\pi_F(\tau^t(A))=\e^{\i tH}\pi_F(A)\e^{-\i tH},\qquad 
\pi_F(\vartheta^\varphi(A))=\e^{\i\varphi N}\pi_F(A)\e^{-\i\varphi N}.
$$

The Hamiltonian $h$ of a single fermion confined
in a finite volume $\Lambda\subset\RR^d$ typically has
purely discrete spectrum and $\e^{-\beta h}$ is trace
class for any  $\beta>0$. Using the identity
\beq
\det( \one+A)=\tr(\Gamma(A)),
\label{detdef}
\eeq
we conclude that
$\tr(\e^{-\beta(H-\mu N)})=\det(\one+\e^{-\beta(h-\mu)})$
for any $\beta>0$ and $\mu\in\RR$. Hence
$\e^{-\beta(H-\mu N)}$ is also trace class and
the Gibbs grand canonical ensemble at inverse temperature $\beta$
and chemical potential $\mu$ is a gauge-invariant \ul{Fock state} with
density matrix
$$
\rho_{\beta\mu}=\frac{\e^{-\beta(H-\mu N)}}
{\mathrm{tr}(\e^{-\beta(H-\mu N)})}.
$$
It is the unique \ul{$\beta$-KMS state} on $\CAR(\fh)$ for the dynamics
$t\mapsto\tau^t\circ\vartheta^{-\mu t}$.
Using again identity~\eqref{detdef},
a simple calculation shows that the \ul{characteristic function}
of this state (see Section~\ref{ssec:Characteristic functions}) is given by
\beq
E_{\beta\mu}(u)=\det({ \one}+(u-{\one})f_{\beta\mu}(h)),
\label{KMSchar}
\eeq
where
$$
f_{\beta\mu}(\varepsilon)=\frac{1}{{ \one}+\e^{\beta(\varepsilon-\mu)}}
$$
is the Fermi-Dirac distribution function.

Since $u-\one$ is finite rank, the characteristic function~\eqref{KMSchar}
still makes sense in the infinite volume limit, despite the fact
that the Boltzmann weight $\e^{-\beta(H-\mu N)}$ is no longer  trace
class in this limit.  One can show directly that~\eqref{KMSchar} is the
characteristic function of the unique $\beta$-KMS state for the group
$t\mapsto\tau^t\circ\vartheta^{-\mu t}$.  Its restriction to the
gauge-invariant sub-algebra $\CAR_0(\fh)$ is therefore a
$(\tau,\beta)$-KMS state.

More generally, one has the following result. 
\begin{theorem}Let $T$ be a selfadjoint operator on the Hilbert space $\fh$
satisfying $0\le T\le {\one}$. Then 
\[E(u)=\det({ \one}+(u-{\one})T)\]
 is the
characteristic function of a gauge-invariant state $\omega_T$ on $\CAR(\fh)$.
$\omega_T$ is called the gauge-invariant quasi-free state generated by $T$.
Equivalent ways to characterize this state are:
\begin{enumerate}%[(i)]
\item For all $f_1,\ldots,f_n\in\fh$ and $g_1,\ldots,g_m\in\fh$ one has
 $$
\omega_T(a^\ast(f_1)\cdots a^\ast(f_n)a(g_m)\cdots a(g_1))=\delta_{nm}\det\{(g_i|Tf_j)\}.
$$
\item For $f\in\fh$ set $\varphi(f)=2^{-1/2}(a(f)+a^\ast(f))$.
The Wick theorem
\begin{eqnarray*}
\omega_T(\varphi(f_1)\cdots\varphi(f_{2n+1}))&=&0,\\
\omega_T(\varphi(f_1)\cdots\varphi(f_{2n}))&=&
\sum_{\pi\in\cP_n}\sign(\pi)\prod_{j=1}^n
\omega_T\left(\varphi(f_{\pi(2j-1)})\varphi(f_{\pi(2j)})\right),
\end{eqnarray*}
holds. In the last expression, the sum runs over the set
$\cP_n$ of pairings,
\ie permutations $\pi$ of $\{1,\ldots,2n\}$ such that
$\pi(2j-1)<\pi(2j)$ and $\pi(2j-1)<\pi(2j+1)$. Moreover,
$\sign(\pi)$ denotes the signature of $\pi\in\cP_n$.
\end{enumerate}
\end{theorem}

The state $\omega_T$ also has an information-theoretic
characterization: it has maximal entropy among all the
gauge-invariant states $\nu$ such that $\nu(a^\ast(f)a(g))=(g|Tf)$
in  the  following sense. For a finite dimensional subspace
$\fk\subset\fh$ define the entropy
$$
S(\nu|\fk)=-\tr(\rho\log\rho)
$$
where $\rho$ is the density matrix of the restriction of $\nu$ to the
finite dimensional algebra $\CAR(\fk)$. Then
$$
S(\omega_T|\fk)=\max_{\nu\in E_T}S(\nu|\fk),
$$
where $E_T$ denotes the set of gauge-invariant states
$\nu$ such that $\nu(a^\ast(f)a(g))=(g|Tf)$. This follows from
a simple adaptation of the proof of~\cite[Proposition~1a]{Lanford1972}.

We refer the reader to~\cite{Araki1970/71,Bratteli1981} for more information
on quasi-free states on $\CAR(\fh)$.

%%%%%%%%%%%%%%%%%%%%%%%%%%%%%%%%%%%%%%%%
\subsection{Araki-Wyss representation}
%%%%%%%%%%%%%%%%%%%%%%%%%%%%%%%%%%%%%%%%

Let $\omega_T$ be the gauge-invariant quasi-free state on $\CAR(\fh)$
generated by $T$. The associated GNS representation, which we denote
by  $(\cH_T,\pi_T,\Omega_T)$, was first
constructed by Araki and Wyss in~\cite{Araki1964a}. It can be described
as follows:
\begin{enumerate}%[(i)]
\item $\cH_T=\aFock(\fh_1\oplus\fh_2)\subset\aFock(\fh\oplus\fh)$
where $\fh_1=(\Ran({\one}-T))^\mathrm{cl}$ and $\fh_2=(\Ran T)^\mathrm{cl}$.
\item $\Omega_T=\Omega$, the Fock vacuum
vector.
\item The $\ast$-morphism $\pi_T$ is given by
$$
\pi_T(a(f))=a\left(\sqrt{{\one}-T} f\oplus 0\right)+a^\ast\left(0\oplus\overline{\sqrt{T}f}\right),
$$
where $\overline{\argdot}$ denotes an arbitrary complex conjugation
on $\fh$.
\end{enumerate}

Using the \ul{exponential law for fermions}
$U:\aFock(\fh_1\oplus\fh_2)\to\aFock(\fh_1)\otimes\aFock(\fh_2)$,
an equivalent representation with cyclic vector
$U\Omega_T=\Omega\otimes\Omega$
is obtained. It is explicitly given by
$$
U\pi_T(a(f))U^\ast=
a\left(\sqrt{{\one}-T}f\right)\otimes { \one}+\Theta\otimes
a^\ast\left(\overline{\sqrt{T} f}\right),
$$
where $\Theta=\Gamma(-{\one})=(-1)^N$.  In the limiting cases $T=0$ and
$T={\one}$, which correspond to the \ul{vacuum state} $\mathbf{vac}$ and to
the \ul{filled Fermi sea}  $\mathbf{full}$, the Araki-Wyss
representation degenerates to the Fock and anti-Fock representations
$\pi_F$ and $\pi_{AF}$.

The reader should consult~\cite{Araki1970/71,Derezinski2006a} for an
introduction to quasi-free representations of $\CAR(\fh)$.

%%%%%%%%%%%%%%%%%%%%%%%%%%%%%%%%%%%%%%%%%%%%%
\subsection{Enveloping von Neumann algebra}
%%%%%%%%%%%%%%%%%%%%%%%%%%%%%%%%%%%%%%%%%%%%%

The following theorem summarizes some interesting
features of the \ul{enveloping von Neumann algebra}
$$
\fM_T=\pi_T(\CAR(\fh))''
$$
of a gauge-invariant quasi-free state $\omega_T$.

\begin{theorem}
\begin{enumerate}%[(i)]
\item $\omega_T$ is primary, \ie its enveloping von Neumann algebra
is a \ul{factor}. $\fM_T$ is of type
\begin{description}
\item[$I$] if either  $\fh$ is finite dimensional or
$\fh$ is infinite dimensional  and  $T=0$ or $T={\one}$.
\item[$II$] if\, $\fh$ is infinite dimensional and $T={\one}/2$.
\item[$III_\lambda$] if\, $\fh$ is infinite dimensional
and $T=({\one}+\lambda^{\pm1})^{-1}$ for some $\lambda\in]0,1[$.
\item[$III_1$] if the continuous spectrum of $T$ is not empty.
\end{description}
\item  $\omega_T$ is \ul{modular}, \ie the cyclic vector $\Omega_{T}$
is separating for $\fM_T$, if and only if
$\Ker T=\Ker({\one}-T)=\{0\}$.
\item $\omega_T$ and $\omega_S$ are \ul{quasi-equivalent} if
and only if the operators $T^{1/2}-S^{1/2}$ and
$({\one}-T)^{1/2}-({\one}-S)^{1/2}$ are Hilbert-Schmidt.
\end{enumerate}
\end{theorem}

If $\Ker\,T=\Ker(\one-T)=\{0\}$, then \ul{modular theory} applies to
$\fM_T$, see Section~\ref{sec:Tomita--Takesaki theory}.
On $\cH_{T}$ there exist an anti-unitary involution $J$
(the modular conjugation) and a positive operator $\Delta$
(the modular operator) such that
$$
J\Delta^{1/2}A\Omega_{T}=A^\ast\Omega_{T}
$$
for all $A\in\fM_T$. These operators are explicitly given
by
$$
J=(-1)^{N(N-1)/2}\Gamma(j),\quad
\Delta=\Gamma(\e^s\oplus\e^{-\overline{s}}),
$$
where $j:f\oplus g\mapsto\overline{g\oplus f}$
and $s=\log T({\one}-T)^{-1}$.

%%%%%%%%%%%%%%%%%%%%%%%%%%%%%%%%%%%%%%%%%%%%%%%%%%%%%%%
\section{Fock and non-Fock states on CAR-algebras}
\label{sec:Fock and non-Fock states on CAR-algebras}
%%%%%%%%%%%%%%%%%%%%%%%%%%%%%%%%%%%%%%%%%%%%%%%%%%%%%%%

In the formalism of \ul{second quantization} a system of fermions is
described by \ul{creation and annihilation operators} $a^\ast(f)$,
$a(f)$ on the antisymmetric \ul{Fock space} $\aFock(\fh)$ over the
one-particle Hilbert space $\fh$.  For systems confined in a finite
volume, this Hilbert space description is sufficient and states of
finite positive density can be represented by \ul{density matrices} acting on 
$\aFock(\fh)$. The situation changes when one takes  the thermodynamic
(infinite volume) limit. In this case, there is no density matrix in Fock space describing a positive density state of an infinitely extended fermionic system. A more sophisticated description is therefore required,  see
Section~\ref{sec:The C*-algebra approach}.

\subsection{Gauge invariant states on $\CAR(\fh)$}
\label{ssec:Gauge invariant states}

Global \ul{$U(1)$-gauge symmetry} is a fundamental property of
quantum mechanics. Its implementation on $\CAR(\fh)$ is given by the
gauge group $\mathbb R\ni\varphi\mapsto\vartheta^\varphi$, the group
of \ul{Bogoliubov automorphisms} defined by
$$
\vartheta^\varphi(a^\ast(f))
=a^\ast(\e^{\i\varphi}f)=\e^{\i\varphi}a^\ast(f),\quad
\vartheta^\varphi(a(f))
=a(\e^{\i\varphi}f)=\e^{-\i\varphi}a(f).
$$
As a Banach space, $\CAR(\fh)$ has a direct sum decomposition into
charge sectors
$$
\CAR(\fh)=\bigoplus_{n\in\ZZ}\CAR_n(\fh),
$$
where $\CAR_n(\fh)$ is the closed linear span of monomials of the form
$$
a^\ast(f_1)\cdots a^\ast(f_j)a(g_k)\cdots a(g_1)
$$
with $j-k=n$. In terms of the gauge-group, one has
$$
\CAR_n(\fh)=
\{A\in\CAR(\fh)\,|\,\vartheta^\varphi(A)=\e^{\i
  n\varphi}A\}.
$$
If $A\in\CAR_n(\fh)$ and $B\in\CAR_m(\fh)$, then
$AB\in\CAR_{n+m}(\fh)$ and  $A^\ast\in\CAR_{-n}(\fh)$. In particular, the
zero charge sector $\CAR_0(\fh)$ is a $C^\ast$-subalgebra generated by
$\one$ and elements of the form $a^\ast(f)a(g)$. Physical observables of  a fermionic system are gauge invariant and hence elements of
$\CAR_0(\fh)$.

A state $\omega$ on $\CAR(\fh)$ is gauge-invariant if
$\omega\circ\vartheta^\varphi=\omega$ for all $\varphi\in\RR$. A state
$\omega_0$ on $\CAR_0(\fh)$ has a unique extension to a
gauge-invariant state $\omega$ on $\CAR(\fh)$ given by
$$
\omega\left(\bigoplus_nA_n\right)=\omega_0(A_0).
$$
Thus, a gauge-invariant state on $\CAR(\fh)$ is completely determined
by its restriction to the gauge-invariant sub-algebra $\CAR_0(\fh)$.
When dealing with fermionic systems, it is often convenient to
work on the full algebra $\CAR(\fh)$ and  restrict the gauge-invariant  states.

%%%%%%%%%%%%%%%%%%%%%%%%%%%%%%%%%%%%%%%
\subsection{Characteristic functions}
\label{ssec:Characteristic functions}
%%%%%%%%%%%%%%%%%%%%%%%%%%%%%%%%%%%%%%%

Denote by $\cU$ the group of unitaries $u$ on $\fh$ such that
$u-{\one}$ is finite-rank. For each $u\in\cU$ there exist a  finite-rank selfadjoint operator $k$ such that
\beq
k=\sum_{j=1}^n\kappa_jf_j(f_j|\argdot),\quad u=\e^{\i k}.
\label{urep}
\eeq
Moreover, the unitary
$$
U(u)=\e^{\i\sum_j\kappa_ja^\ast(f_j)a(f_j)}\in\CAR(\fh),
$$
only depends on $u$, not on the particular choice of the
representation~\eqref{urep}.
The Araki-Wyss characteristic function of a gauge-invariant state
$\omega$ on $\CAR(\fh)$ is defined as
$$
\begin{array}{rlcc}
E:&\cU&\to &\mathbb C\\
&u&\mapsto &\omega(U(u)).
\end{array}
$$
It satisfies
\begin{enumerate}
\item For any $u_1,\ldots,u_N\in\cU$ and $z_1,\ldots,z_N\in\CC$,
$$
\sum_{j,k=1}^N E(u_j^\ast u_k)\bar z_j z_k\ge0.
$$
\item For any $u,v\in\cU$, $f\in\fh$ and $\lambda\in\RR$
$$
\frac{E(u\e^{\i\lambda(f|\argdot)f}v)-E(uv)}
{\e^{\i\lambda\|f\|^2}-1},
$$
is independent of $\lambda$.
\end{enumerate}
Reciprocally, any function $E:\cU\to\CC$  satisfying the above
two conditions is the characteristic function of a unique gauge-invariant
state $\omega$ on $\CAR(\fh)$, see~\cite{Araki1964a}.

%%%%%%%%%%%%%%%%%%%%%%%%%%%%%%%%%%%%%%%%%%%%%%%%%%%
\subsection{Vacuum state and Fock representation}
%%%%%%%%%%%%%%%%%%%%%%%%%%%%%%%%%%%%%%%%%%%%%%%%%%%

The vacuum state $\mathbf{vac}(\argdot)$ on $\CAR(\fh)$ describes
the system in absence of any fermion.
If $\{e_i\,|\,i\in I\}$ denotes an arbitrary orthonormal
basis of $\fh$, then $n_i=a^\ast(e_i)a(e_i)$ is the number of fermions
in the state $e_i$, and we  have $\mathbf{vac}(\prod_{i\in J}n_i)=0$
for any finite $J\subset I$ (note that $[n_i,n_j]=0$). It 
follows immediately that the characteristic function of the vacuum state is
$E_\mathbf{vac}(u)=1$.

The GNS representation associated to the
vacuum state is the Fock representation $(\cH_F,\pi_F,\Omega_F)$
where $\cH_F=\aFock(\fh)$ is the fermionic Fock space over $\fh$,
$\pi_F(a(f))=a_F(f)$ is the annihilation operator on $\aFock(\fh)$, and
$\Omega_F$ is the Fock vacuum vector. For $f_i,g_j\in\fh$ one has
\begin{align*}
\mathbf{vac}(a(g_1)\cdots a(g_m)&a^\ast(f_n)\cdots a^\ast(f_1))\\
=(a_F^\ast(g_m)\cdots a_F^\ast(g_1)\Omega_F|&a_F^\ast(f_n)\cdots a_F^\ast(f_1)\Omega_F)
=\delta_{nm}\det\left\{(g_i|f_j)\right\}.
\end{align*}
Special features of the Fock representation are:
\begin{enumerate}
\item $\pi_F(\CAR(\fh))$ is irreducible, \ie any bounded operator on
$\aFock(\fh)$ commuting with all $a^\#_\mathrm{F}(f)$ is a multiple
of the identity. Equivalently, the enveloping von Neumann algebra
$\pi_F(\CAR(\fh))''$ is the $C^\ast$-algebra of all bounded
operators on $\aFock(\fh)$.
\item The second quantization $\Gamma(U)$ of a unitary operator
$U$ on $\fh$ provides a unitary implementation of the associated
Bogoliubov automorphism $\gamma(a(f))=a(Uf)$,
$$
\pi_F(\gamma(a(f)))=\Gamma(U)\pi_F(a(f))\Gamma(U)^\ast.
$$
In particular, the gauge group $\vartheta^t$ is implemented by a
strongly continuous unitary group whose generator
$N=\d\Gamma({\one})$ is the number operator.
\end{enumerate}

A Fock state on $\CAR(\fh)$ is a state $\omega$ which is normal
with respect to the vacuum state $\mathbf{vac}$. Such a state is therefore defined
by $\omega(A)=\mathrm{tr}(\rho\pi_F(A))$
where $\rho$ is a density matrix on $ \aFock(\fh)$.
The GNS representation of a Fock state $\omega$ is a
direct sum of  Fock representations, \ie there exists a Hilbert space
$\cK$ such that $\cH_\omega=\cH_F\otimes\cK$
and $\pi_\omega(A)=\pi_F(A)\otimes \one$. Typical examples of Fock
states are finite volume, grand-canonical Gibbs ensembles
$$
\rho=\frac{\e^{-\beta(H_\Lambda-\mu N_\Lambda)}}
{\mathrm{tr}(\e^{-\beta(H_\Lambda-\mu N_\Lambda)})},
$$
for Fermi gases with stable interactions, see~\cite{Bratteli1981}.
Thermodynamic limits of such states yield non-Fock states with finite
density. It is usually impossible to describe explicitly the GNS
representations of these infinite volume \ul{KMS states}. Notable
exceptions are the ideal Fermi gases which lead to the
\ul{Araki-Wyss} re\-pre\-sen\-tations.

Since there
exists a selfadjoint (and hence densely defined) number operator
$N=\d\Gamma({\one})$ on the Fock space $\cH_F$, Fock states describe
systems with a finite number of fermions.  A number operator
can be tentatively defined in the GNS representation of any state $\omega$
as follows. For any finite $J\subset I$ denote by $n_J$
the quadratic form associated to the
operator $\sum_{i\in J}\pi_\omega(n_i)$.
For $\Psi\in\cH_\omega$ set $n_\omega(\Psi)=\sup_J n_{J}(\Psi)$.
It can be shown that $n_\omega$ is a closed, non-negative quadratic
form on the domain $D_\omega=\{\Psi\in\cH_\omega\,|\,n_\omega(\Psi)<\infty\}$.
If this domain is dense, then $n_\omega$ is the
quadratic form of a selfadjoint number operator $N_\omega$,
and the state $\omega$ is a Fock state, see~\cite[Section~5.2.3]{Bratteli1981}
for details.

%%%%%%%%%%%%%%%%%%%%%%%%%%%%%%%%%%%%%%
\subsection{Anti-Fock representation}
%%%%%%%%%%%%%%%%%%%%%%%%%%%%%%%%%%%%%%

A state $\mathbf{full(\argdot)}$ describing a completely filled Fermi
sea must satisfy, for any orthonormal basis $\{e_i\,|\,i\in I\}$ and
any finite $J\subset I$, $\mathbf{full}(\prod_{i\in J}(1-n_i))=0$.
It can be obtained using the particle-hole duality. Denote by
$\overline{\argdot}$ an arbitrary complex conjugation on $\fh$
and define the $\ast$-automorphism $\alpha$ by
$\alpha(a(f))=a^\ast(\bar f)$.  Since
$1-n_i=a(e_i)a^\ast(e_i)=\alpha(a^\ast(\bar e_i)a(\bar e_i))$, we
can set $\mathbf{full}=\mathbf{vac}\circ\alpha$.
It follows that
\begin{eqnarray*}
\mathbf{vac}(a(\bar f_1)\cdots a(\bar f_n)a^\ast(\bar g_m)\cdots a^\ast(\bar g_1))
=  \delta_{nm}\det\left\{(g_i|f_j)\right\}.
\end{eqnarray*}
For $u\in\cU$
one has
$$
\alpha(U(u))=\det(u) U(\bar u),
$$
and so  the characteristic function of the filled Fermi sea is
$E_\mathbf{full}(u)=\det(u)$. The corresponding GNS
representation is the anti-Fock representation
$(\cH_F,\pi_{AF},\Omega_F)$ where $\pi_{AF}=\pi_F\circ\alpha$.

If $\fh$ is finite dimensional, then the states $\mathbf{vac}$ and
$\mathbf{full}$ are mutually normal, and the Fock and anti-Fock
representations are equivalent. By fixing an orthonormal basis
$\{e_1,\ldots,e_n\}$ and setting $a_J=\prod_{i\in J}a(e_i)$,
the unitary operator defined by
$U A_J^\ast\Omega_F=A_{I\setminus J}^\ast\Omega_F$ intertwines
$\pi_F$ and $\pi_{AF}$. If $\fh$ is infinite dimensional these two
representations are inequivalent and $\mathbf{full}$ is not
a Fock state.

%%%%%%%%%%%%%%%%%%%%%%%%%%%%%%%%%%%%%%%%%%
\subsection{Jordan-Wigner representation}
%%%%%%%%%%%%%%%%%%%%%%%%%%%%%%%%%%%%%%%%%%

The equivalence of the Fock and anti-Fock representations of CAR
algebras over finite dimensional spaces is a consequence of a more
general fact which we discuss briefly in this last
section. We refer the reader to~\cite{Derezinski2006a} for a more detailed
discussion.

If $\fh$ is finite dimensional, then $\CAR(\fh)$ is $\ast$-isomorphic to the
full matrix algebra $\mathrm{Mat}(2^{\dim\fh})$. An explicit
representation is provided by the Jordan-Wigner transformation
described below. Since it maps fermions into quantum spins, this
transformation is  useful in many applications
to statistical mechanics.

Let $\{e_1,\ldots,e_n\}$ be an orthonormal basis of $\fh$ and denote
by $\sigma^{(1)}$, $\sigma^{(2)}$, $\sigma^{(3)}$ the usual Pauli matrices.
On the $n$-fold tensor product $\cH=\CC^2\otimes\cdots\otimes\CC^2\simeq\CC^{2^n}$ 
define
$$
\sigma_k^{(\alpha)}=\one\otimes\cdots\otimes\sigma^{(\alpha)}\cdots\otimes \one,
$$
where $\sigma^{(\alpha)}$ acts on the $k$-th copy of $\CC^2$.  Clearly,
this operators generate the full matrix algebra $\cB(\cH)\simeq\mathrm{Mat}(2^n)$. 
One easily checks that the operators
$$
a_k=\sigma_1^{(3)}\cdots\sigma_{k-1}^{(3)}(\sigma_k^{(1)}-\i\sigma_k^{(2)})/2,
$$
satisfy $[a_k,a_l]_+=0$ and $[a_k,a_l^\ast]_+=\delta_{k,l}$.  The
Jordan-Wigner representation of $\CAR(\fh)$ is defined by
$$
a_\mathrm{JW}\left(\sum_kz_ke_k\right)=\sum_k\bar z_ka_k.
$$
The inversion formulas
$$
\sigma_k^{(3)}=2a_k^\ast a_k-{\one},\quad
\sigma_k^{(1)}=\sigma_1^{(3)}\cdots\sigma_{k-1}^{(3)}(a_k+a_k^\ast),
\quad\sigma_k^{(2)}=\i\sigma_1^{(3)}\cdots\sigma_{k-1}^{(3)}(a_k-a_k^\ast),
$$
show that $\CAR(\fh)$ is isomorphic to $\mathrm{Mat}(2^n)$.  For fermionic systems, the Jordan–Wigner representation plays  the same distinguished role as the Schrödinger representation in the CCR context: when 
$\dim\fh<\infty$, all irreducible representations of 
$\CAR(\fh)$ are unitarily equivalent to it.

%%%%%%%%%%%%%%%%%%%%%%%%%%%%%%%
\part{Non-equilibrium systems}
%%%%%%%%%%%%%%%%%%%%%%%%%%%%%%%

%%%%%%%%%%%%%%%%%%%%%%%%%%%%%%%%
\section{Quantum Koopmanism}
\label{sec:Quantum Koopmanism}
%%%%%%%%%%%%%%%%%%%%%%%%%%%%%%%%

The Koopman-von Neumann spectral approach to ergodic theory is a
powerful tool in the study of statistical properties of dynamical
systems. Its extension to quantum dynamical systems -- the spectral
theory of Liouvilleans -- is at the center of many recent results in
quantum statistical mechanics, see Section~\ref{sec:NESS in quantum statistical mechanics}
and~\cite{Bach2000,Derezinski2003,Froehlich2004,Jaksic1996a}.

Let $(\cO,\tau)$ be a \ul{$C^\ast$- or $W^\ast$-dynamical system} equipped
with a $\tau$-invariant state $\omega$, assumed to be normal in the
$W^\ast$-case. By the \ul{GNS-representation
$(\cH_\omega,\pi_\omega,\Omega_\omega)$, to the triple
$(\cO,\tau,\omega)$ we associate} $(\cO_\omega,\tilde\tau,\tilde\omega)$, a
$W^\ast$-dynamical system on the \ul{enveloping von Neumann algebra}
$\cO_\omega=\pi_\omega(\cO)''$ with a normal invariant state
$\tilde\omega(A)=(\Omega_\omega|A\Omega_\omega)$.  The
$W^\ast$-dynamics $\tilde\tau$ is given by
$$
\tilde\tau^t(A)=\e^{\i tL_\omega}A\e^{-\i tL_\omega},
$$
where $L_\omega$ is the $\omega$-Liouvillean, see
Section~\ref{ssec:Liouvilleans}.

We shall say that
$(\pi_\omega,\cO_\omega,\cH_\omega,L_\omega,\Omega_\omega)$ is the
normal form of $(\cO,\tau,\omega)$. 

%%%%%%%%%%%%%%%%%%%%%%%%%%%%%%%%%%%%%%%%%%%%%%%%%%%%%%%%%%%%%
\subsection{Ergodic properties of quantum dynamical systems}
\label{sec-ergodic-property}
%%%%%%%%%%%%%%%%%%%%%%%%%%%%%%%%%%%%%%%%%%%%%%%%%%%%%%%%%%%%%

Let $\fM$ be a \ul{von Neumann algebra} acting on a Hilbert space
$\cH$.  The support $s_\omega$ of a \ul{normal state} $\omega$ on
$\fM$ is the smallest orthogonal projection $P\in\fM$ such that
$\omega(P)=1$. A normal state $\omega$ is faithful if and only if
$s_\omega={\one}$.   It is easy to see that the support of the state $\omega(A)=(\Omega|A\Omega)$
is the orthogonal projection on the closure of the subspace
$\fM'\Omega$.

\medskip\noindent\textbf{Notation.}\, We write $\nu\ll\omega$ whenever $\nu$ is a
\ul{$\omega$-normal} state such that $s_\nu\le s_\omega$.

\remark If $\fM$ is Abelian, then  any $\omega$-normal state
$\nu$ satisfies $\nu\ll\omega$. This explains why the support
condition is absent in the classical ergodic theory (the reader may
consult~\cite{Pillet2006} for a detailed discussion of this point).
In most applications to statistical mechanics, $\omega$ is faithful,
and any $\omega$-normal state $\nu$ satisfies $\nu\ll\omega$.

\begin{definition}\label{def:ergoprops}
Let $(\fM,\tau)$ be a $W^\ast$-dynamical system and $\omega$ a 
normal $\tau$-invariant state.
\begin{enumerate}%[(i)]
\item $(\fM,\tau,\omega)$ is ergodic if
$$
\lim_{T\to\infty}\frac{1}{T}\int_0^T\omega(As_\omega \tau^t(B))\,\d t=\omega(A)\omega(B),
$$
holds for any $A,B\in \fM$.
\item $(\fM,\tau,\omega)$ is weakly mixing if
$$
\lim_{T\to\infty}\frac{1}{T}\int_0^T\left|\omega(As_\omega\tau^t(B))-\omega(A)\omega(B)\right|\,\d t=0,
$$
holds for all $A,B\in \fM$.
\item $(\fM,\tau,\omega)$ is mixing or returns to equilibrium if
$$
\lim_{t\to\infty}\omega(As_\omega\tau^t(B))=\omega(A)\omega(B),
$$
for any $A,B\in\fM$

\item If $\omega$ is an invariant state of the $C^\ast$-dynamical system
$(\cO,\tau)$, we say that $(\cO,\tau,\omega)$ is ergodic (resp. mixing, weakly mixing) if
$(\cO_\omega,\tilde\tau,\tilde\omega)$ is ergodic (resp. mixing, weakly mixing).
\end{enumerate}
\end{definition}

Clearly, we have the implication (3)$\Rightarrow $(2)$\Rightarrow$ (1).

\remark Ergodicity~(1) is equivalent to the statement that 
$$
\lim_{T\to\infty}\frac{1}{T}\int_0^T\nu(\tau^t(A))\,\d t=\omega(A),
$$
holds for all $A\in\fM$ and all states $\nu\ll\omega$. 
The mixing property~(3) is equivalent to the statement that 
$$
\lim_{t\to\infty}\nu(\tau^t(A))=\omega(A),
$$
holds for all $A\in\fM$ and all states $\nu\ll\omega$, see~\cite{Robinson1976,Jaksic1996a}.

%%%%%%%%%%%%%%%%%%%%%%%%%%%%%%%%%%%%%%%%%%%%%%%%%%%%%%%%%%%%%
\subsection{Spectral characterization of ergodic properties}
%%%%%%%%%%%%%%%%%%%%%%%%%%%%%%%%%%%%%%%%%%%%%%%%%%%%%%%%%%%%%

We refer to~\cite{Pillet2006} for proofs of the results in this section.

The following theorem is the quantum version of the well-known
Koopman-von Neumann spectral characterizations,
\cite{Arnold1968,Koopman1931,vonNeumann1932b}.

\begin{theorem} Let $(\cO,\tau)$ be a $C^\ast$- or $W^\ast$-dynamical system
equipped with a $\tau$-invariant state $\omega$, assumed to be normal in the
$W^\ast$-case. Denote by $(\pi_\omega,\cO_\omega,\cH_\omega,L_\omega,\Omega_\omega)$
its normal form, and by $\cK_\omega$ the closure of
$\pi_\omega(\cO)'\Omega_\omega$.
\begin{enumerate}%[(i)]
\item The subspace $\cK_\omega$ reduces the operator $L_\omega$.
Denote by $\fL_\omega$ the restriction $L_\omega\bigr|_{\cK_\omega}$.
\item $(\cO,\tau,\omega)$ is ergodic if and only if
$\Ker(\fL_\omega)$ is one-dimensional.
\item $(\cO,\tau,\omega)$ is weakly mixing if and only if $0$ is the
only eigenvalue of $\fL_\omega$ and $\Ker(\fL_\omega)$ is one dimensional.
\item $(\cO,\tau,\omega)$ is mixing if and only if
$$
\wlim_{t\to\infty}\e^{\i t\fL_\omega}=\Omega_\omega(\Omega_\omega|\argdot).
$$
\item If the spectrum of $\fL_\omega$ on $\{\Omega_\omega\}^\perp$
is purely absolutely continuous, then $(\cO,\tau,\omega)$ is mixing.
\end{enumerate}
\end{theorem}

Note that $\cK_\omega$ is the range of the support of $\tilde\omega$,
 that is, $\cK_\omega=\pi_{\tilde\omega}\cH_\omega$. Thus,
if $\tilde\omega$ is faithful, then $\fL_\omega=L_\omega$.

Like the classical Koopman operator, the reduced Liouvillean $\fL_\omega$ of an
ergodic quantum dynamical system exhibits several special  spectral properties.

\begin{theorem} Assume, in addition to the hypotheses of the previous theorem, that
$(\cO,\tau,\omega)$ is ergodic. Then the following hold:
\begin{enumerate}%[(i)]
\item The point spectrum of $\fL_\omega$ is a subgroup $\Sigma$
 of the additive group $\RR$.
\item The eigenvalues of $\fL_\omega$ are simple.
\item The spectrum of $\fL_\omega$ is invariant under translations in $\Sigma$,
 that is, $\mathrm{spec}(\fL_\omega)+\Sigma=\mathrm{spec}(\fL_\omega)$.
\item If\, $\Psi$ is a normalized eigenvector of $\fL_\omega$, then
$(\Psi|\pi_\omega(A)\Psi)=\omega(A)$ for all $A\in\cO$.
\item If $\omega$ is a $(\tau,\beta)$-\ul{KMS state} then
  $\Sigma=\{0\}$ and the system is weakly mixing.
\end{enumerate}
\end{theorem}

%%%%%%%%%%%%%%%%%%%%%%%%%%%%%%%%%%%%%%%%%%
\section{Nonequilibrium steady states}
\label{sec:Nonequilibrium steady states}
%%%%%%%%%%%%%%%%%%%%%%%%%%%%%%%%%%%%%%%%%%

Equilibrium states lie at the heart of equilibrium statistical mechanics. Much
of the success of this theory stems from the fact that these states can be
constructed and characterized without explicit reference to the underlying
dynamics. This is remarkable, given that the very notion of an equilibrium
state—and the traditional justification of the theory following
Boltzmann—ultimately relies on the dynamics.

Nonequilibrium statistical mechanics has a fundamentally different status. It is
conceptually and technically more demanding, since a detailed understanding of
the dynamics is required even to capture the most elementary nonequilibrium
properties of a system. Among the vast manifold of nonequilibrium states, the
simplest are the steady states that arise when the system is subjected to weak,
stationary external forces. Such forcing may be implemented by an external field
or by maintaining a gradient of intensive thermodynamic parameters across the
system—for instance, a constant temperature drop. Despite significant progress in recent years, a coherent and comprehensive theory of steady states remains out of reach, if such a theory  exists at all.
%%%%%%%%%%%%%%%%%%%%%%%%%%%%%%%%%%%%%%%%%%%%%%%%%%%%%%%%%
\subsection{Phenomenological theory and linear response}
%%%%%%%%%%%%%%%%%%%%%%%%%%%%%%%%%%%%%%%%%%%%%%%%%%%%%%%%%

To appreciate recent rigorous results in the field, a basic knowledge
of nonequilibrium thermodynamics is required. This section is a
condensed introduction to the phenomenological theory of
nonequilibrium steady states (NESS). The interested reader should
consult~\cite{Groot1962} or~\cite{Callen1985} for detailed expositions.

Consider bringing into contact two systems $\cS_1$, $\cS_2$, each of
which is in thermal equilibrium. Denote by 
\[S_\alpha(A_{\alpha
  1},A_{\alpha 2},\ldots)\]
   the entropy of $\cS_\alpha$ as a function
of its extensive thermodynamic parameters.  The combined system
$\cS_1+\cS_2$, being otherwise isolated, has the values of $A_j=A_{1
  j}+A_{2 j}$  fixed, and the entropy of the joint system is given
by $S=\sum_\alpha S_\alpha(A_{\alpha 1},A_{\alpha 2},\ldots)$.  The
conditions for joint thermal equilibrium are thus
$$
X_j=\frac{\partial S}{\partial A_{1j}}=
\frac{\partial S_1}{\partial A_{1 j}}
-\frac{\partial S_2}{\partial A_{2 j}}
=I_{1 j}-I_{2 j}=0,
$$
where $I_{\alpha j}$ denotes the intensive thermodynamic parameter
conjugate to $A_{\alpha j}$. If the combined system is not in thermal
equilibrium, then the $X_j\not=0$ act as thermodynamic forces. Their effect
is to generate fluxes of the extensive quantities;  these fluxes are given by 
$$
\Phi_j=\frac{\d A_{1 j}}{\d t}.
$$
The rate of entropy production is then
\beq
\frac{\d S}{\d t}=\sum_j\frac{\partial S}{\partial A_{1 j}}
\frac{\d A_{1 j}}{\d t}=\sum_j X_j\Phi_j.
\label{EPdiscrete}
\eeq
Near equilibrium, the forces $X_j$ are weak, and the first-order
perturbation theory in these forces -- also called linear response
theory -- becomes a good approximation. Writing the fluxes as
\beq
\Phi_j=\sum_k L_{kj}X_k+\text{higher order terms}
\label{linresp}
\eeq
defines the {\it kinetic} or {\it transport coefficients} $L_{kj}$.
The matrix $L=(L_{kj})$ is called the {\it Onsager matrix.} It depends only on
the intensive parameters $I_{1 j}$. One of the most basic problems of
nonequilibrium statistical mechanics is the calculation of Onsager's
matrix starting from a microscopic description of the system.

In linear response theory the entropy production rate becomes
$$
\frac{\d S}{\d t}\simeq\sum_k L_{kj}X_kX_j.
$$
The second law of thermodynamics implies that the symmetric part
of $L$ is positive. Moreover, if the systems are time-reversal
invariant, then the Onsager reciprocity relations
$$
L_{kj}=L_{jk},
$$
hold, \ie the matrix $L$ is symmetric. If the systems are not time
reversal invariant, these relations have to be appropriately modified.
For example, if an external magnetic field $B$ is applied, then the
matrix $L$ depends parametrically on $B$, and the Onsager-Casimir
relations
$$
L_{kj}(B)=L_{jk}(-B),
$$
hold.

\medskip\remarks\textbf{1.} Applying the above discussion to small volume elements of a
macroscopic body, it is possible to obtain a phenomenological
description of the local structure of a nonequilibrium state under the
so-called local thermodynamic equilibrium (LTE) hypothesis.
The reader should consult the above mentioned references for details.

\noindent\textbf{2.} Some variational characterizations of NESS have been
proposed, most notably the principle of minimal entropy production.
However, the status of such principles is still controversial, and their
validity seems to be only approximate and limited to some special
systems. See~\cite{Maes2007} for more information on this topic.

%%%%%%%%%%%%%%%%%%%%%%%%%%%%%%%%%
\subsection{Microscopic theory}
%%%%%%%%%%%%%%%%%%%%%%%%%%%%%%%%%

A microscopic theory of NESS must be based on the asymptotic analysis of the
dynamics which describes the evolution of the system. A non-isolated system is
driven out of equilibrium by forces exerted upon it by its environment. Under
appropriate conditions the system eventually settles in a steady state. Denote
by $\langle\argdot\rangle_0$ the initial state, in the sense of statistical
mechanics, of the system and its environment. If
$t\mapsto\langle\argdot\rangle_t$ denotes the evolution of this state, then the
limit
$$
\langle\argdot\rangle_+=\lim_{t\to\infty}\langle\argdot\rangle_t,
$$
or more generally
$$
\langle\argdot\rangle_+=\lim_{t\to\infty}\frac{1}{t}\int_0^t
\langle\argdot\rangle_s\,\d s,
$$
defines a stationary state.

On physical grounds, one expects more -- namely, that the limiting state
$\langle\argdot\rangle_+$ is  insensitive to local perturbations  of
the initial state $\langle\argdot\rangle_0$. To elaborate this
essential point, let us consider a smooth dynamical system on a compact
phase space $X\subset\RR^n$. A physically natural statistics of the
configurations of the system is provided by the normalized Lebesgue
measure $\d x$ on $X$: initial configurations sampled according to $\d
x$ can be considered typical. This choice is independent of the dynamics, and
any distribution $\rho(x)\d x$ with strictly positive density $\rho$
would serve the same purpose. Accordingly, we expect that all these  initial
states  lead to  the same limiting state
$\langle\argdot\rangle_+$. 
For systems with infinitely many degrees of freedom, no analogue of Lebesgue measure exists, so a reference state is required to sample initial configurations.
The purpose of this state is to specify
the thermodynamic state of the system.  Suppose, for example, that the
system is driven out of equilibrium by several infinite reservoirs.
Specifying an equilibrium state (a Gibbs measure or a KMS state) for
each reservoir sets the temperature and other relevant thermodynamic
parameters of each reservoir. States with different thermodynamic
parameters are mutually singular and will lead to different NESS. 
Local perturbations of these equilibrium states will relax
to equilibrium under the reservoir dynamics. Therefore, as in the
finite dimensional case, we expect the limiting state
$\langle\argdot\rangle_+$ to be insensitive to such local
perturbations. In the language of \ul{$C^\ast$-dynamical systems}, the
NESS should only depend on the \ul{folium} of the initial state.

The minimal goal of a microscopic theory of NESS is a mathematical
derivation of the main results of linear response theory: Onsager
reciprocity relations and fluctuation-dissipation relations (e.g. the
Green-Kubo formula for the kinetic coefficients $L_{kj}$); see
Section~\ref{sec:Linear response theory}.
A more ambitious program is to explore the largely unknown domain of
far from equilibrium steady-state thermodynamics.
Setting aside the obvious distinction between classical and quantum systems,
recent progress in these directions can be grouped into two categories,
depending on the chosen description of the microscopic dynamics: Hamiltonian or
Markovian. The Hamiltonian description is, of course, the more fundamental, as
microscopic dynamics is inherently Hamiltonian.
However, the facts that
\begin{itemize}
\item various routes (scaling limits, coarse graining, restriction to
specific degrees of freedom, etc\dots) lead from Hamiltonian
dynamics to Markov processes;
\item Hamiltonian dynamics is much more difficult to control than
Markovian evolution;
\end{itemize}
explain why most  of the available results belong to the second category. In
fact, the Markovian  time evolution was already at the heart
of the pioneering works of Onsager and his followers, see~\cite{Groot1962}.

A large body of works have been devoted to the study of interacting
particle systems, see~\cite{Kipnis1999,Bertini2006}.  We do not consider such models here, and instead focus on a small mechanical system 
$\cS$ with finitely many degrees of freedom, driven out of equilibrium by external forcing.

In the framework of classical mechanics, two approaches are possible.
In what we will call {\it the canonical approach}, the system $\cS$
interacts with infinitely extended reservoirs $\cR_1$, $\cR_2$ \dots
Apart from its physical appeal, the advantage of the canonical
approach is that it allows for a Hamiltonian description of the
coupled system $\cS+\cR_1+\cdots$. A mathematical disadvantage is the
necessity to deal with infinitely many degrees of freedom. As a result,
only very simple reservoirs (e.g., ideal gases or free fields) are 
considered. These simple systems also act as almost ideal reservoirs
in the sense that, at least in the dimension larger than two, the local
properties of their internal state do not change too much as a result
of the coupling to the system. Moreover, it is possible to design the
coupling of these reservoirs to the system in such a way that the
dynamics of $\cS$ becomes essentially Markovian, see~\cite{Eckmann1999b}.
This is a significant feature, one that is seldom realized in other 
approaches. We refer to~\cite{Rey-Bellet2006} for an introduction to 
recent results in the canonical approach to classical NESS.

It is important to realize that, in the canonical approach, the
reservoirs serve two complementary purposes. First, they have a chaotic
internal dynamics. The forces they exert on the system $\cS$ act as a
source of randomness. Under the action of these forces, the dynamics
of $\cS$ becomes itself chaotic. Second, the dynamics of the reservoir
is dissipative: Any local perturbation gets carried away to spatial
infinity. This allows the system $\cS$ to relax from large
fluctuations in its internal state by transferring energy, momentum,
\dots to the reservoirs.

In {\it the microcanonical approach} the same effects of the environment on the
motion of the system are obtained in a different way. Fluctuations of the system
$\cS$ are generated by a non-conservative external force. Under the action of
this force the system would constantly heat-up. To avoid this effect and allow
the system to settle in a stationary state, a dissipative force is added -- a
so-called thermostat.  A popular example is the Gaussian thermostat which keeps
the energy (or the kinetic energy) of $\cS$ constant. The advantage of the
microcanonical approach is that it leads to a dynamical system on a finite
dimensional compact manifold (a surface of constant energy of $\cS$).  In
particular, numerical experiments are much easier to perform on such a system
than on the infinite dimensional systems obtained in the canonical approach.
This fact has recently led to the numerical discovery of an unexpected
fluctuation relation in far from equilibrium NESS~\cite{Evans1993}. This
relation has been turned into a mathematical statement by combining the
microcanonical approach with an axiomatic setup -- the so called {\it chaotic
hypothesis}~\cite{Gallavotti1995b}. We refer to~\cite{Ruelle1999,Dorfman1999}
for detailed introductions to this approach. A proof of the fluctuation relation
for the canonical NESS of an anharmonic chain of oscillators has also been
obtained in~\cite{Rey-Bellet2002a}. More recently the fluctuation relation has
been observed in real experiments in various systems of very different nature,
confirming its apparent universality.

Equivalence of the equilibrium ensembles (microcanonical, canonical,
grand-canonical,\dots) is a cornerstone of equilibrium statistical
mechanics. It is conjectured
that a similar equivalence holds for NESS. In the thermodynamic
limit, \ie as the size of $\cS$ becomes large, one expects that
the statistical properties of macroscopic observables in a NESS
do not depend on the construction of the NESS. Microcanonical
NESS with different kinds of thermostats, as well as canonical
NESS with different kinds of reservoirs, should yield the same
statistics. To our knowledge the only available result in this direction
is~\cite{Ruelle2000a}.

Due to the intrinsic Hamiltonian nature of quantum mechanics the
canonical approach is the only way to define and study the NESS of a
quantum system (at least within a purely quantum description). We
refer to Section~\ref{sec:NESS in quantum statistical mechanics} for a detailed
discussion of quantum NESS.

%%%%%%%%%%%%%%%%%%%%%%%%%%%%%%%%%%%%%%%%%%%%%%%%%%%
\section{NESS in quantum statistical mechanics}
\label{sec:NESS in quantum statistical mechanics}
%%%%%%%%%%%%%%%%%%%%%%%%%%%%%%%%%%%%%%%%%%%%%%%%%%%

In this section  we describe the construction of canonical
non-equilibrium steady states (NESS) for a small quantum system $\cS$
coupled to several extended reservoirs $\cR_1$,\dots,$\cR_M$; see
Section~\ref{sec:Nonequilibrium steady states}. We shall work in the framework of
\ul{$C^\ast$-dynamical systems}, and denote by $\cO_0$ the $C^\ast$-algebra
of $\cS$, which we assume to be finite dimensional.  Each reservoir
$\cR_j$ is described by a $C^\ast$-algebra $\cO_j$. For simplicity, we
assume that the algebra of the joint system $\cS+\cR_1+\cdots+\cR_M$
is the \ul{$C^\ast$-tensor product}
$\cO=\cO_\cS\otimes\cO_\cR=\otimes_{0\le a\le M}\cO_a$. The following
is easily adapted to more general cases, e.g., fermionic algebras.

For $0\le a\le M$, let $(\cO_a,\tau_a)$ be the $C^\ast$-dynamical
system describing the isolated subsystem $a$. The dynamics of the decoupled
joint system is $\tau=\otimes_{0\le a\le M}\tau_a$.
The dynamics $\tau_V$ of the coupled joint system is
a local perturbation of $\tau$ induced by
$$
V=\sum_{1\le j\le M}V_j,\quad V_j=V_j^\ast\in\cO_0\otimes\cO_j,
$$
where $V_j$ is the interaction between $\cS$ and $\cR_j$,
see Section~\ref{sec:Quantum dynamical systems}.

\begin{definition} Let $\omega$ be a state on $\cO$. We say that
$\omega_+$ is a NESS of $\tau_V$ associated to the reference
state $\omega$ if there exists a net $t_\alpha\to\infty$ such that
$$
\omega_+(A)=\lim_\alpha\frac{1}{t_\alpha}\int_0^{t_\alpha}
\omega\circ\tau_V^t(A)\,\d t,
$$
for all $A\in\cO$. We denote by $\Sigma_+(\tau_V,\omega)$ the set
of these NESS.
\end{definition}

If the algebra ${\cal O}$ is separable, the  net can be replaced by a sequence. 

A few remarks are in order:
\begin{enumerate}%[(i)]
\item By definition, the elements of $\Sigma_+(\tau_V,\omega)$ are
  $\tau_V$-invariant states on $\cO$. Moreover, if $\omega$ is such a state,
  then $\Sigma_+(\tau_V,\omega)=\{\omega\}$.
\item 
The limit $\omega_+$ may turn out to be  a KMS state for $\tau_V$. This
occurs trivially if $\omega$ is such a state, but is also expected
when $\omega$ is (normal relative to) a KMS state
for the decoupled dynamics $\tau$. This is {\sl return to equilibrium},
see Definition~\ref{def:ergoprops}(3).
In this case $\omega_+$, will be $\omega$-normal. Consequently, in this special case the terminology NESS may be misleading, as the state actually corresponds to an equilibrium situation. In genuine
nonequilibrium cases $\omega_+$ is expected to be singular
with respect to $\omega$.
\item Entropy production plays a central role in nonequilibrium
statistical mechanics. We refer to Section~\ref{sec:Entropy production} for
a discussion of related properties of NESS. Let us simply  mention
here that NESS have a non-negative entropy production rate.
\item Since the set of all states on $\cO$ is weak-$\ast$ compact,
$\Sigma_+(\tau_V,\omega)$ is nonempty.
\item If the perturbation $V$ is time-dependent, then natural
  nonequilibrium states (NNES) are defined in a similar way as limit
  points
$$
\omega_+^t(A)=
\lim_\alpha\frac{1}{t_\alpha}\int_{-t_\alpha}^t\omega\circ\tau_V^{s\to t}(A)
\,\d s.
$$
They satisfy $\omega_+^t\circ\tau_V^{t\to r}=\omega_+^r$, see~\cite{Ruelle2000}.
\end{enumerate}

As stressed in Section~\ref{sec:Nonequilibrium steady states}, a NESS should be
insensitive to local perturbations of the initial state $\omega$. The
following result shows that this is indeed the case under a rather weak
ergodic hypothesis (see also~\cite{Jaksic2002b}).

\begin{theorem}[\cite{Aschbacher2006}]\label{thm:WAAM}
Assume that $\omega$ is a \ul{factor state} on $\cO$
  and that, for any $\omega$-normal state $\eta$,
$$
\lim_{t\to\infty}\frac{1}{t}\int_0^t\eta([\tau_V^s(A),B])\,\d s=0
$$
holds for all $A$, $B$ in a dense subset of $\cO$
(weak asymptotic Abelianness in mean). Then $\Sigma_+(\tau_V,\eta)
=\Sigma_+(\tau_V,\omega)$ holds for all $\omega$-normal
states $\eta$.

\end{theorem}

In typical applications the reference state $\omega$ is
 specified by the requirement that its restrictions to the
subalgebras $\cO_a$ are $\beta_a$-\ul{KMS states}\footnote{chemical
potentials can also be
prescribed by an appropriate definition of $\tau$} for the corresponding
dynamics $\tau_a$. This means that
$\omega$ is a KMS state at inverse
temperature $-1$ for the dynamics $\sigma_\omega^t=
\otimes_a\tau_a^{-\beta_a t}$. In particular, $\omega$ is
\ul{modular} and $\sigma_\omega$ is its modular group,
see Section~\ref{sec:Tomita--Takesaki theory}.
The group $\sigma_\omega$ plays an important and somewhat
unexpected role in the mathematical theory
of linear response, see Section~\ref{sec:Linear response theory}.

Accordingly, we shall assume in the remaining part of this paragraph
that $\omega$ is modular, and we denote by
$(\cH_\omega,\pi_\omega,\Omega_\omega)$ the
corresponding \ul{GNS representation} of $\cO$. The \ul{enveloping
von Neumann algebra} $\pi_\omega(\cO)''$ is in \ul{standard
form}, and we denote by $J$ the \ul{modular conjugation}.
If $L$ is the \ul{standard Liouvillean} of $\tau$, then
$L_V=L+\pi_\omega(V)+J\pi_\omega(V)J$ is the standard
Liouvillean of $\tau_V$. The spectral analysis of $L_V$
yields interesting information on the structure of
$\Sigma_+(\tau_V,\omega)$, see~\cite{Aschbacher2006}.

\begin{theorem} Assume that the state $\omega$ is modular.
\begin{enumerate}%[(i)]
\item If\, $\Ker\,L_V=\{0\}$, then there is no $\omega$-normal
$\tau_V$-invariant state. In particular, any NESS in
$\Sigma_+(\tau_V,\omega)$ is purely $\omega$-singular.
\item If the assumptions of Theorem~\ref{thm:WAAM} hold and
if $\Ker\,L_V\not=\{0\}$, then $\Ker\,L_V$ is one-dimensional and there
exists a unique $\omega$-normal $\tau_V$-invariant state
$\omega_V$. Moreover, $\Sigma_+(\tau_V,\omega)=\{\omega_V\}$.
\end{enumerate}
\end{theorem}

As already mentioned, case~\textit{(i)} in the above theorem is the
expected behavior out of equilibrium while case~\textit{(ii)} 
describes a typical equilibrium situation.

To our knowledge, there are two approaches to the construction
of NESS which we now describe.

%%%%%%%%%%%%%%%%%%%%%%%%%%%%%%%%%%%%%
\subsection{The scattering approach}
%%%%%%%%%%%%%%%%%%%%%%%%%%%%%%%%%%%%%

The first approach was proposed by Ruelle in~\cite{Ruelle2000} and relies on the
scattering theory of $C^\ast$-dynamical systems, see~\cite{Robinson1973}.  We
also refer to~\cite{Froehlich2003,Jaksic2007a} for related works.

The scattering approach assumes the existence of the
strong limit
\beq
\alpha_V=\slim_{t\to\infty}\tau^{-t}\circ\tau_V^t.
\label{moller}
\eeq
This limit defines an isometric $\ast$-endomorphism
of $\cO$ such that $\alpha_V\circ\tau_V^t=\tau^t\circ\alpha_V$,
often  called M\o ller morphism. $\alpha_V$ is
injective but its range $\cO_+$, a $\tau$-invariant $C^\ast$-subalgebra
of $\cO$, can be strictly smaller than $\cO$. One immediately obtains

\begin{proposition} Assume that the M\o ller morphism~\eqref{moller}
exists and that $\omega$ is $\tau$-invariant. Then, for all $A\in\cO$,
$$
\lim_{t\to\infty}\omega\circ\tau_V^t(A)=\omega_+(A),
$$
where $\omega_+=\omega\circ\alpha_V$. In particular, one
has $\Sigma_+(\tau_V,\omega)=\{\omega_+\}$.
\label{scattering}
\end{proposition}

If the previous proposition applies, then $\alpha_V$
provides an isomorphism between the
the coupled dynamical system $(\cO,\tau_V,\omega_+)$
and the decoupled one
$(\cO_+,\tau|_{\cO_+},\omega|_{\cO_+})$. 
Ergodic properties of the latter are therefore  inherited
by the former. The following proposition is a simple
consequence of this fact, see~\cite{Aschbacher2006}.

\begin{proposition} Assume that the assumptions of
Proposition~\ref{scattering} hold.
\begin{enumerate}%[(i)]
\item If $\omega|_{\cO_+}$ is ergodic for $\tau|_{\cO_+}$, then
$\Sigma_+(\tau_V,\eta)=\{\omega_+\}$ for any $\omega$-normal
state $\eta$.
\item If $\omega|_{\cO_+}$ is mixing for $\tau|_{\cO_+}$, then
$$
\lim_{t\to\infty}\eta\circ\tau_V^t(A)=\omega_+(A)
$$
holds for all $A\in\cO$ and any $\omega$-normal state $\eta$.
\end{enumerate}
\end{proposition}

For a finite system coupled to infinite reservoirs, we expect that
$\cO_+=\cO_\cR$, and  that the coupled system out of equilibrium
inherits the ergodic properties of the reservoirs.

At present the only known technique to study $C^\ast$-scattering  is the
basic Cook’s method, which requires   rather restrictive assumptions. We refer to~\cite{Ruelle2000,Froehlich2003,Aschbacher2007,Jaksic2007a} for
more details and examples.
The Hilbert-space techniques, which we will describe in the next
subsection, are often  a more flexible alternative.

%%%%%%%%%%%%%%%%%%%%%%%%%%%%%%%%%%%%%%
\subsection{The Liouvillean approach}
%%%%%%%%%%%%%%%%%%%%%%%%%%%%%%%%%%%%%%

This alternative to the scattering approach has been proposed
in~\cite{Jaksic2002a}, where the NESS of a $N$-level quantum system
coupled to ideal Fermi reservoirs is constructed. For this kind of
systems it has not yet been possible to obtain the propagation
estimates needed to construct the  M\o ller morphism. In fact it
is not clear that the scattering approach applies in this case.

In the Liouvillean approach, NESS are
related to resonances of a new kind of generator of the
dynamics in the GNS representation called  the $C$-Liouvillean.
The main advantage of this method is that the
required analysis can be performed in a Hilbert space setting.
The technical difficulties are related to the fact that the
$C$-Liouvillean is not selfadjoint on the GNS Hilbert space. 
Here we describe only the strategy and refer the reader to~\cite{Jaksic2002a} for details of the implementation.

We assume that $\omega$ is modular and work directly in the
GNS representation $(\cH_\omega,\pi_\omega,\Omega_\omega)$,
identifying $\cO$ with $\pi_\omega(\cO)$. Recall that $\sigma_\omega$
is the modular group of $\omega$, $J$ is  the modular
conjugation, and $L$, $L_V$ are  the standard Liouvilleans of $\tau$,
$\tau_V$. Denote by $\Delta_\omega$ the \ul{modular operator}.

\begin{definition} If $t\mapsto\sigma_\omega^t(V)$
is analytic in the strip $\{z\in\CC\mid|\Im z|<1/2\}$ and bounded
continuous in its closure then the $C$-Liouvillean of $\tau_V$ is
the closed operator defined on the domain of $L$ by
$$
K_V=L+V-J\sigma_\omega^{-\i/2}(V)J.
$$
\end{definition}

Since $J\sigma_\omega^{-\i/2}(V)J\in\pi_\omega(\cO)'$, one
easily checks that
$\e^{\i tK_V}A\e^{-\i tK_V}=\tau_V^t(A)$.
Moreover, since $L\Omega_\omega=0$, it follows from modular
theory that
$$
K_V\Omega_\omega=V\Omega_\omega-J\Delta_\omega^{1/2}V
\Delta_\omega^{-1/2}J\Omega_\omega
=
V\Omega_\omega-J\Delta_\omega^{1/2}V\Omega_\omega
=(V-V^\ast)\Omega_\omega
=0.
$$
Hence $\omega\circ\tau_V^t(A)=(\Omega_\omega|\e^{\i tK_V}A
\Omega_\omega)=
(\e^{-\i tK_V^\ast}\Omega_\omega|A\Omega_\omega)$ where
$K_V^\ast=L+V-J\sigma_\omega^{\i/2}(V)J$.

Suppose that there exists a  triplet of rigged Hilbert spaces 
$\cK\subset\cH_\omega\subset\cK'$ and a dense subalgebra
$\tilde\cO\subset\cO$ such that $\tilde\cO\Omega_\omega\subset\cK$
and 
$$
\wstarlim_{t\to\infty}\frac{1}{t}\int_0^t
\e^{-\i sK_V^\ast}\Omega_\omega\,\d s=\Psi\in\cK'
$$
holds in $\cK'$. Then the functional $\tilde\cO\ni
A\mapsto(\Psi|A\Omega_\omega)$
extends by continuity to a state $\omega_+$
on $\cO$ and we can conclude that $\Sigma_+(\tau_V,
\omega)=\{\omega_+\}$.
Note that if $\Psi\in\cH_\omega$ then $\omega_+$ is
$\omega$-normal. Thus, we expect that $\Psi\not\in\cH_\omega$
in genuine nonequilibrium situations. Under appropriate conditions,
one can show that $\Psi$ is a zero-resonance vector of
$K_V^\ast$, \ie that there exists an extension of $K_V^\ast$
to $\cK'$ for  which $\Psi$ is a zero eigenvector. In~\cite{Jaksic2002a}, and more
recently in~\cite{Merkli2007a}, spectral deformation techniques have been
used to gain perturbative control on the resonances of $K_V^\ast$.
This yields a convergent expansion for the NESS $\omega_+$ in
powers of the coupling $V$ which, to lowest order, coincide
with the weak coupling (van Hove) limit studied in~\cite{Spohn1978b}.
It also gives the convergence $\nu\circ\tau_V^t(A)\to\omega_+(A)$
for all $\omega$-normal states $\nu$ and all $A\in\cO$
with a precise estimates on the
exponential rate of convergence for dense sets of
such $\nu$ and $A$.

%%%%%%%%%%%%%%%%%%%%%%%%%%%%%%%%%%
\section{Entropy production}
\label{sec:Entropy production}
%%%%%%%%%%%%%%%%%%%%%%%%%%%%%%%%%%

Entropy, as defined by Clausius, is the  cornerstone of equilibrium
thermodynamics. Its statistical interpretation by Boltzmann is the key to our
microscopic understanding of equilibrium. 

A notion of
entropy production (rate) has emerged from recent developments in classical and
quantum statistical mechanics of nonequilibrium steady states.  Taking this
notion seriously, it does not seem possible to define the entropy of a
nonequilibrium steady state~\cite{Ruelle2003a}. As argued
in~\cite{Gallavotti2004a}, if such an entropy exists, then it is most likely to
take the value $-\infty$, because a system in such a state loses entropy at a
constant rate.

The purpose of this section is to introduce the notion of entropy production for
nonequilibrium steady states of a small quantum system in contact with thermal
reservoirs. Throughout, we shall freely use the concepts and notation introduced in Section~\ref{sec:NESS in quantum statistical mechanics}.

%%%%%%%%%%%%%%%%%%%%%%%%%%%%%%%%
\subsection{Relative entropy}
\label{sec-relative-entropy}
%%%%%%%%%%%%%%%%%%%%%%%%%%%%%%%%

The relative entropy  of two density matrices $\rho$ and $\rho'$ is defined,
in analogy with the relative entropy of two probability measures, by
$$
\Ent(\rho'|\rho)=\tr(\rho'(\log\rho-\log\rho')).
$$
It has been generalized by Araki to arbitrary states on a von Neumann
algebra~\cite{Araki1975/76,Araki1977}. To describe the general definition, we need to
introduce the notion of relative modular operator.

Let $\fM$ be a von Neumann algebra acting on a Hilbert space $\cH$
and let $\Psi,\Phi\in\cH$ be two unit vectors. Denote by $s_\Psi$ the support
of the state $\psi(A)=(\Psi|A\Psi)$, \ie the orthogonal projection
on the closure of $\fM'\Psi$ (see Section~\ref{sec:Tomita--Takesaki theory}).
Denote by $s'_\Psi$ the  support of the
  state on $\fM'$ defined by $\psi'(B)=(\Psi|B\Psi)$,
  $B\in\fM'$, which coincides with
the orthogonal projection onto the closure of $\fM\Psi$.
Since $A,B\in\fM$ and $A\Psi=B\Psi$ imply $A s_\Psi=B s_\Psi$,
formula
$$
A\Psi\mapsto s_\Psi A^\ast\Phi
$$
defines a closable antilinear operator on 
$\fM\Psi$, which is a dense subspace of $s'_\Psi\cH$. The operator has values in $s_\Psi\cH$. Denote by $S_{\Phi|\Psi}$
its closure.  The selfadjoint operator
$\Delta_{\Phi|\Psi}=S_{\Phi|\Psi}^\ast S_{\Phi|\Psi}$ is called
the relative modular operator of the pair $(\Phi,\Psi)$.

\begin{definition}Let $\omega$ be a \ul{modular state} on the $C^\ast$-algebra
$\cO$.
Denote by $(\cH,\pi,\Psi_\omega)$ the induced \ul{GNS re\-presen\-ta\-tion} and
by $\cH_+$ its \ul{natural cone}. For any \ul{$\omega$-normal} state $\nu$ on 
$\cO$, let $\Psi_\nu$ be its unique vector representative in $\cH_+$ (see 
Theorem~\ref{thm:NormalStates}). The entropy of a state $\nu$ relative 
to $\omega$ is defined by
$$
\Ent(\nu|\omega)=\left\{
\begin{array}{ll}
(\Psi_\nu|\log\Delta_{\Psi_\omega|\Psi_\nu}\Psi_\nu),&\text{if $\nu$ is $\omega$-normal},\\[4pt]
-\infty&\text{otherwise}.
\end{array}
\right.
$$
\end{definition}

\remarks\textbf{1.} We have restricted the above definition to modular $\omega$ for
simplicity. To obtain a completely general definition,  it suffices to pass to a \ul{standard
representation} of the \ul{enveloping von Neumann algebra} $\cO_\omega$ if $s_\nu\le s_\omega$,
and to set $\Ent(\nu|\omega)=-\infty$ otherwise.

\noindent\textbf{2.} We use the notation $\Ent(\argdot|\argdot)$ of~\cite{Bratteli1981,Ohya1993}
which differs by sign and by the ordering of the arguments from the
original notation in~\cite{Araki1975/76,Araki1977}.

\medskip
The most important properties of relative entropy for our purposes are
\begin{enumerate}%[(i)]
\item $\Ent(\nu|\omega)\le0$, with equality if and only if $\nu=\omega$.
\item For any $C\in\RR$ the set of states $\{\nu\,|\,\Ent(\nu|\omega)\ge C\}$
is a weak-$\ast$ compact subset of the \ul{folium} $\cN_\omega$.
\item $\Ent(\nu\circ\tau|\omega\circ\tau)=\Ent(\nu|\omega)$ for any $\tau\in\Aut(\fM)$.
\end{enumerate}
The reader should consult~\cite{Ohya1993} for a more exhaustive list and a detailed discussion.

%%%%%%%%%%%%%%%%%%%%%%%%%%%%%%%%%%%%%%%%%%%
\subsection{The entropy balance equation}
%%%%%%%%%%%%%%%%%%%%%%%%%%%%%%%%%%%%%%%%%%%

The change in relative entropy due to the action of an inner $\ast$-automorphism
is given by the following result.

\begin{theorem}[\cite{Jaksic2003}]\label{BasicEB}
Let $\omega$ be a modular state on the $C^\ast$-algebra $\cO$. Denote by
$\delta_\omega$ the infinitesimal generator of its \ul{modular group}
 on the enveloping algebra of $\cO$.
For any unitary $U\in\cO$, set $\tau_U(A)=U^\ast AU$. Then 
$$
\Ent(\nu\circ\tau_U|\omega)=\Ent(\nu|\omega)-\i\,\nu\left(U^\ast\delta_\omega(U)\right)
$$
for any state $\nu$ on $\cO$ and any unitary $U\in\Dom(\delta_\omega)$.
\end{theorem}

Using Property~(iii) of the relative entropy, a direct application of this
theorem to \ul{local perturbations} of quantum dynamical systems (see
Section~\ref{ssec:Perturbation theory}) yields

\begin{corollary}Let $(\cO,\tau)$ be a $C^\ast$- or $W^\ast$-dynamical system
equipped with a modular invariant state $\omega$. Denote by
$\delta_\omega$ the generator of the modular group of $\omega$.
For any local perturbation $\tau_V$ induced by $V=V^\ast\in\Dom(\delta_\omega)$
one has
\beq
\Ent(\nu\circ\tau_V^t|\omega)=
\Ent(\nu|\omega)-\int_0^t\nu\circ\tau_V^s(\delta_\omega(V))\,\d s.
\label{LocalEB}
\eeq
\end{corollary}

\remark In the case of a time-dependent local
perturbation $V(t)$ such that $t\mapsto V(t)$ and
$t\mapsto\delta_\omega(V(t))$ are continuous in the natural topology
of $\cO$, Theorem~\ref{BasicEB} yields
$$
\Ent(\nu\circ\tau_V^{s\to t}|\omega)=
\Ent(\nu|\omega)-\int_s^t\nu\circ\tau_V^{s\to u}(\delta_\omega(V(u)))\,\d s.
$$
To our knowledge, this formula was first obtained in~\cite{Ojima1988} for
a $(\tau,\beta)$-KMS state $\omega$. In this special case $\delta_\omega=-\beta\delta$
where $\delta$ is the infinitesimal generator of $\tau$.

\medskip
Assume that $\omega_+\in\Sigma_+(\tau_V,\omega)$, \ie that
$\omega_+$ is an NESS of the perturbed dynamics
(see Section~\ref{sec:NESS in quantum statistical mechanics}).
Then there exists a net $t_\alpha\to\infty$ such that
$$
\omega_+(A)=\lim_\alpha\frac{1}{t_\alpha}\int_0^{t_\alpha}\omega\circ\tau_V^s(A)\,\d s.
$$
The entropy balance formula~\eqref{LocalEB} and Property~(i) of the relative entropy yield
$$
0\le-\lim_\alpha\frac{\Ent(\omega\circ\tau_V^{t_\alpha}|\omega)}{t_\alpha}
=\omega_+(\delta_\omega(V)),
$$
from which, given the following definition, the next proposition follows.

\begin{definition}\label{EPdef}
\ben
\item We define the entropy production observable of the
local perturbation $V$ relative to the reference state $\omega$ by
$\sigma(\omega,V)=\delta_\omega(V)$.
\item The entropy production rate of a NESS
$\omega_+\in\Sigma_+(\tau_V,\omega)$ is $\mathrm{Ep}(\omega_+)=\omega_+(\sigma(\omega,V))$.
\een
\end{definition}

\begin{proposition}
The entropy production rate of a NESS is non-negative.
\end{proposition}

For quantum spin systems, our definition formally agrees with Ruelle's
proposal~\cite{Ruelle2001,Ruelle2002}. It is also closely related
to the definition of entropy production used in~\cite{Spohn1978b}.

%%%%%%%%%%%%%%%%%%%%%%%%%%%%%%%%%%%%%%%%%%%
\subsection{Thermodynamic interpretation}
%%%%%%%%%%%%%%%%%%%%%%%%%%%%%%%%%%%%%%%%%%%

Let us consider the case of a small system $\cS$, with a finite
dimensional algebra $\cO_0$, coupled to several infinitely extended
reservoirs $\cR_1,\ldots,\cR_M$. We  freely the notation of
Section~\ref{sec:NESS in quantum statistical mechanics}.

Denote by $\delta_a$ the generator of $\tau_{a}^t=\tau^t|_{\cO_a}$ for
$0\le a\le M$. Since $\cO_0$ is finite dimensional, one has
$\delta_0=\i[H_\cS,\argdot]$ for some Hamiltonian $H_\cS$.
Observables describing the energy fluxes out of the reservoirs can be
obtained in the following way. The total energy of the system is the
sum of the energy of each reservoir, of the energy $H_\cS$ of the
small system, and of the interaction energy $V$. Since the total energy
is conserved, the rate at which the energy of the reservoirs decreases
under the coupled dynamics is
$$
\frac{\d}{\d t}\tau_V^t(H_\cS+V)=
\tau_V^t\left(\sum_{1\le j\le M}\delta_j(H_\cS+V)+\i[H_\cS+V,H_\cS+V]\right)
=\tau_V^t\left(\sum_{1\le j\le M}\delta_j(V)\right).
$$
Noting that $\delta_j(V)=\delta_j(V_j)\in\cO_0\otimes\cO_{j}$, we can
identify $\Phi_j=\delta_j(V)$ with the energy flux out of reservoir $\cR_j$.

Suppose now that each reservoir $\cR_j$ is initially at thermal equilibrium at
inverse temperature $\beta_j$, the system $\cS$ being in an arbitrary
$\tau_0$-invariant faithful state. From the observation in the paragraph
following Theorem~\ref{thm:WAAM}, we conclude that the generator of the modular
group of the initial state $\omega$ takes the form
$$
\delta_\omega=-\sum_{1\le j\le M}\beta_j\delta_j+\i[K,\argdot],
$$
for some $K\in\cO_0$ such that $\delta_a(K)=0$ for $0\le a\le M$. It follows that
the entropy production observable is
$$
\sigma(\omega,V)=-\sum_{1\le j\le M}\beta_j\delta_j(V)+\i[K,V]=
-\sum_{1\le j\le M}\beta_j\Phi_j-\delta_V(K),
$$
where $\delta_V=\sum_a\delta_a+\i[V,\argdot]$ is the generator of $\tau_V$.
It is important to realize that the second term in the right-hand side of this identity
is a total derivative. Consequently, its contribution to entropy production remains
uniformly bounded in time
$$
\int_0^t\tau_V^s(\sigma(\omega,V))\,\d s=-\sum_{1\le j\le M}\beta_j\int_0^t\tau_V^s(\Phi_j)\,\d s
+\left(\tau_V^t(K)-K\right).
$$
In particular, since $\omega_+\in\Sigma_+(\tau_V,\omega)$ is $\tau_V$-invariant, this
boundary term does not contribute to the entropy production rate of the NESS, and we
can write
$$
\mathrm{Ep}(\omega_+)=-\sum_{1\le j\le M}\beta_j\,\omega_+(\Phi_j),
$$
which is the familiar phenomenological expression~\eqref{EPdiscrete}.
A similar interpretation is possible in the case of time-dependent
perturbations, see~\cite{Ojima1988}.

For classical, thermostated systems used in the construction of
microcanonical NESS, entropy production is usually defined as the
local rate of phase space contraction $\alpha$ (see~\cite{Gallavotti2004a}).
If $\phi^t$ denotes the phase space flow and $\mu$ the reference measure
(typically Lebesgue's measure), then
$$
\mu_t(f)=\mu(f\circ\phi^t)=\mu(f\,\e^{\int_0^t\alpha\circ\phi^{-s}\,\d s}).
$$
A simple calculation shows that, if $\nu$ is absolutely continuous with respect to $\mu$, then
$$
\Ent(\nu_t|\mu)=\Ent(\nu|\mu)-\int_0^t\nu(\alpha\circ\phi^s)\,\d s.
$$
Comparison with~\eqref{LocalEB} shows perfect agreement with Definition~\ref{EPdef}
(see~\cite{Pillet2001} for a completely parallel treatment of the classical and quantum cases).

%%%%%%%%%%%%%%%%%%%%%%%%%%%%%%%%%%%%%%%%%%%%%%%%%%%%%%
\subsection{Strict positivity of entropy production}
%%%%%%%%%%%%%%%%%%%%%%%%%%%%%%%%%%%%%%%%%%%%%%%%%%%%%%

We have seen that $\mathrm{Ep}(\omega_+)\ge0$ for a NESS $\omega_+$. One expects
more, namely $\mathrm{Ep}(\omega_+)>0$. Strict positivity of entropy production is
a delicate dynamical problem. It is related to the singularity of the NESS with respect
to the reference state, as indicated by the following result.

\begin{theorem}[\cite{Jaksic2002b}]
If $\omega_+\in\Sigma_+(\tau_V,\omega)$ is $\omega$-normal, then
$\mathrm{Ep}(\omega_+)=0$. Moreover, if
$$
\sup_{t>0}\left|\int_0^t\left\{
\omega\circ\tau_V^s(\sigma(\omega,V))-\omega_+(\sigma(\omega,V))
\right\}\,\d s\right|<\infty,
$$
then $\mathrm{Ep}(\omega_+)=0$ implies that $\omega_+$ is $\omega$-normal.
\end{theorem}

Strict positivity of entropy production has been proved in a number of
models.  We refer to the original articles~\cite{Spohn1978b,Jaksic2002a,Froehlich2003,Aschbacher2006b}.
The strict positivity of entropy production for generic perturbations has been studied in~\cite{Jaksic2007b}.

%%%%%%%%%%%%%%%%%%%%%%%%%%%%%%%%%%%%
\section{Linear response theory}
\label{sec:Linear response theory}
%%%%%%%%%%%%%%%%%%%%%%%%%%%%%%%%%%%%

Linear response theory is a special instance of first-order
perturbation theory. Its purpose is to describe the response of 
mechanical system to external forces in the regime of weak forcing. Of
particular interest is the response of systems that  are driven out of
some dynamical equilibrium by non-conservative mechanical forces or
thermal forces,  such as temperature or density gradients. This article is
a short introduction to some mathematical results in quantum
mechanical linear response theory. We refer to~\cite{Kubo1991} for an
introduction to the physical aspects of the subject.

%%%%%%%%%%%%%%%%%%%%%%%%%%%%%%%%%%%%%%%%%%
\subsection{Finite time linear response}
%%%%%%%%%%%%%%%%%%%%%%%%%%%%%%%%%%%%%%%%%%

The first problem of linear response theory is the determination of
the response of the system to the action of  driving forces over a
finite interval of time. We shall consider separately the simple case
of mechanical forcing and the more delicate thermal drives.

The unperturbed system is a \ul{$C^\ast$- or $W^\ast$-dynamical system}
$(\cO,\tau)$ equipped with a \ul{modular} invariant state $\omega$. We
denote by $\delta$ the generator of the dynamics $\tau$, by $\sigma$
the modular group of $\omega$, and by $\zeta$ the  generator of $\sigma$.
$\cO_\mathrm{self}$ further denotes the set of selfadjoint elements of
$\cO$. We note that the important role of the modular structure in
linear response theory was already apparent in~\cite{Naudts1975}.

Since $\omega$ is $(\sigma,-1)$-\ul{KMS}, for any $A,B\in\cO$ there
exists a function $z\mapsto F(A,B;z)$ which is analytic in the strip
$\{z\in\CC\mid-1<\mathrm{Im}\,z<0\}$, bounded and continuous on its closure,
and such that $F(A,B;\theta)=\omega(A\sigma^\theta(B))$ and
$F(A,B;\theta-\i)=\omega(\sigma^\theta(B)A)$ for $\theta\in\RR$.  With a slight abuse of 
the notation, we denote $F(A,B;z)$ by either
$\omega(A\sigma^z(B))$ or $\omega(\sigma^{-z}(A)B)$ for
$-1\le\mathrm{Im}\,z\le0$. In particular, the canonical correlation of
$A,B\in\cO$ is defined by
$$
\langle A|B\rangle_\omega=\int_0^1\omega(A\sigma_\omega^{-\i\theta}(B))\,\d\theta=
\int_0^1F(A,B;-\i\theta)\,\d\theta.
$$
One easily checks that it defines an inner product on the real vector
space $\cO_\mathrm{self}$, see~\cite{Naudts1975}.

%%%%%%%%%%%%%%%%%%%%%%%%%%%%%%%%%
\subsubsection{Mechanical drive}
%%%%%%%%%%%%%%%%%%%%%%%%%%%%%%%%%

Let $\tau_V^{s\to t}$ be the dynamics on $\cO$ generated by
$\delta+\i[V(t),\argdot]$, \ie the solution of
\begin{align*}
\partial_t\tau_V^{s\to t}(A)&=\tau_V^{s\to t}(\delta(A)+\i[V(t),A]),\\
\partial_s\tau_V^{s\to t}(A)&=\delta(\tau_V^{s\to t}(A))+\i[V(s),\tau_V^{s\to t}(A)],
\end{align*}
which satisfies $\tau_V^{t\to t}(A)=A$. Suppose  that $t\mapsto V(t)$
belongs to $C(\RR,\cO_\mathrm{self})$.

  Standard time-dependent perturbation theory yields
(see Section~\ref{sec:Quantum dynamical systems})
$$
\tau_V^{s\to t}(A)=\tau^{t-s}(A)+\int_{s}^{t}
\i[\tau^{u-s}(V(u)),\tau^{t-u}(A)]\,\d u
+\text{higher order terms}.
$$
Using the invariance of $\omega$ under $\tau$ and the identity
$\omega\circ\tau^{u-s}=\omega$, the first-order term can be rewritten as
\[
(\Delta\omega)^{s\to t}(A)
=\int_s^t \omega\!\left(\i[V(u),\tau^{t-u}(A)]\right)\,\d u,
\]
which motivates the definition of the response function $\cK$:
$$
(\Delta\omega)^{s\to t}
%=\int_s^t\omega(\i[V(u),\tau^{t-u}(X)])\,\d u
=\int_s^t\cK(t-u)V(u)\,\d u.
$$
The operator-valued response function $\cK(u):\cO\to\cO^{\scriptscriptstyle\#}$, 
where $\cO^{\scriptscriptstyle\#}$ denotes the dual of $\cO$, is defined by
\beq
(\cK(u)V)(A)=\omega(\i[V,\tau^u(A)]),
\label{RespK}
\eeq
for $u\ge0$; $\cK(u)=0$ for $u<0$ (causality).

Note that if $\omega$ is \ul{mixing} for $(\cO,\tau)$, then
$\lim_{t\to\infty}(\Delta\omega)^{s\to t}(X)=0$ for any $V\in
C(\RR,\cO_\mathrm{self})$ of compact support and any $X\in\cO$, \ie
the system recovers from infinitesimal localized perturbations. We
refer to~\cite{Verbeure1975} for further connections between the mixing property
and linear response theory.

For $V\in\Dom(\zeta)$, the KMS property implies that  Equ.~\eqref{RespK} can be rewritten as
$$
(\cK(u)V)(A)=\langle\tau^u(A)|\zeta(V)\rangle_\omega.
$$
 Here $\zeta$ denotes  the generator of the modular group
  of $\omega$.
This is a typical fluctuation-dissipation relation: 
the response function $\cK$ on the left-hand side describes dissipation, while the correlation
function on the right-hand side measures fluctuations.

For later reference, let us consider the special case where $\omega$ is
$(\tau,\beta)$-KMS and $V(t)=-\sum_jX_j(t)A_j$. Here the operators 
$A_j\in\cO_\mathrm{self}\cap\Dom(\delta)$ describe the
coupling of the system to external fields, and the functions $X_j$ are the
time-dependent field strengths. One has $\zeta=-\beta\delta$ and the
observable $\Phi_j=\delta(A_j)$ describes the flux conjugate to $A_j$.
The linear response is given by the finite time Green-Kubo formula
\beq
(\Delta \omega)^{s\to t}(A)=\sum_j\beta\int_s^t
\langle\tau^{t-u}(A)|\Phi_j\rangle_\omega\,X_j(u)\,\d u.
\label{spgk}
\eeq

%%%%%%%%%%%%%%%%%%%%%%%%%%%%%%
\subsubsection{Thermal drive}
\label{sssec:Thermal}
%%%%%%%%%%%%%%%%%%%%%%%%%%%%%%

To discuss thermal forcing, we need more structure. We consider the
setting of Section~\ref{sec:NESS in quantum statistical mechanics}:
a small system $\cS$, described by $(\cO_0,\tau_0)$, coupled to infinite reservoirs
$\cR_1,\ldots,\cR_M$ described by $(\cO_j,\tau_j)$, $1\le j\le
M$. The algebra factorizes accordingly $\cO=\otimes_{0\le a\le
  M}\cO_a$, and the dynamics of the decoupled system is
$\tau_\mathrm{dec}=\otimes_{0\le a\le M}\tau_a=\e^{t\delta_\mathrm{dec}}$.
We denote by $\delta_a$ the generator of $\tau_a$ so that
$\delta_\mathrm{dec}=\sum_a\delta_a$.

The dynamics $\tau$ of the coupled system is defined as the local
perturbation of $\tau_\mathrm{dec}$ by the coupling $V=\sum_{1\le j\le
  M}V_j$, where $V_j\in\cO_0\otimes\cO_j$. Its generator is
$\delta=\delta_\mathrm{dec}+\i[V,\argdot]$.

For simplicity,  we shall only consider  the case where the fiducial
state $\omega$ is $(\tau,\beta)$-KMS, so that the generator of its
modular group is $\zeta=-\beta\delta$. We study small departures from
this state resulting from imposed temperature gradients. Our discussion can easily be generalized to include other
thermodynamic forces, such as  inhomogeneous electro-chemical potentials.

To start, we assume the following.

\assuming{H1}{HOne} {There exists a unique $(\tau_0,\beta)$-KMS state $\omega$.
  For all sufficiently small 
    $X=(X_1,\ldots,X_M)\in\RR^M$, there 
exist unique states $\omega_{j,X_j}$, for $1\le j\le M$, such that
  $\omega_{j,X_j}$ is $(\tau_j,\beta-X_j)$-KMS. }

We set $\omega_{{\mathrm{dec}},X}:=\omega_0\otimes
\mathop{\otimes}\limits_{1\leq j\leq M}\omega_{j,X_j}$.
    
In such  a state  each reservoir $\cR_j$ is in equilibrium
at inverse temperature $\beta_j=\beta-X_j$. The $X_j$'s are the
thermodynamic forces which drive the system out of equilibrium,
see Section~\ref{sec:NESS in quantum statistical mechanics}.
The conjugate fluxes are the energy
currents $\Phi_j=\delta_j(V)$, see Section~\ref{sec:Entropy production}.
To ensure that they are well-defined, we assume

\assuming{H2}{HTwo}{$V\in\Dom(\delta_j)$ for $1\le j\le M$.}

It follows from Araki's perturbation theory, Theorem~\ref{thm:ArakiPerturb},
that $\omega_{X=0}^{(0)}$ (which is the unique
$(\tau_\mathrm{dec},\beta)$-KMS state) and $\omega$ are mutually
normal. We note, however, that since $\omega_{X=0}^{(0)}\not=\omega$,
expanding $\omega_X^{(0)}\circ\tau^t(A)-\omega(A)$ around $X=0$
generates a spurious zeroth-order term. To avoid this problem, we
shall construct a family of states $\omega_X$ which, on the one hand,
has the same thermodynamic properties as $\omega_X^{(0)}$, and on the
other hand satisfies $\omega_{X=0}=\omega$.

By definition, the modular group of $\omega^{(0)}_X$ is generated by
$-\beta\delta_\mathrm{dec}+\sum_{1\le j\le M}X_j\delta_j$. Denote by
$\sigma_X$ the group of $\ast$-automorphisms generated by
$$
\zeta_X=
-\beta\delta_\mathrm{dec}+\sum_{1\le j\le M}X_j\delta_j-\i\beta[V,\argdot]
=\zeta+\sum_{j=1}^MX_j\delta_j.
$$
Araki's perturbation theory implies that there exists a unique
$(\sigma_X,-1)$-KMS state $\omega_X$ such that $\omega_X^{(0)}$ and
$\omega_X$ are mutually normal. Thus, these two states have the same
thermodynamic properties, and since $\zeta_{X=0}=\zeta$, one also has
$\omega_{X=0}=\omega$.

We say that $A\in\cO$ is centered if $\omega_X(A)=0$ for all sufficiently
small $X\in\RR^n$.

\begin{theorem}[\cite{Jaksic2006c}] Under the Hypothesis~\HOne{} and~\HTwo, the function
$X\mapsto\omega_X\circ\tau^t(A)$ is differentiable at $X=0$
for any centered observable $A\in\cO$, and the finite time
Green-Kubo formula
\beq
\partial_{X_j}\omega_X\circ\tau^t(A)|_{X=0}=\int_0^t
\langle\tau^s(A)|\Phi_j\rangle_\omega\,\d s,
\label{FTKuboForm}
\eeq
holds.
\end{theorem}

\remarks\textbf{1.} Formula~\eqref{FTKuboForm} is limited to centered
observables because, at the current level of generality, we do not have 
control of $\omega_X(A)$ as $X\to0$. If $A\in\cO$
is such that $X\mapsto\omega_X(A)$ is differentiable at $X=0$, then the
above formula still holds after addition of the static contribution
$\partial_{X_k}\omega_X(A)|_{X=0}$ to its right-hand side. We note,
however, that for infinite systems the states $\omega_X$ for distinct
values of $X$ are usually mutually singular. The differentiability of
$\omega_X(A)$ is therefore a delicate question, and is not expected to
hold for general observables $A$.

\noindent\textbf{2.} One can prove that the energy fluxes $\Phi_j$, and more generally
the fluxes conjugate to intensive thermodynamic parameters, are
centered. We refer to~\cite{Jaksic2006c} for more details.

%%%%%%%%%%%%%%%%%%%%%%%%%%%%%%%%%%%
\subsection{The long time problem}
%%%%%%%%%%%%%%%%%%%%%%%%%%%%%%%%%%%
The hard problem of linear response theory concerns the validity of the linear response formulas derived in the previous section in the long-time limit. This issue has been widely discussed in the physics literature. The most famous objection to the validity of linear response was raised by van Kampen in~\cite{Kampen1971}. His argument is based on the fact that the microscopic dynamics of a large system with many degrees of freedom is strongly chaotic. He argued that the time scale over which a perturbative calculation remains valid may be very short, and concluded that finite-time linear response may therefore be physically irrelevant on macroscopic time scales. Discussions of van Kampen’s objection can be found in~\cite{Kubo1991}, and the interested reader may also consult~\cite{Lebowitz1986}.

A mathematical idealization reduces the long time problem to the
interchange of two limits: the zero forcing limit involved in the
derivation of the finite time linear response formulas and the
infinite time limit. To illustrate this point, let us continue the
discussion of Section~\ref{sssec:Thermal} which led  to
formula~\eqref{FTKuboForm}, under the following assumption:

\assuming{H3}{HThree}{For all sufficiently small $X\in\RR^M$ there exists a NESS
$\omega_{X+}$, see Section~\ref{sec:NESS in quantum statistical mechanics},
such that,
\beq
\lim_{t\to\infty}\omega_X\circ\tau^t(A)=\omega_{X+}(A)
\label{Ness}
\eeq
for any $A\in\cO$.
}

\noindent We note that under normal circumstances
one expects more, namely that
$$
\lim_{t\to\infty}\eta\circ\tau^t(A)=\omega_{X+}(A)
$$
holds for any $A\in\cO$ and any $\omega_X$-normal (or equivalently
$\omega_X^{(0)}$-normal) state $\eta$.

We shall say that an observable $A\in\cO$ is regular if the function
$X\mapsto\omega_{X+}(A)$ is differentiable at $X=0$ and
$$
\partial_{X_k}\omega_{X+}(A)|_{X=0}
=\lim_{t\to\infty}\partial_{X_k}\omega_{X}\circ\tau^t(A)|_{X=0}.
$$
If $A$ is a regular  centered observable, then  Equ.~\eqref{Ness} and
Formula~\eqref{FTKuboForm} yield the Green-Kubo formula
\beq
\partial_{X_j}\omega_{X+}(A)|_{X=0}=\int_0^\infty
\langle\tau^s(A)|\Phi_j\rangle_\omega\,\d s.
\label{GK1}
\eeq
In particular, if the fluxes $\Phi_k$ are regular, then the transport
coefficients (recall Section~\ref{sec:Nonequilibrium steady states}), defined by
$$
\omega_{X+}(\Phi_k)=\sum_{1\le j\le M} L_{jk}X_j+\mathrm{o}(X),
$$
are given by the formula
$$
L_{jk}=
\int_0^\infty
\langle\tau^s(\Phi_k)|\Phi_j\rangle_\omega\,\d s.
$$

To justify the exchange of limits for a sufficiently large set of
centered observables $A\in\cO$, and in particular for the flux
observables, is a delicate problem requiring a fairly good control on
the dynamics of the system. This was recently achieved for two classes
of systems: $N$-levels systems coupled to free Fermi reservoirs
in~\cite{Jaksic2006a} and locally interacting Fermi gases in~\cite{Jaksic2007a}. In the
first case the NESS was previously constructed in~\cite{Jaksic2002a} using the
Liouvillean approach, see Section~\ref{sec:NESS in quantum statistical mechanics}. In
the second case, the NESS is obtained following Ruelle's scattering
approach. In both cases, the interchange of limits is validated via
the following simple consequence of Vitali's theorem.

\begin{proposition}Suppose that~\HOne{} and~\HThree{} hold and let $A\in\cO$.
  Assume that for some $\epsilon>0$ and any $t\ge0$, the function
  $X\mapsto\omega_X\circ\tau^t(A)$ has an analytic extension to the
  open polydisk $D_\epsilon=\{X\in\CC^M\,|\,\max_j|X_j|<\epsilon\}$.
  If
  $$
  \sup_{X\in D_\epsilon,t\ge0}|\omega_X\circ\tau^t(A)|<\infty
  $$
  holds, then $A$ is regular.
\end{proposition}
It is evident, though sometimes overlooked, that the long-time problem cannot be solved merely by proving that a finite-time linear response formula remains meaningful in the long-time limit. Suppose, for example,
that the system $(\cO,\tau)$ is $L^1$-asymptotically Abelian, namely that 
\beq
\int_0^{\infty}\|[A,\tau^{t}(B)]\|\,\d t<\infty,
\label{mer-sz}
\eeq
holds for a dense set of  $A,B$ in $\cO$. It follows that the linear response to the
perturbation $V(t)=-\sum_jX_j(t)A_j$ such that $x=\sum_j\sup_t|X_j(t)|<\infty$, 
\beq
(\Delta\omega)^{t}(A)=
\lim_{s\to-\infty}(\Delta\omega)^{s\to t}(A)
=\sum_j\int_0^\infty\omega(\i[A_j,\tau^u(A)])X_j(t-u)\,\d u,
\label{theeasystuf}
\eeq
where the integrals are absolutely convergent.\footnote{We assume that \eqref{mer-sz} holds for $A_j$'s and $A$.} This, however, does not
mean that (i) the natural nonequilibrium
state
$$
\omega^V_t(A)=\lim_{s\to-\infty}\omega\circ\tau_V^{s\to t}(A)
$$
exists, and (ii) that
$$
\omega^V_t(A)-\omega(A)
=(\Delta\omega)^{t}(A)+\text{o}(X).
$$
In fact both (i) and (ii) require a precise control of the perturbed
dynamics $\tau_V$ whereas Equ.~\eqref{theeasystuf} only involves the
unperturbed $\tau$. If (i) and (ii) hold, and if $\omega$ is
$(\tau,\beta)$-KMS, then, by Equ.~\eqref{spgk}, the infinite time
Green-Kubo formula
$$
(\Delta\omega)^{t}(A)=
\sum_j\beta\int_0^\infty
\langle\tau^s(A)|\Phi_j\rangle_\omega\, X_j(t-s)\,\d s,
$$
holds.

%%%%%%%%%%%%%%%%%%%%%%%%%%%%%%%%%%%%%%%%%%%%%%%%%%%%%%%%%%%%%%%%%%%%%%%%
\subsection{Time-reversal invariance and Onsager reciprocity relations}
%%%%%%%%%%%%%%%%%%%%%%%%%%%%%%%%%%%%%%%%%%%%%%%%%%%%%%%%%%%%%%%%%%%%%%%%

A time-reversal of $(\cO,\tau)$ is an involutive  antilinear
$\ast$-automorphism $\Theta$ of $\cO$ such that
$\tau^t\circ\Theta=\Theta\circ\tau^{-t}$ for any $t\in\RR$. A state
$\nu$ on $\cO$ is time-reversal invariant if
$\nu\circ\Theta(A)=\nu(A^\ast)$ for all $A\in\cO$. An observable
$A\in\cO_\mathrm{self}$ is even/odd under time-reversal if 
$\Theta(A)=\pm A$.

The following proposition is a simple consequence
of the KMS condition, see~\cite{Jaksic2006c}.

\begin{proposition}
  Assume that $(\cO,\tau)$ is equipped with a time-reversal $\Theta$.
  Let $\omega$ be a time-reversal invariant, mixing, $(\tau,\beta)$-KMS
  state.  If $A, B\in\cO_\mathrm{self}$ are both even or odd under
  time-reversal, then
$$
\int_0^t
\langle\tau^s(A)|B\rangle_\omega\,\d s
=\frac{1}{2}\int_{-t}^t\omega(A\tau^{s}(B))\,\d s+ \mathrm{o}(1)
$$
in the limit $t\to\infty$.
\end{proposition}

\remark If $\omega$ is the unique $(\tau,\beta)$-KMS
state, then it is automatically time-reversal invariant.

\medskip
To apply this proposition to the Green-Kubo formula~\eqref{GK1} we assume:

\assuming{H4}{HFour}{$(\cO,\tau)$ is equipped with a time-reversal $\Theta$ and
$\omega$ is a time-reversal invariant, mixing $(\tau,\beta)$-KMS state.
Moreover, the couplings $V_j$ are even under time-reversal.
}

\begin{corollary}\label{cor:GK2Cor}
 Under Hypothesis~\HOne, \HTwo, \HThree{} and \HFour{}, the
Green-Kubo formula~\eqref{GK1} can be written as
$$
\partial_{X_j}\omega_{X+}(A)|_{X=0}=\frac{1}{2}\int_{-\infty}^\infty
\omega(A\tau^{s}(\Phi_j))\,\d s
$$
for regular, centered observables $A\in\cO_\mathrm{self}$ that are odd under
time-reversal. In particular, if the fluxes $\Phi_k$ are regular, then
the transport coefficients are given by
\beq
L_{jk}=\frac{1}{2}
\int_{-\infty}^\infty
\omega(\Phi_k\tau^{s}(\Phi_j))\,\d s.
\label{GK2}
\eeq
\end{corollary}

If $\omega$ is a mixing $(\tau,\beta)$-KMS state, then the stability condition
$$
\lim_{T\to\infty}\int_{-T}^T\omega([A,\tau^t(B)])\,\d t=0
$$
holds for any $A,B\in\cO$, see~\cite[Section~5.4.2]{Bratteli1981}.
An important consequence of this fact and Equ.~\eqref{GK2} is

\begin{corollary} Under the assumptions of Corollary~\ref{cor:GK2Cor} the
transport coefficients satisfy the Onsager reciprocity relations
$$
L_{jk}=L_{kj}.
$$
\end{corollary}

%%%%%%%%%%%%%%%%%%%%%%%%%%%%%%%%%%%%%%%%%%%
\part{Small systems coupled to reservoirs}
%%%%%%%%%%%%%%%%%%%%%%%%%%%%%%%%%%%%%%%%%%%

%%%%%%%%%%%%%%%%%%%%%%%%%%%%%%%%%%%%
\section{Coupling to reservoirs}
\label{sec:Coupling to reservoirs}
%%%%%%%%%%%%%%%%%%%%%%%%%%%%%%%%%%%%

In this article we introduce  classes of models used to describe a
(typically small) quantum system interacting with a large external system,
often called a ``reservoir'' or a ``heat bath''.

A typical small system that we have in mind is a ``quantum dot'' -- an atom or a
molecule. We will often  restrict ourselves  to finitely many  eigenstates of such a  "quantum dot".
Mathematically, it is  described by a Hilbert space $\cK$, often assumed to
be finite dimensional.

We will assume  that the reservoir is as simple as possible -- in
practice it means that it is  a \ul{free Bose or Fermi gas}.

We will  refrain from presenting  detailed physical examples of systems
considered in our article -- they are easy to provide. We will only  describe
their general mathematical structure.

%%%%%%%%%%%%%%%%%%%%%%%%%%%%%%%%%%%%%%%%%%%%%%%%%%%%%%%%%%%%%%%%%%%%%
\subsection{Models of a small quantum system coupled to a reservoir}
%%%%%%%%%%%%%%%%%%%%%%%%%%%%%%%%%%%%%%%%%%%%%%%%%%%%%%%%%%%%%%%%%%%%%

In the bosonic case we consider the Hilbert space
$\cK{\otimes}\Gamma_\s(L^2(\Xi))$, where $L^2(\Xi)$ describes the 1-particle
space and $\Gamma_\s(L^2(\Xi))$ is the corresponding \ul{bosonic Fock space}.
$a^*(\xi)$, $a(\xi)$ will denote the \ul{creation/annihilation operators}
satisfying the usual commutation relations. They are, strictly speaking, operator-valued distrubitions,
which become (unbounded) operators when
smeared  with appropriate test functions.

A typical Hamiltonian describing the interaction of a small system with a
bosonic reservoir has the form
\beq
H= K{\otimes}\one+\one\otimes\int \omega(\xi)a^*(\xi)a(\xi)\d\xi
+\int v(\xi){\otimes}a^*(\xi)\d\xi+\int v(\xi)^*{\otimes}
a(\xi)\d\xi.
\label{hammi0}
\eeq
Here $\omega$ is the 1-particle excitation spectrum, $K$ is the Hamiltonian of the
small system, and the function $\xi\mapsto v(\xi)$ describes the coupling  of
the bosons with the small system.

One of the most popular choices  is $\cK=\CC^2$, $K=\sigma_3$, $v(\xi)=\sigma_1
f(\xi)$ for some function $f$, where $\sigma_3$, $\sigma_1$ are the usual
\ul{Pauli matrices}. With this choice, $H$ is called the \ul{spin-boson
Hamiltonian}.

One often adds a perturbation to $H$ quadratic in the creation/annihilation
operators. Adding higher powers may  make the definition of $H$ as a
selfadjoint operator problematic.

Motivated by~\cite{Fierz1938}, we shall call \ul{Pauli-Fierz operators} the
operators of the form~\eqref{hammi0}; see e.g.~\cite{Derezinski2003}, Note,
however, that this terminology is not universally accepted. Other names  found in
the literature to denote~\eqref{hammi0} or similar classes of operators include
the generalized spin-boson Hamiltonian, the \ul{Nelson Hamiltonian}, the
\ul{Fröhlich Hamiltonian}, and  the \ul{polaron Hamiltonian}.

A natural example of an operator similar to $H$ (with  quadratic terms in the
perturbation) is the Hamiltonian of  non-relativistic  \ul{QED}, which some authors
also call  the Pauli-Fierz Hamiltonian~\cite{Spohn2004}. This usage differs slightly from the meaning above. The QED in
the \ul{dipole approximation} is also of the form~\eqref{hammi0} (without
quadratic terms in the perturbation). In this case the bosonic reservoir
describes \ul{photons}, $\omega(\xi)=|\xi|$ and $L^2(\Xi)$ is the space of
divergence-free vector fields on $\RR^3$. Hence one  can take
$\Xi=\RR\times\{1,2\}$, where $\{1,2\}$ labels two photon polarizations. 

Operators of the form~\eqref{hammi0} are also used to describe interaction of a
small system with lattice vibrations of a crystal lattice;  in this case the bosonic
reservoir consists of \ul{phonons}.

If the reservoir is fermionic, then the  Hilbert space of the composite system is
$\cK{\otimes}\Gamma_\a(L^2(\Xi))$, where $\Gamma_\a(L^2(\Xi))$ is the
corresponding \ul{fermionic Fock space} over the 1-particle space $L^2(\Xi)$. We
denote the fermionic  creation/annihilation operators by $a^*(\xi)$, $a(\xi)$ as well.

A typical Hamiltonian describing a small system interacting with a fermionic
reservoir considered in the literature has the form
\[
\begin{split}
H&:= K{\otimes}\one+\one\otimes\int \omega(\xi)a^*(\xi)a(\xi)\d\xi\\
&\quad +\lambda\sum_{n,m}\int
v_{n,m}(\xi_1,\dots,\xi_n,\xi_m',\dots,\xi_1'){\otimes}
a^*(\xi_1)\cdots
a^*(\xi_n)
a(\xi_m')\cdots a(\xi_1')
\d\xi_1\cdots\d\xi_n\d\xi_m'\cdots\d\xi_1'.
\end{split}
\]
In physically motivated models the interaction typically contains only terms with $n+m$ even.

A typical example of fermionic reservoir is metal containing conductance
electrons.

Often one  assumes that the small system is also described by a bosonic or
fermionic Fock space and all the terms in the Hamiltonian are quadratic.
In this case $\e^{\i tH}$ can be described in terms of the corresponding
1-particle dynamics, which greatly simplifies  analysis.

%%%%%%%%%%%%%%%%%%%%%%%%%%%%%%%%%%%%%%%%%%%%%%%%%%%%%%%%%%%
\section{Coupling to reservoirs of a positive density}
\label{sec:Coupling to reservoirs of a positive density}
%%%%%%%%%%%%%%%%%%%%%%%%%%%%%%%%%%%%%%%%%%%%%%%%%%%%%%%%%%

Consider a small quantum system interacting with a bosonic or fermionic
reservoir of the form described in Section~\ref{sec:Coupling to reservoirs}.
Its Hilbert space is\footnote{The subscript $\s/\a$ refers to the bosonic/fermionic Fock space.} $\cK{\otimes}\Gamma_{\s/\a}(L^2(\Xi))$, and its Hamiltonian
in the bosonic case has the form
\beq
H:= K{\otimes}\one+\one\otimes\int \omega(\xi)a^*(\xi)a(\xi)\d\xi
+\int v(\xi){\otimes}a^*(\xi)\d\xi+\int v(\xi)^*{\otimes}a(\xi)\d\xi.
\label{Hammi}
\eeq
In the fermionic case
\beq
\begin{split}
H& := K{\otimes}\one+\one\otimes\int \omega(\xi)a^*(\xi)a(\xi)\d\xi\\[4pt]
&+\sum_{n,m}\int v_{n,m}(\xi_1,\dots,\xi_n,\xi_m',\dots,\xi_1'){\otimes}
a^*(\xi_1)\cdots
a^*(\xi_n)a(\xi_m')\cdots a(\xi_1')\d\xi_1\cdots\d\xi_n\d\xi_m'\cdots\d\xi_1',
\end{split}
\label{Hammi1}
\eeq
where in physically motivated models the interaction contains only terms with $n+m$ even.

If the function $\omega(\xi)$ is positive, we say that the reservoir is
at zero temperature (or at zero density). We will explain how to modify the
Hamiltonians~\eqref{Hammi} and~\eqref{Hammi1} when the reservoir is at a
positive temperature, or, more generally, at a positive density.

We shall adopt the  $W^*$-algebraic framework  here (although, at least in the
fermionic case, one could also use the $C^*$-algebraic framework).

%%%%%%%%%%%%%%%%%%%%%%%%%%%%%%%%%%%%%%%%%%%%%
\subsection{The semi-standard representation}
%%%%%%%%%%%%%%%%%%%%%%%%%%%%%%%%%%%%%%%%%%%%%

We will use the formalism of Araki-Woods representations of the CCR and
Araki-Wyss representations of the CAR described in more detail in
Section~\ref{sec:Free Bose and Fermi gases -- the algebraic approach}.
Fixing a nonnegative function $\xi\mapsto\gamma(\xi)$, let us define
$$
\rho(\xi) :=(\gamma(\xi)^{-1}\mp1)^{-1},
$$
where the minus sign corresponds
to the bosonic case and the plus sign to the fermionic case.
We consider the bosonic/fermionic Fock space
$\Gamma_{\s/\a}\left(L^2(\Xi)\oplus L^2(\Xi)\right)$.
Consider the ``left creation/annihilation operators corresponding to
$\gamma$'', $a_{\gamma, \l}^\ast(\xi)$, $a_{\gamma, \l}(\xi)$.

The algebra of observables of the coupled system is defined as the
$W^*$-algebra  $\fM_\gamma\subset\cB(\cK\otimes\Gamma_{\s/\a}(L^2(\Xi)\oplus L^2(\Xi))$
generated by the operators of the small system $\cB(\cK)\otimes1$ and
by the reservoir creation/annihilation operators $a_{\gamma,\l}^*(\xi)$,
$a_{\gamma,\l}(\xi)$. On $\cK\otimes\Gamma_\s(L^2(\Xi)\oplus L^2(\Xi))$ we define
the operator
\beq
L_\gamma^\semi := K{\otimes}\one+\one\otimes\int\omega(\xi)\left(a_{\l}^*(\xi)a_{\l}(\xi)-a_{\r}^*(\xi)a_{\r}(\xi)\right)\d\xi
+\int \left(v(\xi){\otimes}a_{\gamma,\l}^*(\xi)+v(\xi)^*{\otimes}a_{\gamma,\l}(\xi)\right)\d\xi,
\label{Hammi2}
\eeq
which is the analog of~\eqref{Hammi} for general $\gamma$. On
$\cK\otimes\Gamma_\a(L^2(\Xi)\oplus L^2(\Xi))$ we define
\beq
\begin{split}
L_\gamma^\semi& := K{\otimes}\one+\one\otimes\int \omega(\xi)\left(a_\l^*(\xi)a_\l(\xi)-a_\r^*(\xi)a_\r(\xi)\right)\d\xi\\[4pt]
&+\sum_{n,m}\int v_{n,m}(\xi_1,\dots,\xi_n,\xi_1,\dots,\xi_m'){\otimes}
a_{\gamma,\l}^*(\xi_1)\cdots a_{\gamma,\l}^*(\xi_n)a_{\gamma,\l}(\xi_m')\cdots a_{\gamma,\l}(\xi_1')
\d\xi_1\cdots\d\xi_n\d\xi_m'\cdots\d\xi_1',
\end{split}
\label{Hammi3}
\eeq
which is the analog of~\eqref{Hammi1} for general $\gamma$. $L_\gamma^\semi$ is
sometimes called the semi-Liouvillean. Let 
\[
\tau_{\gamma,t}(A) := \e^{\i tL_\gamma^\semi}A\e^{-\i tL_\gamma^\semi}.
\]
It is easy to show that $\tau_{\gamma,t}$ preserves the algebra $\fM_\gamma$, and we
obtain a $W^*$-dynamical system $(\fM_\gamma,\tau_\gamma)$.

The special case  $\gamma=0$ corresponds to the
zero temperature state of the reservoir. In this case, the $W^*$-dynamical
system $(\fM_\gamma,\tau_\gamma)$ is  equivalent to the system
\[
\left(\cB(\cK\otimes\Gamma_{\s/\a}(L^2(\Xi))),\e^{\i tH}(\argdot)\e^{-\i tH}\right).
\]

If $\gamma(\xi)=\e^{-\beta\omega(\xi)}$, then we say that the reservoir is
thermal. In this case, the state $\psi$ on the algebra $\fM_\gamma$ defined by
\[
\psi(A\otimes B)=\frac{\tr\left(A\e^{-\beta K}\right)}{\tr\left({\e^{-\beta K}}\right)}(\Omega|B\Omega),
\qquad A\in\cB(\cK),\quad B\in\cB(\Gamma_{\s/\a}(L^2(\Xi){\oplus}L^2(\Xi))),
\]
is \ul{$\beta$-KMS } for the dynamics $\tau$ with no interaction term.

%%%%%%%%%%%%%%%%%%%%%%%%%%%%%%%%%%%%%%%%%
\subsection{The standard representation}
%%%%%%%%%%%%%%%%%%%%%%%%%%%%%%%%%%%%%%%%%

Recall from Section~\ref{sec:The C*-algebra approach}
 that
every $W^*$ algebra has a distinguished representation called the
\ul{standard repre\-sentation}. In such a representation every $W^*$-dynamics
has a distinguished unitary implementation generated by a selfadjoint
operator called the \ul{Liouvillean}.

The standard representation  of $\fM$ acts as
\[
\pi:\fM_\gamma\to\cB\left(\cK{\otimes}\bar\cK{\otimes}\Gamma_{\s/\a}(L^2(\Xi)\oplus L^2(\Xi))\right),
\]
where an element of $\fM_\gamma$ is tensor multiplied by the identity on
$\bar\cK$ (the complex conjugate of $\cK$). The Liouvillean
corresponding to~\eqref{Hammi2} is the selfadjoint operator
on $\cK{\otimes}\bar\cK\otimes\Gamma_\s(L^2(\Xi)\oplus L^2(\Xi))$ defined by
\beq
\begin{split}
L_\gamma := &K\otimes\one\otimes\one-\one\otimes\overline{K}\otimes\one
+\one\otimes\one\otimes\int\omega(\xi)\left(a_{\l}^*(\xi)a_{\l}(\xi)-a_{\r}^*(\xi)a_{\r}(\xi)\right)\d\xi\\[4pt]
&+\int\left(v(\xi)\otimes\one\otimes a_{\gamma,\l}^*(\xi)+v(\xi)^*\otimes\one\otimes a_{\gamma,\r}(\xi)\right)\d\xi
-\int\left(\one\otimes\bar v(\xi)\otimes a_{\gamma,\l}^*(\xi)+\one\otimes\bar v(\xi)^*{\otimes}a_{\gamma,\r}(\xi)\right)
\d\xi.
\end{split}
\label{Hammi2a}
\eeq
The Liouvillean corresponding to~\eqref{Hammi3} is the selfadjoint operator
on $\cK{\otimes}\overline{\cK}\otimes\Gamma_\a(L^2(\Xi)\oplus L^2(\Xi))$ defined by
\beq
\begin{split}
L_\gamma& :=K\otimes\one\otimes\one-\one\otimes\overline{K}\otimes\one
+\one\otimes\one\otimes\int\omega(\xi)\left(a_{\l}^*(\xi)a_{\l}(\xi)-a_{\r}^*(\xi)a_{\r}(\xi)\right)\d\xi\\
&+\sum_{n,m}\int v_{n,m}(\xi_1,\dots,\xi_n,\xi_1,\dots,\xi_m')\otimes\one\otimes
a_{\gamma,\l}^*(\xi_1)\cdots a_{\gamma,\l}^*(\xi_n)a_{\gamma,\l}(\xi_m')\cdots a_{\gamma,\l}(\xi_1') \d\xi_1\cdots\d\xi_n\d\xi_m'\cdots\d\xi_1'\\
&-\sum_{n,m}\int\one\otimes\bar v_{n,m}(\xi_1,\dots,\xi_n,\xi_1,\dots,\xi_m')\otimes
a_{\gamma,\r}^*(\xi_1)\cdots a_{\gamma,\r}^*(\xi_n)a_{\gamma,\r}(\xi_m')\cdots a_{\gamma,\r}(\xi_1')
\d\xi_1\cdots\d\xi_n\d\xi_m'\cdots\d\xi_1'.
\end{split}
\label{Hammi3a}
\eeq
Above we used the ``right creation/annihilation operators corresponding to $\gamma$'',
denoted by  $a_{\gamma, \r}^\ast(\xi)$, $a_{\gamma, \r}(\xi)$ and described in
Section~\ref{sec:Free Bose and Fermi gases -- the algebraic approach}.

Note that the Liouvillean $L_\gamma$ and the semi-Liouvillean
$L_\gamma^\semi$ describe the same dynamics, but in different representations.
This is expressed by 
\[
\pi\left(\e^{\i tL_\gamma^\semi}A\e^{-\i tL_\gamma^\semi}\right)
=\e^{\i t L_\gamma}\pi(A)\e^{-\i t L_\gamma},\qquad A\in\fM_\gamma.
\]
Note also that the operators~\eqref{Hammi2}, \eqref{Hammi2a}, as well
as~\eqref{Hammi3}, \eqref{Hammi3a}, are special cases of
operators of the form~\eqref{Hammi} and~\eqref{Hammi1}, where we allow the
1-particle excitation spectrum to have negative values.

%%%%%%%%%%%%%%%%%%%%%%%%%%%%%%%%%%
\subsection{Composite reservoirs}
%%%%%%%%%%%%%%%%%%%%%%%%%%%%%%%%%%

Sometimes one considers two (or more) reservoirs, $\cR_1$ and $\cR_2$,
interacting with a single small system. This can be taken into account by 
 assuming that $\Xi$ is the union of two disjoint sets $\Xi_1$ and
$\Xi_2$ describing the 1-particle spaces of $\cR_1$ and  $\cR_2$.
Then the composite reservoir is the tensor product of the reservoir
$\cR_1$ and the reservoir $\cR_2$, since, using the
\ul{exponential property of Fock spaces}, we have the
natural identification
\[
\Gamma_{\s/\a}(L^2(\Xi_1\cup\Xi_2))\simeq
\Gamma_{\s/\a}(L^2(\Xi_1){\oplus}L^2(\Xi_2))\simeq
\Gamma_{\s/\a}(L^2(\Xi_1)){\otimes}\Gamma_{\s/\a}(L^2(\Xi_2)).
\]
Physically, an especially important case arises when the reservoirs
$\cR_1$ and $\cR_2$ are thermal but correspond to two distinct inverse
temperatures.

%%%%%%%%%%%%%%%%%%%%%%%%%%%%%%%%%%%%%%%%%%%%%%%%%%%
\section{Completely positive Markov semigroups}
\label{sec:Completely positive Markov semigroups}
%%%%%%%%%%%%%%%%%%%%%%%%%%%%%%%%%%%%%%%%%%%%%%%%%%%

Let  $\cB(\cK)$  denote the algebra of bounded operators on a Hilbert space
$\cK$, which we interpret as the algebra  of  observables of the small quantum system.
We say that a transformation $\Lambda$ on $\cB(\cK)$ is positive if the image
of a positive operator is positive. We say that $\Lambda$ is \ul{completely positive}
if its extension to $n\times n$ matrices with entries in $\cB(\cK)$ is positive
for any $n$. We say  it is Markov if it preserves the identity operator.

One can also consider the  transformation $\Lambda_\#$ acting on $\cB^1(\cK)$ -- the
trace class operators on $\cK$ -- defined by
\[
\tr\left(\rho\Lambda(A)\right)=\tr\left(\Lambda_\#(\rho)A\right),\ \ A\in \cB(\cK),\ \ \rho\in
\cB^1(\cK),
\]
so that $\Lambda$ is the adjoint of $\Lambda_\#$. Clearly, $\Lambda$ is Markov iff
$\Lambda_\#$ preserves the trace (in this case $\Lambda_\#$ preserves the
set of \ul{density matrices}).

Both $\Lambda$ and $\Lambda_\#$ can be used to describe dynamics of a quantum
system. One says that $\Lambda$ describes it in the \ul{Heisenberg picture} and
$\Lambda_\#$ in the  \ul{Schrödinger picture}.

Let $t\mapsto\e^{tM}$ be a completely positive Markov semigroup on $\cB(\cK)$. It is well-known that $M$ can be written in
the so-called  \ul{Lindblad form}:
\beq
M(A)=\i[\Upsilon,A]-\frac12\sum_{i=1}^n[W_i^*W_i,A]_++\sum_{i=1}^nW_i^*AW_i,
\label{lind}
\eeq
where $\Upsilon$ is a selfadjoint operator and $W_i$, $i=1,\dots, n$ is
a family of operators, see, e.g., \cite{Alicki2007}. We also have the corresponding
semigroup in the Schrödinger picture $\e^{t M_\#}$. Its generator has the form
$$
M_\#(\rho)=-\i[\Upsilon,\rho]-\frac12\sum_{i=1}^n[W_i^*W_i,\rho]_++\sum_{i=1}^nW_i\rho W_i^*.
$$
Completely positive Markov semigroups  are often used in physics as
approximate dynamics of a system weakly interacting with environment;  Section~\ref{Weak coupling limit for the reduced dynamics}. 
 
%%%%%%%%%%%%%%%%%%%%%%%%%%%%%%%%%%%%%%%%
\subsection{Detailed Balance Condition}
%%%%%%%%%%%%%%%%%%%%%%%%%%%%%%%%%%%%%%%%

We say that a density matrix $\rho$ is invariant under  $\e^{tM}$ iff
$\e^{tM_\#}\rho=\rho$ for any $t>0$, or equivalently, $M_{\#}(\rho)=0$. It is
easy to see that for a finite dimensional $\cK$ such an invariant density matrix
always exists.

In the literature there exist a number of definitions of the
Detailed Balance Condition for quantum systems that slightly differ
from one another. One of them  goes as follows. Let $\rho$ be a
nondegenerate density matrix. Introduce the following inner product
on the operators on $\cK$:
\beq
(A|B)_\rho := \tr\left(\rho^{\frac12}A^*\rho^{\frac12}B\right).
\label{dbc}
\eeq
Let $\e^{tM}$ be a completely positive Markov semigroup. We will
say that $M$ satisfies the \ul{Detailed Balance Condition} for $\rho$
iff $\rho$  is invariant under  $\e^{tM}$ and there exists a selfadjoint
operator $\Upsilon$ such that $M-\i[\Upsilon,\argdot]$ is selfadjoint for
the inner product~\eqref{dbc}.

In another version of the Detailed Balance Condition one uses the inner product
\beq
\tr(\rho A^*B)
\label{dbc1}
\eeq
instead of~\eqref{dbc}, see~\cite{Spohn1978b,Derezinski2006} and references therein.

%%%%%%%%%%%%%%%%%%%%%%%%%%%%%%%%%%%%%%%%%%%%%%%%%%%%%%%%%%
\subsection{Weak coupling limit for the reduced dynamics}
\label{Weak coupling limit for the reduced dynamics}
%%%%%%%%%%%%%%%%%%%%%%%%%%%%%%%%%%%%%%%%%%%%%%%%%%%%%%%%%%

Consider a quantum Hamiltonian
\beq
H_\lambda := K{\otimes}\one+\one\otimes\int \omega(\xi)a^*(\xi)a(\xi)\d\xi
+\lambda\int v(\xi){\otimes}a^*(\xi)\d\xi+\lambda\int v(\xi)^*{\otimes}a(\xi)\d\xi
\label{hammia1}
\eeq
on a Hilbert space $\cK{\otimes}\Gamma_\s(L^2(\Xi))$. One could consider
also fermionic reservoirs, see Section~\ref{sec:Coupling to reservoirs}
for a discussion of the notation and motivation.

Suppose that we are interested in the measurements performed only at the small
system. Such measurements are described by $A\otimes\one$, where $A\in\cB(\cK)$.

Let $\Omega$ denote the vacuum vector. We assume that the initial state of the
reservoir is given by the state $(\Omega\mid\argdot\,\Omega)$, and the state of the
small system is given by a density matrix $\rho\in\cB^1(\cK)$. Then the
expectation value of the measurement of the observable $A{\otimes}\one$ at time $t$
is equal to
\[
\tr\bigl(\rho\otimes|\Omega)(\Omega|\bigr)\bigl(\e^{\i tH_\lambda}(A\otimes\one)\e^{-\i tH_\lambda}\bigr).
\]
These expectation values can be encoded in the following family of completely positive Markov maps:
\[
\cB(\cK)\ni A\mapsto I_\cK^*\e^{\i tH_\lambda}(A{\otimes}\one )\e^{-\i tH_\lambda}I_\cK\in \cB(\cK),
\]
where $I_\cK$ denotes the isometric embedding of $\cK$ into $\cK\otimes\Omega$.
The following remarkable fact can be proven.
\bet
Assume some relatively mild conditions on the interaction. Then there exists a completely
positive Markov semigroup $\e^{t M}$ on $\cB(\cK)$ such that
\beq
\e^{t M}(A)=\lim_{\lambda\to0}\e^{-\i \frac{t}{\lambda^2}K}I_\cK^*\e^{\i \frac{t}{\lambda^2}H_\lambda}(A\otimes1)\e^{-\i \frac{t}{\lambda^2}H_\lambda}I_\cK\e^{\i \frac{t}{\lambda^2}K}.
\label{weak}
\eeq
The semigroup $\e^{tM}$ commutes with the free dynamics of the small system
$\e^{\i tK}\argdot\e^{-\i tK}$. If the reservoir is at the inverse temperature
$\beta$ (see Section~\ref{sec:Coupling to reservoirs of a positive density}),
 then $M$ satisfies the Detailed Balance Condition for  $\e^{-\beta
  K}/\tr \e^{-\beta K}$ (the Gibbs state
of the small system), both in the sense of~\eqref{dbc} and~\eqref{dbc1}.
\eet

The above theorem is due to E.~B.~Davies. It indicates that for small $\lambda$
one can approximate the dynamics reduced to the small system  with a
completely positive Markov semigroup.

The limit considered in~\eqref{weak} goes under the name of the \ul{weak
coupling limit} or \ul{van Hove limit}. The latter name is unfortunately
ambiguous -- it is used in statistical physics with a different meaning.

See~\cite{Spohn1978b} for use of Davies generator in context  of non-equilibrium open quantum systems.

%%%%%%%%%%%%%%%%%%%%%%%%%%%%%%%%%%%%%%%
\subsection{Quantum Langevin dynamics}
\label{Quantum Langevin dynamics}
%%%%%%%%%%%%%%%%%%%%%%%%%%%%%%%%%%%%%%%

Suppose that $\e^{tM}$ is a completely positive Markov semigroup on $\cB(\cK)$
with the generator $M$ written in the Lindblad form~\eqref{lind}.
Introduce the Hilbert space $\cZ=\cK\otimes\Gamma_\s(\CC^n\otimes L^2(\RR))$ and  the operator
\beq
Z:= \Upsilon\otimes\one+\sum_{j=1}^n\one{\otimes}\int\xi a_j^*(\xi) a_j(\xi)\d\xi
+\sum_{j=1}^n\int \left(W_j\otimes a_j^*(\xi)+ W_j^*\otimes a_j(\xi)\right)\d\xi.
\label{lange}
\eeq
Strictly speaking, \eqref{lange} is only a formal expression whose precise
meaning as a selfadjoint operator on $\cZ$ can be given using an appropriate
regularization and a limiting procedure. Then one can show that
\begin{eqnarray}
I_\cK^*\e^{\i tZ}(A\otimes1)\e^{-\i tZ}I_\cK&=&\e^{tM}(A),\label{lange1}\\[4pt]
I_\cK^*\e^{\i tZ}I_\cK&=&\e^{\i t\Gamma},\label{lange2}
\end{eqnarray}
where
$\Gamma=\Upsilon+\frac{\i}{2}\sum_{j=1}^n W_j^*W_j$ and where $I_\cK$
denotes the isometric embedding of $\cK$ onto
$\cK\otimes\Omega\subset\cZ$.

Relation~\eqref{lange2} means that $\e^{\i tZ}$ is a \ul{unitary dilation}
of the semigroup $\e^{\i t\Gamma}$. More interestingly, \eqref{lange1} means that
the Heisenberg dynamics given by $\e^{\i tZ}$ reduced to the small system
coincides with $\e^{t M}$.

The unitary dynamics generated by $Z$ is called a \ul{quantum Langevin dynamics}
or quantum stochastic dynamics for the semigroup $\e^{tM}$. Its construction was
first given by Hudson and Parthasarathy. They used a quantum version of the so-called
\ul{stochastic calculus}; see~\cite{Fagnola2006} and references therein.

There exist versions of the weak coupling limit not just for the reduced
dynamics, but also for the dynamics of the full system. They say that in an
appropriate version of the weak coupling limit, but without reducing the
dynamics to the small system, the dynamics generated by a Hamiltonian of the
form~\eqref{hammia1} converges to a quantum Langevin  dynamics. First results of
this type, under the name of the stochastic limit, are due to Accardi, Frigerio
and Lu; there exists also a somewhat different approach under the name of
extended weak coupling limit; see~\cite{Derezinski2008a} and references therein.

%%%%%%%%%%%%%%%%%%%%%%%%%%%%%%%%%%%%%%%%%%%%%%%%%%%%%%%%%%%%%%%%%%%%%%%%%%%%%%%%%%
\section{Scattering theory for a small system interacting with quantum fields}
\label{sec:Scattering theory for a small system interacting with quantum fields}
%%%%%%%%%%%%%%%%%%%%%%%%%%%%%%%%%%%%%%%%%%%%%%%%%%%%%%%%%%%%%%%%%%%%%%%%%%%%%%%%%%
The aim of scattering theory is to describe a complicated dynamics by a
simpler one. In its most basic form, applicable for example to one- and
two-body Schr\"odinger Hamiltonians, one starts from a pair of dynamics:
the ``full'' one $\e^{-\i t H}$ and the ``free'' one $\e^{-\i t H_0}$.
One then defines \ul{wave operators} and \ul{scattering operators}.
See~\cite{Yafaev2010,Reed1979}.

Scattering theory for small quantum systems interacting with quantum
fields uses a different formalism. One starts from a ``full'' dynamics
$\e^{-\i t H}$ and then seeks to describe its asymptotic behavior in
terms of a simpler dynamics, which must be constructed.

A typical example of a Hamiltonian for which one can try to develop
scattering theory is
\beq
H :=  K{\otimes}1+1\otimes\int \omega(\xi)a^*(\xi)a(\xi)\d\xi
+\int v(\xi){\otimes}a^*(\xi)\d\xi+\int v(\xi)^*{\otimes}
a(\xi)\d\xi
\label{hammi4}
\eeq
on $\cB(\cK){\otimes}\Gamma_\s(L^2(\RR^d))$. (One could consider also similar
fermionic systems). We stress that  the free dynamics is not given a priori
given. In particular, the dynamics generated by the first two terms of the
right hand side of~\eqref{hammi4} should not be viewed as the free dynamics
in the sense of scattering theory. Therefore one needs a different formalism.

Let us describe a version of scattering theory for the Hamiltonian $H$ given
by~\eqref{hammi4}, which is  adapted to the case when the small system is well
localized in space (confined). In our presentation we follow mostly~\cite{Derezinski2004a}.
Note that this formalism of scattering theory is quite different from that
used in the case of e.g. Schrödinger operators -- it is much closer to the
\ul{LSZ formalism} or the \ul{Haag-Ruelle theory} used in
quantum field theory.

We first construct the so-called \ul{incoming/outgoing} creation/annihilation
operators. In the fermionic case they are given as
\begin{align}
\int f(\xi)a^{\pm*}(\xi)\d\xi
& := \slim_{t\to\pm\infty}\e^{\i tH}\int\e^{-\i t\omega(\xi)}f(\xi)a^*(\xi)\d\xi\ \e^{-\i tH},\label{bou1}\\[4pt]
\int \bar f(\xi)a^{\pm}(\xi)\d\xi
& := \slim_{t\to\pm\infty}\e^{\i tH}\int\e^{\i t\omega(\xi)}\bar f(\xi)a(\xi)\d\xi\ \e^{-\i tH},\label{bou2}
\end{align}
where ${\rm s}-\lim$ denotes the \ul{strong limit}.
In the bosonic case the definition is similar except that we need to take
bounded functions of~\eqref{bou1} and~\eqref{bou2}.

The incoming/outgoing creation/annihilation operators exist under quite weak
assumptions and their existence can be usually proven by a version of the
\ul{Cook method}. They satisfy the usual commutation/anticommutation relations.

To proceed further one introduces the space of asymptotic Fock vacua
$\cK^\pm$, consisting of vectors annihilated  by the  asymptotic annihilation operators
$a^\pm(\xi)$. One can show that bound states of $H$ belong to both
$\cK^\pm$. Then one introduces \ul{wave operators} $W^\pm$ -- isometric
operators from the asymptotic space $\cK^\pm\otimes\Gamma_{\s/\a}(L^2(\Xi))$ into
the physical space  $\cK\otimes\Gamma_{\s/\a}(L^2(\Xi))$, defined by the
relations
\begin{eqnarray*}
W^\pm a(\xi)= a^\pm(\xi)W^\pm,&\ 
W^\pm a^*(\xi)= a^{\pm*}(\xi)W^\pm,&\
W^\pm\Psi\otimes\Omega=\Psi,\ \ \Psi\in\cK^\pm.\end{eqnarray*}
($\Omega$, as usual, denotes the vacuum vector). Finally, one introduces the 
\ul{scattering operator} $S=W^{+*}W^-$, which can be used to compute
scattering cross-sections.

Here is an example of a rigorous result about scattering theory in
the above formalism, see~\cite{Derezinski2004a} and references therein:
\bet\label{thm:theo}
Assume that the 1-particle excitation spectrum is positive and has a mass gap,
e.g. $\omega(\xi)=\sqrt{m^2+\xi^2}$, $m>0$ (and some additional minor technical
assumptions). Then
\begin{enumerate}%[(i)]
\item the wave operators $W^\pm$ are unitary;
\item the spaces of asymptotic vacua $\cK^\pm$ coincide with the space
spanned by eigenvectors of $H$.
\end{enumerate}
\eet

If the above formalism is applicable and Theorem~\ref{thm:theo} holds, then we
have a clear physical picture of the dynamics $\e^{\i tH}$ for large times:
all vectors  can be understood as linear combinations of bound
states of $H$ accompanied by ``asymptotic bosons'' whose 1-particle excitation
spectrum coincides with that of ``free bosons''. Note that
Theorem~\ref{thm:theo}  can be viewed as the analog of \ul{asymptotic completeness} from
scattering theory of $N$-particle Schrödinger operators.

Scattering in the absence of  the energy gap in the excitation spectrum remains poorly understood. Even less understood is scattering theory for
Hamiltonians similar to~\eqref{hammi4} with a translation symmetry~\cite{Froehlich2002}.

%%%%%%%%%%%%%%%%%%%%%%%%%%%%%%%%%%%%%%%%%%%%%%%%%%%%%%%%%%%%%%%%%%%%%%%%%%%%%%%%%%%%
\section{Spectral analysis of small quantum systems interacting with a reservoir}
\label{sec:Spectral analysis of small quantum systems interacting with a reservoir}
%%%%%%%%%%%%%%%%%%%%%%%%%%%%%%%%%%%%%%%%%%%%%%%%%%%%%%%%%%%%%%%%%%%%%%%%%%%%%%%%%%%%

In Quantum Optics one often uses various approximate Hamiltonians to
describe interaction of photons with nonrelativistic matter. Here is
an example:
\beq
H := (p-\alpha^{3/2}A(\alpha x))^2-\frac{z}{4\pi|x|}
+\sum_{\epsilon}\int|\xi|a_\epsilon^*(\xi)a_\epsilon(\xi)\d\xi
\label{hammi6}
\eeq
$H$ acts on $L^2(\RR^3){\otimes}\Gamma_\s(L^2(\RR^3\times\{1,2\})$.
It describes a single electron in the potential of a nucleus
interacting with photons through the minimal coupling.
Note that $p$ is the momentum operator, $\epsilon$ describes
two polarizations of the photon, $\alpha$ is  the fine
structure constant, $z$ is the charge of the ``nucleus'', and $A$ is
the quantized electromagnetic potential defined
with the  help of appropriately cut-off photonic creation and
annihilation operators $a_\epsilon^*(\xi), a_\epsilon(\xi)$.

Various Hamiltonians similar to \eqref{hammi6} are used in
the literature. For instance, one can argue that the  term quadratic
in electromagnetic potentials can be eliminated with the help of the
so-called Pauli-Fierz transformation 
\cite{Derezinski2003}. 
This leads to the following class of Hamiltonians, which we will
consider in this section:
\beq
H := K\otimes\one+\one\otimes\int \omega(\xi)a^*(\xi)a(\xi)\d\xi
+\lambda\int v(\xi)\otimes a^*(\xi)\d\xi+\lambda\int v(\xi)^*\otimes a(\xi)\d\xi
\label{hammi5}
\eeq
$H$ acts on $\cB(\cK){\otimes}\Gamma_\s(L^2(\RR^d))$, where
we will assume that $\cK$ is finite-dimensional.
$\Gamma_\s(L^2(\RR^d))$ denotes the bosonic Fock space over
$L^2(\RR^d)$. Hamiltonians of this form go under various names:
generalized spin-boson Hamiltonians, Pauli-Fierz Hamiltonians, etc.

One can also consider
 small systems interacting with fermionic reservoirs,
see Section~\ref{sec:Coupling to reservoirs}. Similar mathematical
methods work for them.

%%%%%%%%%%%%%%%%%%%%%%%%%%%%%%%%%%%%%%%%%%%%%
\subsection{Absolute continuity of spectrum}
\label{ssec:Absolute continuity of spectrum}
%%%%%%%%%%%%%%%%%%%%%%%%%%%%%%%%%%%%%%%%%%%%%

It is easy to analyze the spectrum of the operators~\eqref{hammi5} for
$\lambda=0$. In typical situations the spectrum is absolutely continuous apart
from some  point spectrum coming from the eigenvalues of $K$. The non-trivial 
case $\lambda\not=0$  is much more
difficult to study.

There are a number of results  saying that for nonzero $\lambda$ most of the
spectrum of $H$ is absolutely continuous as well. Two basic
methods are used to prove this: the \ul{analytic deformation method} and the
\ul{method of positive commutators}, often also called  the \ul{Mourre method}.
Both methods were originally discovered in the context of \ul{$N$-body
Schrödinger operators}, but they can be adapted to  the operators considered
in this article. Note that both methods not only prove the absolute continuity
of the spectrum but also yield important information on the resolvent. The
former method implies that the resolvent can be analytically continued across
the spectrum. The latter method yields the existence of  boundary values of the
resolvent at the spectrum, as a quadratic form between appropriate weighted
spaces. (The last statement goes under the name of \ul{Limiting Absorption
Principle}).

In both methods an important ingredient is the choice of the so-called
\ul{conjugate operator}. In the literature, there are two main choices for  the
conjugate operator. The first choice, introduced in~\cite{Jaksic1996a},
involves the \ul{second quantization} of the \ul{generator of
translation} in the spectral variable. It is used mostly when the spectrum
covers the whole real line and we can ``glue'' the negative and positive
frequencies (the so-called \ul{Jak\v si\'c--Pillet gluing}),
see also~\cite{Derezinski2003}. The second choice is the second quantization
of the \ul{generator of dilations}, used e.g. in~\cite{Bach1999,Bach2000}.

%%%%%%%%%%%%%%%%%%%%%%%%%%%%%%%%%%%%%%%%%%%%%%%%%%%%%%%
\subsection{Point spectrum and the Fermi Golden Rule}
\label{ssec:Point spectrum and the Fermi Golden Rule}
%%%%%%%%%%%%%%%%%%%%%%%%%%%%%%%%%%%%%%%%%%%%%%%%%%%%%%%

There exist also results about the point spectrum of perturbed operators.

Consider the operator~\eqref{hammi5}. Let us assume that $k$ is an
eigenvalue of $K$, and hence an eigenvalue of $H$ for $\lambda=0$.
We attempt  to compute the shift of this eigenvalue using the perturbation
theory. For simplicity we assume that $k$ is a non-degenerate eigenvalue of
$K$. The term linear in $\lambda$ vanishes. The second-order contribution to the eigenvalue  shift is
\beq
\int|v(\xi)|^2(k+\i0-\omega(\xi))^{-1}\d\xi.
\label{fgr}
\eeq
The imaginary part of~\eqref{fgr} equals $-\i\pi|v(k)|^2$. If this quantity is  non-zero,
then it clearly it cannot arise  in the  perturbation expansion of an eigenvalue
of a selfadjoint operator.

The formula~\eqref{fgr} is an instance of the so-called
\ul{Fermi Golden Rule}. It can be readily generalized to the case where $k$
is a degenerate eigenvalue. In that situation, the scalar expression
\eqref{fgr} is replaced by the so-called \ul{level shift operator}.
Heuristically, it is easy to see that, for small but nonzero $\lambda$,
the number of eigenvalues of the perturbed operator in a neighborhood of a
given unperturbed eigenvalue should be  less than the dimension of
the kernel of the level shift operator.

In many situations this heuristic picture can be made rigorous.
More precisely, one can often show that there exists $\lambda_0>0$ such
that, for $0<|\lambda|<\lambda_0$, the above expectations are indeed
fulfilled.

When the interaction is sufficiently analytic, one can study
\ul{resonances}, that is, eigenvalues of suitably analytically deformed
operators. At positive temperatures, resonances can be introduced using
the Jak\v{s}i\'c--Pillet gluing construction together with analyticity with
respect to translations~\cite{Jaksic1996a}. At arbitrary temperature, one
may attempt to define resonances using dilation analyticity. With this
approach, resonances are not isolated eigenvalues of the deformed
operator; rather, they are located at the tips of cuspidal regions of its
spectrum. Such resonances have been analyzed using an iterative method
based on the \ul{renormalization group}
in~\cite{Bach1999,Bach2000}.

Especially satisfactory results are available in the case of the
\ul{ground state}. Under quite general conditions, and for arbitrary
values of the coupling constant, one can show that the
Hamiltonian~\eqref{hammi5} or~\eqref{hammi6} has an eigenvalue at the bottom
of its spectrum~\cite{Griesemer2001,Georgescu2004}.

%%%%%%%%%%%%%%%%%%%%%%%%%%%%%%%%%%%
\subsection{Return to equilibrium}
\label{ssec:Return to equilibrium}
%%%%%%%%%%%%%%%%%%%%%%%%%%%%%%%%%%%

Assume that the 1-particle excitation spectrum of 
the bosonic reservoir is given by $\omega(\xi)=|\xi|$, and  fix a non-negative  function $\gamma$. Consider the $W^*$-dynamical
system $(\fM_\gamma,\tau_\gamma)$ introduced in
Section~\ref{sec:Coupling to reservoirs of a positive density}.
We would like to analyze its normal stationary states.

The stationary states of the decoupled system (the case 
$\lambda=0$) are easy to analyze: they are all described by density matrices of the small system that commute with 
$K$, tensored with the vacuum state of the reservoir. We expect that, upon switching on the perturbation, most of these states disappear, since they correspond to embedded eigenvalues of the generator of the dynamics. Only the KMS state is stable.

This expectation can be turned into two rigorous theorems, that we describe
below.  We start with a theorem describing the thermal case:

\bet\label{thm:thm1}
Assume that the reservoir is thermal at inverse temperature $\beta$.
Suppose that the interaction is sufficiently regular and
effective (effective means that it couples the reservoir to the small system
sufficiently well). Then there exists $\lambda_0>0$ such that for
$0<|\lambda|\leq\lambda_0$ the system $(\fM_\gamma,\tau_\gamma)$
has a unique normal stationary state $\psi$. This state
is $\beta$-KMS for $\tau_\gamma$. It satisfies the property of
\ul{return to equilibrium}: for any normal state $\phi$
and $A\in\fM_\gamma$
\[
\lim_{|t|\to\infty}\phi\left(\tau_{\gamma,t}(A)\right)=\psi(A).
\]
\eet
In the nonthermal case, by analogous methods one can prove

\bet
Assume that the reservoir consists of two parts, which are thermal at
distinct   temperatures. Suppose that the interaction is sufficiently regular and
effective. Then there exists $\lambda_0>0$ such that for
$0<|\lambda|\leq\lambda_0$ the system $(\fM_\gamma,\tau_\gamma)$
has no normal stationary states.
\eet

In  order to prove these theorems it is convenient to use the so-called
Liouvillean, which is a certain naturally defined selfadjoint operator
implementing the dynamics $\tau_\gamma$ in the so-called
\ul{standard representation} of the $W^*$-algebra $\fM_\gamma$, see
Section~\ref{sec:The C*-algebra approach}.

Note that for $\lambda=0$, the Liouvillean has a degenerate eigenvalue at
$0$ and it has non-zero eigenvalues corresponding to differences of the
eigenvalues of $K$ -- the so-called \ul{Bohr frequencies} of the small system.
To prove Theorem~\ref{thm:thm1} one needs to show that for small nonzero
$\lambda$ the spectral properties of the Liouvillean change: it has a
one-dimensional kernel and its spectrum away from zero is absolutely
continuous.

This is achieved in two steps. The theory of KMS states, mentioned in
Section~\ref{sec:The C*-algebra approach}, shows that the
dimension of the kernel of the Liouvillean is at least one. Then one
studies spectral properties of the Liouvillean, as sketched in
Sections~\ref{ssec:Absolute continuity of spectrum}
and~\ref{ssec:Point spectrum and the Fermi  Golden Rule}.
In particular, an argument based  on the Fermi Golden Rule shows that
the remaining eigenvectors of the Liouvillean do not survive when we
switch on the perturbation, and the dimension of the kernel of the Liouvillean
is at most $1$. There are several methods of a proof of this
theorem that work under various assumptions -- the original one goes back
to~\cite{Jaksic1996a}. See also~\cite{Derezinski2003,Bach2000}.

%%%%%%%%%%%%%%%%%%%%%%%%%%%%%%%%%%%%%%%%%%%%%%%%%%%%%%%%%
\part{Equilibrium quantum statistical mechanics}
\label{sec:Quantum nonequilibrium statistical mechanics}
%%%%%%%%%%%%%%%%%%%%%%%%%%%%%%%%%%%%%%%%%%%%%%%%%%%%%%%%%

Starting with the seminal work~\cite{Haag1967},  the mathematical theory of
equilibrium quantum statistical mechanics based on the KMS-condition has
developed rapidly in  the 1970s, resulting in a structure of rare unity and beauty.
The summary of these developments can be found in the two
volumes~\cite{Bratteli1981, Bratteli1987}; 
see also the monographs~\cite{Haag1996, Israel1979, Simon1993, Thirring2002} and
references therein. In parallel to the classical equilibrium theory of  lattice 
spin systems based on the DLR equation, developed roughly at the same time, 
lattice quantum spin  systems are  the main paradigm of the quantum theory  
based on the KMS-condition\footnote{The equilibrium theory of quantum lattice 
spin systems is easily adapted to  lattice fermions although a pedagogical 
exposition of this adaptation is lacking in the literature. A parallel 
exposition of the spin and fermion case, together with some far reaching 
generalizations, can be found in~\cite{Araki2003}.}.

%%%%%%%%%%%%%%%%%%%%%%%%%%
\section{General setting}
\label{sec-eqsm-gen}
%%%%%%%%%%%%%%%%%%%%%%%%%%

We start with the setting of a quantum system with finite-dimensional Hilbert 
space $\cK$. We will refer to such quantum systems as \textsl{finite quantum systems}. $\cO_\cK$ 
denotes the $C^\ast$-algebra of  all linear maps $A:\cK\to\cK$ and 
$\cS_\cK\subset\cO_\cK$ the set of all density matrices on $\cK$. $\one$
denotes the identity in $\cO_\cK$. The observables of the system are identified
with the elements of $\cO_\cK$ and its physical states with elements of $\cS_\cK$, 
with the usual duality $\nu(A)=\tr(\nu A)$, $\nu\in \cS_\cK$, $A\in \cO_\cK$.  
The number $\nu(A)$ is interpreted as the expectation value of the observable $A$  
if the system is in the state $\nu$. A state $\nu$ is called faithful if $\nu>0$. 
The dynamics is described by the system  Hamiltonian $H=H^\ast\in\cO_\cK$ and the
induced group $\tau=\{\tau^t\,|\, t\in \RR\}$ of $\ast$-automorphisms of $\cO_\cK$,
defined by 
\[
\tau^t(A)=\e^{\i tH}A\e^{-\i tH}.
\]
We will sometimes write $A_t$ for $\tau^t(A)$ and call the map $A\mapsto A_t$ the 
dynamics in Heisenberg picture. In the dual Schrödinger picture, the states evolve 
in time as  $\nu\to\nu_t$, where 
\[
\nu_t=\e^{-\i tH}\nu\e^{\i tH}.
\]
Obviously, $\nu_t(A)=\nu(A_t)$. The time-correlations are quantified by the function 
\[
F_{\nu,A,B}(t)=\nu(AB_t).
\]

A triple $(\cO_\cK,\tau,\omega)$, where $\omega$ is the initial (reference)
state of the system, is called a  finite quantum dynamical system. This system 
is said to be in thermal equilibrium at inverse temperature $\beta\in\RR$ if
its reference state is the Gibbs canonical ensemble 
\beq
\omega_\beta=\e^{-\beta H}/\tr(\e^{-\beta H}).
\label{gibbs}
\eeq
The Gibbs ensemble~\eqref{gibbs} is the unique state in $\cS_\cK$ satisfying the:

\medskip
\noindent\textbf{Finite KMS-condition.} The time-correlation functions 
$F_{\nu,A,B}$ have  the property
\[
F_{\nu,A,B}(t+\i\beta)=F_{\nu,B,A}(t)
\]
for all $A,B\in\cO_\cK$ and $t\in\RR$. 
\bigskip

Another fundamental characterization of the state~\eqref{gibbs} is given by  the following result. 

\medskip
\noindent\textbf{Finite Gibbs variational principle.} Set
$P(\beta)=\log\tr(\e^{-\beta H})$ and denote by 
$S(\nu)=-\tr(\nu\log\nu)$ the von~Neumann entropy of 
$\nu\in\cS_\cK$. Then, for any $\nu\in\cS_\cK$, 
\[
S(\nu)-\beta\nu(H)\leq P(\beta),
\]
with equality iff $\nu=\omega_\beta$.

\medskip
The relative entropy of two states $\nu,\rho\in\cS_\cK$ is defined 
by\footnote{This relative entropy is finite iff $\Ker\rho\subseteq\Ker\nu$,
otherwise it takes the value $-\infty$.} 
\beq 
\Ent(\nu|\rho)=\tr(\nu(\log\rho-\log\nu)).
\label{fin-rel-ent}
\eeq
Its basic property is that $\Ent(\nu|\rho)\leq0$ with equality iff 
$\nu=\rho$. One computes 
\[
\Ent(\nu|\omega_\beta)=S(\nu)-\beta\nu(H)-P(\beta),
\]
and so the finite Gibbs variational principle is an immediate
consequence of the above property of relative entropy. 

In the full generality of algebraic quantum statistical mechanics, observables
of a quantum system are described by elements of a $C^\ast$-algebra $\cO$ with 
unit $\one$. For a large part of the general theory, no other structure is 
imposed on $\cO$. The set of selfadjoint elements in $\cO$ is denoted by 
$\cOs$. Let $\cO^\ast$ be the dual of $\cO$ and  $\cS_\cO$ the set 
of positive normalized elements in $\cO^\ast$. We equip $\cO^\ast$ with the
weak$^\ast$-topology, with respect to which $\cS_\cO$ is a compact, convex subset 
of $\cO^\ast$. The physical states are described by elements of
$\cS_\cO$. If the system is in a state $\nu$, then the number 
$\nu(A)$ is interpreted as the expectation value of the observable $A$. 

The notion of Hamiltonian is lost, and the dynamics in the Heisenberg picture is,
from the outset, described by a strongly continuous
\footnote{$\lim_{t\to0}\|\tau^t(A)-A\|=0$ for all $A\in\cO$} group 
$\tau=\{\tau^t\,|\,t\in\RR\}$ of $\ast$-automorphisms of $\cO$. The group 
$\tau$ is called $C^\ast$-dynamics and the pair $(\cO,\tau)$ a $C^\ast$-dynamical 
system. Its dual action $\tau^\ast$ preserves $\cS_\cO$ and describes the dynamics in the 
Schrödinger picture. We write $A_t$ for $\tau^t(A)$ and $\nu_t$ for 
$\tau^{t\ast}(\nu)=\nu\circ\tau^t$. The function $F_{\nu,A,B}(t)$ is defined in the same way as 
in the finite case. 

A state $\nu$ is called $\tau$-invariant (or stationary) if 
$\nu_t=\nu$ for all $t$. The set of all $\tau$-invariant states is denoted by 
$\cS_\tau$ and is always non-empty. A triple $(\cO,\tau,\omega)$, where $\omega$ 
is the initial (reference) state of the system, is called a $C^\ast$-quantum dynamical 
system. 

 A state $\omega\in\cS_\tau$ is called ergodic if 
\[
\lim_{T\to\infty}\frac1{2T}\int_{-T}^T \omega(B^\ast A_tB)\d t
=\omega(A)\omega(B^\ast B)
\]
holds for all $A,B\in\cO$. Upon passing to the GNS representation, it becomes clear that 
this definition is a special case of the one presented in Section~\ref{sec-ergodic-property}.

Time-reversal plays an important role in statistical mechanics. An anti-linear involutive 
$\ast$-automorphism $\Theta$ of $\cO$ is a time-reversal of the $C^\ast$-dynamical system 
$(\cO,\tau)$ if 
\[
\Theta \circ \tau^t=\tau^{-t}\circ \Theta
\]
for all $t\in\RR$. A state $\omega$ is called time-reversal invariant if 
$\omega(\Theta(A))=\omega(A^\ast)$ for all $A\in\cO$. In the sequel we abbreviate 
time-reversal invariance by TRI. 

Since the notion of Hamiltonian is lost, so is the notion of the Gibbs canonical ensemble~\eqref{gibbs}.
The KMS-condition, however, survives and extends to the general setting as: 

\begin{quote} {\bf KMS-condition.} $\nu\in\cS_\cO$ is a $(\tau,\beta)$-KMS state 
if, for all $A,B\in\cO$, the function $\RR\ni t\mapsto F_{\nu,A,B}(t)$ has an analytic extension 
to the strip $0 <\sign(\beta)\Im z<|\beta|$ that is bounded and continuous on its closure and 
satisfies the KMS-boundary condition 
\[
F_{\nu,A,B}(t+\i\beta)=F_{\nu,B,A}(t)
\]
for all $t\in\RR$. 
\end{quote}

We denote by $\cS_{(\tau,\beta)}$ the set of all $(\tau, \beta)$-KMS states. At the current 
level of generality this set might be empty. One always has $\cS_{(\tau,\beta)}\subset\cS_\tau$. 
A quantum dynamical system $(\cO,\tau,\omega)$ is said to be in thermal equilibrium at inverse 
temperature $\beta\in \RR$ (or just thermal) if $\omega$ is a $(\tau,\beta)$-KMS state. 

A state $\nu$ is called modular if there exists $C^\ast$-dynamics
$\varsigma_\nu$ on $\cO$ such that $\nu$ is $(\varsigma_\nu,-1)$-KMS state. 
$\varsigma_\nu$ is called modular dynamics of $\nu$ and is unique if it exists.
We denote by $\delta_\nu$ the generator of $\varsigma_\nu$ with the convention 
$\varsigma_\nu^t= \e^{t \delta_\nu}$. If $\nu$ is a $(\tau, \beta)$-KMS state
and $\beta\not=0$, then
$\varsigma_\nu^t=\tau^{-\beta t}$, or equivalently, $\delta_\nu=-\beta\delta $
where $\delta$ is the generator of $\tau$.  The modular dynamics plays an
important role in non-equilibrium statistical mechanics. In the equilibrium
setting of quantum spin systems,   its  a priori  existence allows for the
formulation of the Gibbs condition as characterization of thermal equilibrium;
we will briefly discuss this topic in the next section.

Given $(\cO,\tau)$, the two basic questions of equilibrium statistical mechanics are:

\assuming{QI}{QOne}{Describe the properties of KMS-states and the structure of the sets       
$\cS_{(\tau,\beta)}$. 
}

\assuming{QII}{QTwo}{Elucidate the dynamical, and in particular the ergodic, properties of thermal 
quantum dynamical systems. 
}

Note that~\QOne{} and~\QTwo{} have trivial answers in the finite
setting: $\cS_{(\tau,\beta)}$ is always  a singleton and no thermal system is
ergodic  if $\dim(\cK)>1$.  It is precisely this triviality that forces consideration of
infinitely extended systems from the outset in the study of~\QOne{} 
and~\QTwo{}.

From a general  perspective, substantial progress has been made on~\QOne{}
and~\QTwo{} in the 1970s. The volumes~\cite{Bratteli1981,Bratteli1987} are 
devoted to presenting these developments. The link between KMS-condition and 
Tomita-Takesaki's modular theory has played a central role in this progress.  
However, from the standpoint of concrete physically relevant models, advancement  
has been much slower, and  comparatively little is known at a mathematically 
rigorous level. 

The original definition of relative entropy of a pair $(\nu,\rho)$ of states on
$\cO$ goes back to Araki and is based on modular theory.  This definition was 
discussed in Section~\ref{sec-relative-entropy} and we will review it again in 
Section~\ref{sec-modular}. The relative entropy is also characterized by the 
Pusz-Woronowicz-Kosaki variational formula
\beq
\Ent(\nu|\rho)=\inf\int_0^\infty\left[\frac1t\nu(x^\ast(t)x(t))
+\rho(y^\ast(t)y(t))-\frac1{1+t}\right]\frac{\d t}{t},
\label{pw-relent}
\eeq
where $y(t)=\one-x(t)$ and the infimum is taken over all countably valued step
functions  $[0,\infty[\,\ni t\mapsto x(t)\in\cO$ vanishing in a neighborhood of 
zero; see~\cite[Section 5]{Ohya1993}. Its basic property is $\Ent(\nu|\rho)\leq0$ 
with equality iff $\nu=\rho$. In the finite case~\eqref{pw-relent} reduces 
to~\eqref{fin-rel-ent}. The relative entropy plays a very important role in both
equilibrium and non-equilibrium quantum statistical mechanics. In the
equilibrium case and in the context of quantum lattice spin systems, relative
entropy -- through its relation with the specific entropy --  plays a central role in
the derivation of the Gibbs variational principle. This topic is discussed in
detail in Section~6.2.3 of~\cite{Bratteli1981}. 

We conclude  this section with 
three general remarks. 

Modular theory and the closely related Araki's perturbation theory of
KMS-structure play a central role in both equilibrium and non-equilibrium
quantum statistical mechanics. A basic introduction to this subject can be
found in~\cite{Bratteli1981, Bratteli1987}; see also~\cite{Derezinski2003a} and
references therein for modern expositions. A pedagogical introduction to the
modular theory in the context of finite quantum systems can be found
in~\cite{Jaksic2010b}  

$W^\ast$-dynamical systems play a distinguished role in modular theory. A
$W^\ast$-dynamical system is a pair $(\fM,\tau)$ where $\fM$ is a $W^\ast$-algebra 
and $\tau=\{\tau^t\,|\,t\in \RR\}$ is a pointwise $\sigma$-weakly continuous group
of $\ast$-automorphisms on $\fM$. We shall refer to such $\tau$ as $W^\ast$-dynamics. 
A triple $(\fM,\tau,\omega)$, where $\omega$ is a normal state on $\fM$, is called a
$W^\ast$-dynamical system. In the general development of non-equilibrium quantum
statistical mechanics, the $C^\ast$-quantum dynamical systems are preferred starting 
point since the central notion of non-equilibrium steady states cannot be naturally
defined in the $W^\ast$-setting. In this context, the $W^\ast$-systems and
modular structure emerge through the GNS-representation of $\cO$ associated
to the reference state $\omega$. This being said, there  are important physical 
models that can be only described by $W^\ast$-dynamical system; a well-known 
example are Pauli-Fierz systems with bosonic reservoirs. It is most natural 
to develop the non-equilibrium statistical mechanics of such systems on a 
case-by-case basis.

The thermodynamic (often abbreviated TD) limit plays a distinguished role in statistical
mechanics. It realizes infinitely extended systems through a limiting
procedure involving only finite quantum system and is central for the
identification of physically relevant objects in the extended setting. The
precise way the TD limit is taken depends on the structure of the specific
physical model under consideration, and often a number of different
approximation routes are possible. This topic is well-understood and discussed
on many places in the literature; see~\cite{Bratteli1981,Bratteli1987,Ruelle1969}
and~\cite{Jaksic2010b} for a pedagogical introduction to the topic. Since the early 
days it is well-known that the modular structure is stable under TD limit~\cite{Araki1974b}; 
this fact plays an important role in the foundations of quantum statistical
mechanics. The customary route in discussions of the structural theory is 
the following:

\noindent\textbf{Step 1.} A  physical notion, introduced in the context of finite
quantum systems, is expressed in a modular form, and through this form is
directly extended, by definition, to a general $C^\ast$ or $W^\ast$-dynamical
system. One basic example of such procedure is the introduction of the
KMS-condition as characterization of thermal equilibrium states. 

\noindent\textbf{Step 2.} In concrete physical models the definitions of 
Step~1 are justified by the TD limit.

Step 2 has been extensively studied in  the early days of quantum statistical
mechanics, and the wealth of obtained results make its implementation in modern
literature most often a routine exercise. For this reason, this step is often 
skipped. There are rare exceptions  to this rule, one of which emerged in the study 
of entropy production of open quantum systems~\cite{Benoist2023a,Benoist2024b}.

%%%%%%%%%%%%%%%%%%%%%%%%%%%%%%%%%%%%%%%
\section{Lattice quantum spin systems}
\label{sec-lattice-eq}
%%%%%%%%%%%%%%%%%%%%%%%%%%%%%%%%%%%%%%%

We follow~\cite{Bratteli1981}; see also~\cite{Israel1979, Simon1993, Ruelle1969}.

\noindent\textbf{$C^\ast$-algebra.} The lattice $G$ is taken to be a countably 
infinite set. At this stage, no additional structure on $G$ is assumed. The
collection of all finite subsets of $G$ is denoted by $\fGfin$. If $(\Lambda_\alpha)$ 
is a net in $\fGfin$, $\Lambda_\alpha\to\infty$ means that $\Lambda_\alpha$ eventually
contains any finite subset of $G$. 

The single spin Hilbert space is finite dimensional and is denoted by $\fh$. To each 
$x\in G$ we associate a copy $\fh_x$ of $\fh$,  and to each $\Lambda\in\fGfin$ the 
Hilbert space 
\[
\cK_\Lambda=\bigotimes_{x\in\Lambda}\fh_x.
\]
In the sequel, $\cO_\Lambda$ stands for $\cO_{\cK_\Lambda}$. The $C^\ast$-algebra 
$\cO_\Lambda$ describes observables of spins located at the points in $\Lambda$. 
For $\Lambda\subset\Lambda^\prime$ one naturally identifies $\cO_\Lambda$ with a 
$C^\ast$-subalgebra of $\cO_{\Lambda^\prime}$. The algebra of local observables is 
\[
\cOloc=\bigcup_{\Lambda\in\fGfin}\cO_\Lambda.
\]
Finally, the $G$-lattice spin system $C^\ast$-algebra $\cO_G$  is the  completion of 
the $\ast$-algebra of local observables. The algebra $\cO_G$ is unital, simple, and 
separable.  For any $G_0\subset G$ one has the natural identification 
$\cO_G=\cO_{G_0}\otimes\cO_{G_0^c}$. Whenever $G$ is  understood, we write $\cO$ for 
$\cO_G$. 

\noindent\textbf{Dynamics.} An interaction is a map $\Phi:\fGfin\to\cOs$ such that 
$\Phi(X)\in\cO_X$. For $\Lambda\in\fGfin$, the local Hamiltonians are defined by 
\[
H_\Lambda(\Phi)=\sum_{X\subset\Lambda}\Phi(X).
\]
They generate the local $C^\ast$-dynamics $\tau_{\Phi,\Lambda}$ on $\cO$, given by 
\[
\tau_{\Phi,\Lambda}^t(A)=\e^{\i t H_\Lambda(\Phi)}A\e^{-\i t H_\Lambda(\Phi)}.
\]
To extend $\tau_{\Phi,\Lambda}$ to a $C^\ast$-dynamics on $\cO$ in the limit 
$\Lambda\to\infty$, one needs a suitable regularity assumption. We choose 
the following one\footnote{See Theorem 6.2.4 in~\cite{Bratteli1981}.}
\assuming{SR}{SR}{For some $\lambda>0$,
\[
\|\Phi\|_\lambda := \sum_{n\geq0}\e^{\lambda n}
\left(\sup_{x\in G}\sum_{\twol{x\in X}{|X|=n+1}}\|\Phi(X)\|\right)<\infty.
\]
}
\begin{theorem} Suppose that~\SR{} holds. Then:
\ben
\item For all $A\in\cO$, the limit 
\[
\tau_\Phi^t(A) :=\lim_{\Lambda\to\infty}\tau_{\Phi,\Lambda}^t(A)
\]
exists and is uniform for $t$ in compacts sets.
\item  $\tau_\Phi=\{\tau_\Phi^t\,|\,t\in\RR\}$ is a $C^\ast$-dynamics on 
$\cO$. We  denote by $\delta_\Phi$ its generator.
\item $\cOloc\subset\Dom(\delta_\Phi)$, and for $A\in\cO_\Lambda$, 
\[
\delta_\Phi(A)=\i\sum_{X\cap\Lambda\not=\emptyset}[\Phi(X),A].
\]
Moreover,  $\cOloc$ is a core for $\delta_\Phi$. 
\item For $A\in\cOloc$ and $n\geq1$, 
$$
\|\delta_\Phi^n(A)\|\leq\frac{2^n n!}{\lambda^n}
\e^{\lambda|\Lambda|}\|\Phi\|_\lambda^n \|A\|,
$$
where $|\Lambda|$ denotes the cardinality of $\Lambda$. 
In particular, for  all $A\in\cOloc$, the map 
\[
\RR\ni t\mapsto\tau_\Phi^t(A)\in\cO
\]
has an analytic extension to the strip $|\Im z|<\frac{\lambda}{2}\|\Phi\|_{\lambda}^{-1}$. 
\een
\end{theorem}
For the proof, we refer the reader to \cite[Lemma 7.6.1 and Theorem 7.6.2]{Ruelle1969} and  \cite[Theorem 6.2.4]{Bratteli1981}.

Until the end of this section we assume that~\SR{} holds. 

\noindent\textbf{KMS-states.} Let $\beta >0$ be the inverse temperature. For 
$\Lambda\in\fGfin$,  the local Gibbs state is defined by 
\beq
\omega_{\Lambda,\beta}=\e^{-\beta H_\Lambda(\Phi)}/\tr(\e^{-\beta H_\Lambda(\Phi)}).
\label{thgivs}
\eeq
Using the identification $\cO=\cO_{\Lambda}\otimes\cO_{\Lambda^c}$, one extends 
$\omega_{\Lambda, \beta}$ (in an arbitrary way) to a state on $\cO$. Denoting 
this extension by the same letter, it is a basic result that any 
weak$^\ast$-limit point of the net $({\omega}_\beta(\Lambda))_{\Lambda \in\fGfin}$ as 
$\Lambda\to\infty$ is a $(\tau_\Phi,\beta)$-KMS state on $\cO$. The 
$(\tau_\Phi,\beta)$-KMS states that arise in this way  are called thermodynamic-limit point $\beta$-KMS states.\footnote{We follow the terminology introduced 
in \cite[Section 6.2.2.]{Bratteli1981}.} This construction, in particular, shows that 
the set $\cS_{(\tau_\Phi,\beta)}$ is non-empty. 

If it happens that $\cS_{(\tau_\Phi,\beta)}$ is a singleton, then all the nets 
$({\omega}_\beta(\Lambda))_{\Lambda\in\fGfin}$ converge, as $\Lambda\to\infty$, 
to the unique $(\tau_\Phi,\beta)$-KMS state. This is known to be the case in the high-temperature regime, \ie for $\beta\|\Phi\|_\lambda$ small enough. 
For concrete estimates, see~\cite[Proposition 6.2.45]{Bratteli1981}, or~\cite{Froehlich2015} for more recent results. 

If $\cS_{(\tau_\Phi,\beta)}$ is not a singleton, one needs to take into account boundary conditions to reach all KMS-states by thermodynamic limit. That is our next topic.

\noindent\textbf{Araki-Gibbs condition.} For any  $\Lambda\in\fGfin$, the so called  
surface energies 
\[
W_\Lambda(\Phi) :=
\sum_{\twol{X\cap\Lambda\not=\emptyset}{X\cap\Lambda^c\not=\emptyset}}\Phi(X)
\]
are in $\cOs$. Let $\beta>0$ and $V_\Lambda=\beta W_\Lambda(\Phi)$.
Suppose that  $\omega$ is  a modular state on $\cO$  and let $\delta_\omega$ 
be the generator of its modular $C^\ast$-dynamics $\varsigma_\omega$, 
$\varsigma_\omega^t=\e^{t\delta_\omega}$. Consider  the perturbed dynamics 
$\varsigma_{\omega,V_\Lambda}$ generated by $\delta_\omega+\i[V_\Lambda,\argdot]$, 
and let $\omega_{V_\Lambda}$ be the $(\varsigma_{\omega,V_\Lambda},-1)$-KMS state 
associated to $\omega$ by Araki's perturbation theory. 
We say that $\omega$ satisfies $(\beta,\Phi)$ Gibbs condition if, for all  $\Lambda\in\fGfin$, the restriction of $\omega_{V_\Lambda}$ to 
$\cO_\Lambda$ is given by~\eqref{thgivs}. The Gibbs condition is the quantum 
counterpart of the DLR equation in the equilibrium theory of classical spin systems. For the  proofs, see \cite[Proposition 6.2.17]{Bratteli1981}.

\begin{theorem} Suppose that $\omega$ is a modular state on $\cO$ and $\beta>0$.
The following statements are equivalent.
\ben
\item $\omega$ is a $(\tau_\Phi,\beta)$-KMS state.
\item $\omega$ satisfies the $(\beta,\Phi)$ Gibbs condition. 
\een
\end{theorem}

\noindent\textbf{Gibbs variational principle.} We will discuss the Gibbs variational
principle only in the translation invariant setting; for a general lattice theory
see~\cite[Section 6.2.3]{Bratteli1981}. Suppose that $G=\ZZ^d$, and let 
$\varphi=\{\varphi^x\,|\,x\in \ZZ^d\}\subset\Aut(\cO)$ denote the natural action
of the translation group $\ZZ^d$ on $\cO$. We denote by $\cSI$ the set of all translation-invariant (that is, $\varphi$-invariant) states on $\cO$. For any 
$\nu\in\cSI$, the limit 
\[
s(\nu)=\lim_{\Lambda\rightarrow\infty}\frac{S(\nu_\Lambda)}{|\Lambda|}
\]
exists, where $\nu_\Lambda$ denotes the restriction of $\nu$ to $\cO_\Lambda$ and 
$S(\nu_\Lambda)$ is its von~Neumann entropy. $s(\nu)$ is the specific entropy of the 
state $\nu$. The entropy map $\cSI\ni\nu\mapsto s(\nu)$ is affine, upper-semicontinuous, and takes values in $[0,\log\dim\fh]$. An interaction 
$\Phi$ is translation invariant if 
\[
\varphi^x(\Phi(X))=\Phi(X+x)
\]
holds for all finite subsets $X\subset \ZZ^d$ and all $x\in\ZZ^d$. For translation invariant interactions,
\[
\|\Phi\|_\lambda=\sum_{X\ni0}\e^{\lambda(|X|-1)}\|\Phi(X)\|.
\]
In what follows we fix a translation invariant $\Phi$ satisfying 
$\|\Phi\|_\lambda<\infty$ for some $\lambda>0$. The specific energy observable 
of $\Phi$ is 
\[
E_\Phi=\sum_{X\ni0}\frac{\Phi(X)}{|X|},
\]
and for any $\omega \in\cSI$, 
\[
\lim_{\Lambda\to\infty}\frac1{|\Lambda|}\omega(H_\Lambda(\Phi))=\omega(E_\Phi).
\]
Finally,  the finite limit 
\[
P(\Phi)=\lim_{\Lambda\to\infty}\frac1{|\Lambda|}
\log \tr\left(\e^{- H_\Lambda(\Phi)}\right)
\]
exists and is called the pressure of $\Phi$. The next theorem is known as 
the Gibbs variational principle.
\begin{theorem} For any $\beta>0$, 
\[
P(\beta\Phi)=\sup_{\nu\in\cSI}(s(\nu)-\beta\nu(E_\Phi)).
\]
Moreover, 
\[
\cSeq(\beta\Phi)=\{\nu\in\cSI\,|\, P(\beta\Phi)=s(\nu)-\beta\nu(E_\Phi)\}
\]
is a non-empty convex compact subset of $\cSI$. The elements of $\cSeq(\beta\Phi)$ 
are called equilibrium states for interaction $\Phi$ at inverse temperature $\beta$. 
\label{gibbs-j}
\end{theorem}
Finally, equilibrium states and KMS states are related by
\begin{theorem} For any $\beta>0$ the following statements are equivalent.
\ben 
\item $\omega\in\cSeq(\beta\Phi)$.
\item $\omega$ is a translation invariant $(\tau_\Phi,\beta)$-KMS state. 
\label{gibbs-j-1}
\een
\end{theorem}
For further discussion of Theorems \ref{gibbs-j} and \ref{gibbs-j-1}, we refer the reader to 
\cite[Proposition 6.2.39 and Theorem 6.2.40]{Bratteli1981}, and \cite[Theorem 6.2.42]{Bratteli1981}.

%%%%%%%%%%%%%%%%%%%%%%%%%%%%%%%%%%%%%%%%%%%%%%%%%%%%%
\part{Non-equilibrium quantum statistical mechanics}
\label{sec-non-eq-re}
%%%%%%%%%%%%%%%%%%%%%%%%%%%%%%%%%%%%%%%%%%%%%%%%%%%%%

The works~\cite{Jaksic2001a,Ruelle2000,Ruelle2001,Pillet2001}\footnote{The
work~\cite{Ho2000}, although focused on the concrete exactly solvable
non-equilibrium $XY$-spin chain, foreshadowed these developments.} initiated
modern developments in non-equilibrium quantum statistical mechanics by
introducing the notions of entropy production observable and non-equilibrium
steady states (NESS). The concept of quantum entropy production observable goes
back at least to~\cite{Pusz1978},\footnote{See the Remark on page 281 of this paper. Another pioneering work on the subject is~\cite{Spohn1978b}.} and was 
re-introduced independently in the literature several times since; 
see~\cite{Ojima1988,Ojima1989,Ojima1991}. The concept of NESS in the quantum 
setting  was novel and  motivated by developments in classical non-equilibrium 
statistical mechanics; see~\cite{Ruelle1999}. The introduction  of the quantum entropy 
production observable in conjunction with NESS has spurred rapid developments in 
non-equilibrium quantum statistical mechanics—both in structural theory and in 
the study of concrete models—that continue to this day. For a non-exhaustive 
list of references, we refer the reader to~\cite{Benoist2023a}.

%%%%%%%%%%%%%%%%%%%%%%%%%%
\section{General setting}
\label{non-eq-genset}
%%%%%%%%%%%%%%%%%%%%%%%%%%

We follow~\cite{Jaksic2001a}. Let $(\cO,\tau,\omega)$ be a $C^\ast$-quantum 
dynamical system whose reference state $\omega$ is not $\tau$-invariant. 
The NESS of $(\cO,\tau,\omega)$ are the limit points of the net 
\[
\left\{\frac{1}{T}\int_0^T \omega\circ\tau^t\d t\right\}_{T>0}
\]
as $T\uparrow\infty$. The set $\cS_{\tau+}(\omega)$ of NESS is non-empty and 
its elements are $\tau$-invariant.

To introduce the entropy production observable, we assume that the state 
$\omega$ is modular. Recall that  $\delta_\omega$ denotes the generator of 
the modular dynamics $\varsigma_\omega$. We further assume that the generator 
$\delta$ of $\tau$ has the form 
\[
\delta=\delta_\mathrm{fr}+\i[V,\argdot],
\]
where $V\in\cOs$ and $\delta_\mathrm{fr}$ generates a "free" $C^\ast$-dynamics 
$\tau_\mathrm{fr}$ for which $\omega\in\cS_{\tau_\mathrm{fr}}$. Lacking a better terminology, we shall say that such a system $(\cO,\tau,\omega)$ is locally perturbed (abbreviated LP). If $V\in\Dom(\delta_\omega)$, the entropy production 
observable of a LP system is defined by 
\[
\sigma :=\delta_\omega(V).
\]
When we wish to indicate the dependence of $\tau$ and $\sigma$ on $V$ we will 
write $\tau_V$, $\sigma_V$. 

The starting point of the theory is the entropy balance equation
\beq 
\Ent(\omega_t|\omega)=-\int_0^t \omega_s(\sigma)\d s.
\label{ep-oce}
\eeq
Since the relative entropy is non-positive,  it follows  that
\[
\omega_+(\sigma)\geq 0
\]
for any NESS $\omega_+\in\cS_{\tau +}(\omega)$. 
 
The two basic questions of non-equilibrium quantum statistical mechanics are:

\assuming{QIII}{QThree}{Describe the structure of the set $\cS_{\tau+}$
and the properties of its elements. 
}

\assuming{QIV}{QFour}{Elucidate  dynamical mechanisms that 
ensure  $\omega_+(\sigma)>0$ for all $\omega_+\in\cS_{\tau+}$. 
}

There are two general approaches to~\QThree. The first is the 
$C^\ast$-scattering method that goes back to Robinson~\cite{Robinson1973} and 
in the context of NESS was further developed by Ruelle~\cite{Ruelle2000}; 
see also~\cite{Aschbacher2006}. In this method the goal is to prove that 
for $A\in\cO$ the limit 
$$
\gamma^+(A):=\lim_{t\to\infty}\tau_\mathrm{fr}^{-t}\circ\tau^t(A)
$$
exists. In this case $\gamma^+\in\Aut(\cO)$, $\lim_{t\to\infty}\omega_t=\omega_+=\omega\circ\gamma^+$,
and $\cS_{\tau+}=\{\omega_+\}$. A sufficient condition that ensures the 
existence of the M\o ller morphism $\gamma^+$ is that there exists a dense 
subset $\cA\subset\cO$ such that for all $A\in\cA$, 
\[
\int_0^\infty\|[\tau^t(A),V]\|\d t<\infty. 
\]
Conditions of this type are called $L^1$-asymptotic Abelianness and have 
been extensively used in the literature; see~\cite{Bratteli1981}. For some 
physically important models  where this condition has been verified, 
see~\cite{Botvich1983,Froehlich2003,Jaksic2007a}.

The second general approach to~\QThree{} was developed in~\cite{Jaksic2002a} 
 based on the spectral theory of quantum transfer operators. It is a quantum 
extension of Ruelle transfer operator method; see~\cite{Baladi2000}. For a 
pedagogical introduction to quantum transfer operators in finite quantum 
system setting, see~\cite{Jaksic2010b}. For modern developments, see~\cite{Benoist2024a}.

Regarding~\QFour, we refer the reader to~\cite{Aschbacher2006,Jaksic2002b} 
and references therein. Both questions~\QThree{} and~\QFour{} remain 
poorly understood and much work remains to be done.

A special class of LP systems, the so-called open  quantum systems, play a 
privileged role in the study of non-equilibrium quantum statistical mechanics 
and we proceed to describe them.

Consider finitely many, say $M$, thermal reservoirs $\cR_j$ described by 
$C^\ast$-quantum dynamical systems $(\cO_j,\tau_j,\omega_j)$. We denote 
by $\delta_j$ the generator of $\tau_j$. The reservoir $\cR_j$ is assumed to be
in thermal equilibrium at inverse temperature $\beta_j>0$, that is, we assume 
that $\omega_j$ is a $(\tau_j,\beta_j)$-KMS state on $\cO_j$. In the absence of
interaction, the combined reservoir system $\cR=\cR_1+\cdots+\cR_M$  is 
described by the quantum dynamical system  $(\cO_\cR,\tau_\cR,\omega_\cR)$, where\footnote{Whenever the meaning is clear within the context, we write $A$ for $A\otimes\one$ and $\one\otimes A$, $\delta_j$ for $\delta_j\otimes\Id$, 
$\Id\otimes\delta_j$, etc.} 
\begin{align*}
\cO_\cR&:=\cO_1\otimes\cdots\otimes\cO_M,\\[4pt]
\tau_\mathrm{fr}&:=\tau_1\otimes\cdots\otimes \tau_M,\\[4pt]
\omega_\cR&:=\omega_1\otimes\cdots\otimes\omega_M.
\end{align*}
There are  two ways to couple the reservoirs. In the first case of directly coupled 
reservoirs, $\tau_\mathrm{fr}=\tau_\cR$, the interaction is described by  
$V\in\cO_{\cR,\mathrm{sa}}$, and the interacting dynamics $\tau$ is generated by 
$\delta=\delta_\cR+\i[V,\argdot]$. 

The reservoirs can be also coupled through a finite quantum system $\cS$ with Hilbert 
space $\cK_\cS$. Let $(\cO_\cS,\tau_\cS,\omega_\cS)$ be a finite-dimensional quantum system 
dynamical describing $\cS$\footnote{We abbreviate $\cO_{\cK_\cS}$ by $\cO_\cS$}, where 
we assume that $\omega_\cS>0$. The generator of $\tau_\cS$ is 
$\delta_\cS=\i[H_\cS,\argdot]$, where $H_\cS$ is the Hamiltonian of $\cS$. In 
the absence of interaction, the joint system $\cS+\cR$ is described by the 
$C^\ast$-quantum dynamical system $(\cO,\tau_\mathrm{fr},\omega)$ 
where 
\[
\cO=\cO_\cS\otimes\cO_\cR,\qquad\tau_\mathrm{fr}=\tau_\cS\otimes\tau_\cR,
\qquad \omega=\omega_\cS\otimes\omega_\cR.
\]
The interaction of $\cS$ with $\cR_j$ is described by a selfadjoint $V_j\in\cO_\cS\otimes\cO_j$, and the full interaction by 
$V :=\sum_jV_j$. The interacting dynamics $\tau$ is generated by 
$\delta:=\delta_\mathrm{fr}+\i[V,\argdot]$. We assume that 
$V_j\in\Dom(\delta_j)$ for all $j$, which ensures that the entropy production 
observable is well-defined. 

In what follows,  we will always take 
\beq
\omega_\cS=\one/\dim\cK_\cS
\label{s-choice}
\eeq
for the reference state of $\cS$. This choice is made for convenience. It 
is easy to show that none of the asymptotic results of this paper depend 
on the choice of $\omega_\cS$ as long as $\omega_\cS$ is faithful. With the 
choice~\eqref{s-choice},
$$
\sigma=-\sum\beta_j{\cal J}_j,
$$
where the observable ${\cal J}_j:=\delta_j(V_j)$ describes the energy flux 
out of the $j$-th reservoir. 

The above description of open quantum system needs to be adapted in the case of 
fermionic systems. The modifications are straightforward; 
see~\cite{Aschbacher2006, Jaksic2007a}.

%%%%%%%%%%%%%%%%%%%%%%%%%%%%%%%%%%%%%%%%%%%%
\section{Open lattice quantum spin systems}
\label{sec-olqss}
%%%%%%%%%%%%%%%%%%%%%%%%%%%%%%%%%%%%%%%%%%%%

We follow~\cite{Ruelle2001}, and consider the lattice quantum spin system
of Section~\ref{sec-lattice-eq}, the interaction $\Phi$ satisfying Assumption~\SR. We further assume:

\begin{quote}
\textbf{(a)} For some finite $M$, 
\[
G=S\cup\left(\bigcup_{j=1}^M R_j\right)
\]
where $S$ is  a finite set and each $R_j$ is an infinite set. The sets 
$S,R_1,\ldots,R_M$ are assumed to be disjoint.
\end{quote}
We adapt the notation of Section~\ref{sec-lattice-eq} to that of 
Section~\ref{non-eq-genset} by setting $\cO_{\cR_j}=\cO_{R_j}$ for 
$j\in\{1,\ldots,M\}$, $\cO_\cS=\cO_S$, and
$\cK_\cS :=\otimes_{x\in S}\fh_x$. One then has the decomposition 
\[
\cO=\cO_\cS\otimes\cO_\cR=\cO_\cS\otimes\cO_{\cR_1}\otimes\cdots\otimes\cO_{\cR_M}.
\]

In the framework of Section~\ref{non-eq-genset}, the subsystems described by the 
subalgebras $\cO_{\cR_j}$ act as reservoirs, and the next assumption forbids direct 
coupling between these reservoirs. 
\begin{quote}
\textbf{(b)} For any $X\in\fGfin$, if $X\cap R_i\not=\emptyset$, $X\cap R_j\not=\emptyset$ for some $i\not=j$ , then $\Phi(X)=0$. 
\end{quote}

We denote by $\fG_{j,\fin}$ the collection of all finite subsets of $R_j$, and by 
$\Phi_j$ the restriction of $\Phi$ to $\fG_{j,\fin}$. $\tau_{\cR_j}$ denotes the 
$C^\ast$-dynamics on $\cO_{\cR_j}$ generated by $\Phi_j$ and $\delta_{\cR_j}$ its generator. The coupling of the finite subsystem $\cS$ with the $j$-th reservoir
is given by
\[
V_j :=\sum_{X\cap S\not=\emptyset, X\cap R_j\not=\emptyset}\Phi(X),
\]
and we set $V:=\sum_j V_j$. The "free" $C^\ast$-dynamics $\tau_\mathrm{fr}$ 
on $\cO$ is generated by 
\[
\delta_\mathrm{fr}=\delta_\cS+\sum_j\delta_{\cR_j},
\]
where $\delta_\cS=\i[H_S(\Phi),\argdot]$. The fully coupled dynamics
$\tau=\tau_\Phi$ on $\cO$ is generated by
\[
\delta_\Phi=\delta_\mathrm{fr}+\i[V,\argdot].
\] 
The model is completed by a  choice of $(\tau_{\cR_j},\beta_j)$-KMS state 
$\omega_{\beta_j}$ on $\cO_{\cR_j}$ for $j=1,\ldots,M$, and by 
taking\footnote{$\omega_\cS$ is given by~\eqref{s-choice}.} 
\[
\omega=\omega_\cS\otimes\omega_{\beta_1}\otimes\cdots\otimes \omega_{\beta_M}
\]
for the reference state of $(\cO,\tau)$. $(\cO,\tau,\omega)$ is 
an example of open quantum system discussed in Section~\ref{non-eq-genset}. 

Assumption~\SR{} ensures that 
$$
\sum_j \sum_Y \sum_{X\cap S\not=\emptyset, X\cap R_j\not=\emptyset}\|[\Phi(Y),\Phi(X)]\|<\infty.
$$
Since $\delta_\omega=-\sum_j\beta_j\delta_{\cR_j}$, we deduce that 
$V\in\Dom(\delta_\omega)$, the entropy production observable being given by 
$$
\sigma=\delta_\omega(V)=\sum_j\sum_{Y\subset R_j}\sum_{X\cap S\not=\emptyset, X\cap R_j\not=\emptyset}
-\i \beta_j[\Phi(Y),\Phi(X)].
$$

We now turn to the discussion of the thermodynamic limit of open 
lattice quantum spin systems. One starts with $\Lambda\in\fGfin$ of the form 
\[
\Lambda= S\cup\left(\bigcup_{j=1}^M\Lambda_j\right),
\]
where $\Lambda_j\in {\fG}_{j,\fin}$, and 
\[
\cO_\Lambda=\cO_\cS\otimes\cO_{\Lambda_1}\otimes\cdots\otimes\cO_{\Lambda_M}.
\]
The  finite volume dynamics $\tau_\Lambda$ is generated by the Hamiltonian 
\[
H_\Lambda(\Phi)=H_{\mathrm{fr},\Lambda}+V_\Lambda,
\]
where 
\[
H_{\mathrm{fr},\Lambda}:=H_S(\Phi)+\sum_j H_{\Lambda_j}(\Phi_j),
\qquad
V_\Lambda:=\sum_jV_{j,\Lambda},
\]
with 
\[ 
V_{j,\Lambda}:=\sum_{\twol{X\subseteq S\cup \Lambda_j}{X\cap S\not=\emptyset, X\cap \Lambda_j\not=\emptyset}} \Phi(X).
\]
The finite volume reference state has a product structure 
\beq
\omega_\Lambda=\omega_\cS\otimes\omega_{\Lambda_1}\otimes\cdots\otimes\omega_{\Lambda_M},
\label{ru-gh}
\eeq
where 
\beq
\omega_{\Lambda_j}=\e^{-\beta_j H_{\Lambda_j}(\Phi_j)}/
\tr(\e^{-\beta_j H_{\Lambda_j}(\Phi_j)}).
\label{ru-spin}
\eeq

The first basic question regarding the thermodynamic limit in the present 
setting concerns the entropy production observable. In a finite volume $\Lambda$, 
the entropy production observable is given  by 
\beq 
\sigma_\Lambda=-\i \sum_j\beta_j [H_{\Lambda_j}(\Phi_j),V_{j,\Lambda}]
=\i\sum_j\beta_j[H_\Lambda(\Phi),H_{\Lambda_j}(\Phi_j)].
\label{ru-ep-gh}
\eeq
The finite volume entropy balance equation takes the form
\[
\Ent(\omega_\Lambda\circ\tau_\Lambda^t|\omega_\Lambda)
=-\int_0^t\omega_\Lambda(\tau_\Lambda^s(\sigma_\Lambda))\d s,
\]
and of course has an elementary proof.
\begin{theorem}
\ben 
\item $\lim_{\Lambda \rightarrow \infty} \sigma_\Lambda=\sigma$.
\item Suppose that each $\omega_{\beta_j}$ is a thermodynamic-limit point $\beta_j$-KMS 
state on $\cO_{\cR_j}$. Let 
\[
\Lambda_\alpha=S\cup\left(\bigcup_{j=1}^M\Lambda_{j,\alpha}\right),
\]
be a net in $\fGfin$ such that for each $j$, 
\[
\lim _{\Lambda_{j,\alpha}\to\infty}
{\omega}_{\Lambda_{j,\alpha}}=\omega_{\beta_j}.
\]
Then 
\begin{align*}
\Ent(\omega\circ\tau^t|\omega)
&=\lim_{\Lambda_\alpha\to\infty}\Ent(\omega_{\Lambda_\alpha}\circ 
\tau_{\Lambda_\alpha}^t|\omega_{\Lambda_\alpha})\\[2mm]
&=-\lim_{\Lambda_\alpha\to\infty}\int_0^t 
\omega_{\Lambda_\alpha}(\tau_{\Lambda_\alpha}^s(\sigma_{\Lambda_\alpha}))\d s\\[2mm]
&=-\int_0^t\omega(\tau^s(\sigma))\d s.
\end{align*}
\een
\label{jj-3}
\end{theorem}

In~\cite{Ruelle2001}, Ruelle considered a somewhat more general setting by 
allowing for  suitable "boundary terms" in~\eqref{ru-spin}. The reason for 
introducing these "boundary terms" is to reach a broader class of
reservoir KMS-states than the "free boundary condition"~\eqref{ru-spin} allows. 
To address this issue more generally, one  can start with an arbitrary
$(\tau_{\cR_j},\beta_j)$-KMS state $\omega_{\beta_j}$ on $\cO_{\cR_j}$, set 
$\omega_{\Lambda_j}=\omega_{\beta_j}\upharpoonright_{\cO_{\Lambda_j}}$, and
take~\eqref{ru-gh} defined with these new $\omega_{\Lambda_j}$. The entropy
production observable $\sigma_\Lambda$ remains defined by~\eqref{ru-ep-gh}. 
In this case, the total entropy production over the time-interval
$[0, t]$, 
\[
-\int_0^t \omega_\Lambda(\tau_\Lambda^s(\sigma_\Lambda))\d s,
\]
cannot be directly linked to the relative entropy and \textsl{does not need} to be non-negative. However, since
\[
-\lim_\Lambda\int_0^s\omega_\Lambda(\tau_\Lambda^t(\sigma_\Lambda))\d t
=-\int_0^s\omega_t(\sigma)\d t=\Ent(\omega_s|\omega),
\]
the basic properties of entropy production are restored in the thermodynamic limit.
For further discussion of this topic, we refer the reader to~\cite[Section 3.4]{Benoist2024b}.

%%%%%%%%%%%%%%%%%%%%%%%%%%%%%%%%%%%%%%%%%%%%%
\section{Equilibrium versus non-equilibrium}
%%%%%%%%%%%%%%%%%%%%%%%%%%%%%%%%%%%%%%%%%%%%%

There are good reasons why our understanding of equilibrium quantum statistical
mechanics vastly surpasses its non-equilibrium counterpart. Perhaps the most important reasons concern the role of dynamics, the variational principle, and the thermodynamic limit. Although thermal equilibrium states have dynamical characterization in
terms of the KMS-condition, in the concrete setting of lattice spin or
fermionic systems the KMS condition is equivalent to the Gibbs variational
principle and Gibbs condition. The latter two, in conjunction with 
thermodynamic limit, provide  powerful tools in the study of the question~\QOne{} 
and sharply separate it from the much less understood question~\QTwo. There 
is no such separation in the non-equilibrium case, and a detailed understanding of 
the dynamics is central to both~\QThree{} and~\QFour. 

Another important reason concerns the relation of~\QFour{} with the
singularity of NESS. If $\omega_+=\omega_\mathrm{n}+\omega_\mathrm{s}$ 
is the decomposition of a NESS into its $\omega$-normal part $\omega_\mathrm{n}$ 
and its $\omega$-singular part $\omega_\mathrm{s}$, then $\omega_\mathrm{n}(\sigma)=0$,
and so $\omega_+(\sigma)>0$ iff $\omega_\mathrm{s}(\sigma)>0$. Moreover, if the  following boundedness condition holds, 
\[
\sup_{T>0}\left|\int_0^T(\omega(\tau^t(\sigma))-\omega_+(\sigma))\d t\right|<\infty,
\]
then $\omega_+(\sigma)>0$ iff $\omega_\mathrm{s}\not=0$; see \cite[Theorem 1.1]{Jaksic2002a}. Although physically natural, 
the fact that the entropy production of NESS is carried by its singular part makes its 
study particularly  delicate. This emergence of singularity is specific to 
non-equilibrium: if $\omega$ is a $(\tau_\mathrm{fr},\beta)$-KMS state, then 
the NESS is unique and $\omega$-normal. These points are discussed in detail in~\cite{Aschbacher2006, Jaksic2002b}. 

All the reasons discussed above are common to the classical and quantum case, underscoring that the conceptual difficulty of non-equilibrium statistical mechanics is not a specifically quantum phenomenon.

%%%%%%%%%%%%%%%%%%%%%%%%%%%%%%%%%%%%%%%%
\part{Quantum phase space contraction}
%%%%%%%%%%%%%%%%%%%%%%%%%%%%%%%%%%%%%%%%

In open (classical or quantum) systems entropy production is directly related to
the heat generation. The phase-space contraction reflects the loss of microscopic 
information as the system evolves, in the large-time limit, to a singular NESS.
These two  notions  are closely related. Much of this section follows~\cite{Benoist2024a}.

%%%%%%%%%%%%%%%%%%%%%%%%%%%%
\section{Classical setting}
\label{sec-classical}
%%%%%%%%%%%%%%%%%%%%%%%%%%%%

We consider a classical dynamical system   $(\cX,\phi)$ where $\cX$ is a compact metric space and  $\phi=\{\phi^t\,|\,t\in\RR\}$  is a group of homeomorphisms of $\cX$ such 
that the map 
\[
\RR\times\cX\ni (t,x)\mapsto\phi^t(x)\in\cX
\]
is continuous. We denote by $C(\cX)$ the vector space of all continuous complex-valued 
functions on $\cX$ and  equip it with the sup norm 
$\|f\|_\infty :=\sup_{x\in\cX}|f(x)|$. The observables are functions $f\in C(\cX)$ 
and they evolve in time as $f\mapsto f_t := f\circ\phi^t$. The states are Borel 
probability measures on $\cX$ and we write $\nu(f)=\int_\cX f\d\nu$. They evolve in 
time as $\nu\mapsto\nu_t :=\nu\circ\phi^{-t}$. A state $\nu$ is $\phi$-invariant 
if $\nu_t=\nu$ for all $t$. The classical relative entropy of two states $\nu$ and 
$\rho$ is defined by 
\[
\Ent(\nu|\rho)=\int_\cX\log\frac{\d\rho}{\d\nu}\d\nu
\]
if $\rho\ll\nu$, and is set to $-\infty$ otherwise. Its basic property is 
$\Ent(\nu|\rho)\leq 0$ with equality iff $\nu=\rho$. 

A time-reversal of $(\cX,\phi)$ is an involutive homeomorphism $\vartheta:\cX\to\cX$ 
such that
\[
\vartheta\circ\phi^t=\phi^{-t}\circ\vartheta
\]
for all $t\in\RR$. A state $\nu$ is called time-reversal invariant (TRI) if 
$\nu\circ\vartheta=\nu$. 

The starting point is a classical dynamical system $(\cX,\phi,\omega)$, assuming 
that the initial (reference) state $\omega$ is not $\phi$-invariant. 
The system is TRI if $\omega$ is TRI for some time-reversal of $(\cX,\phi)$. 

We set the regularity assumptions. First, throughout this section we assume 
\assuming{C1}{COne}{For all $t\in\RR$ the measures $\omega$ and $\omega_t$ 
are equivalent,\ie have the same sets of measure zero.
}
Let 
\[
\Delta_{\omega_t|\omega} :=\frac{\d\omega_t}{\d\omega}, \qquad \ell_{\omega_t|\omega}:= \log \Delta_{\omega_t|\omega}.
\]
\assuming{C2}{CTwo}{$\ell_{\omega_t|\omega}\in C(\cX)$ for all $t\in \RR$.}
We set 
\[
c^t :=\ell_{\omega_t|\omega}\circ \phi^t,
\]
and denote by $Q_t$ the law of $c^t$ w.r.t.\;$\omega$. For $\alpha\in\CC$, let
\[
\fF_t(\alpha):=\int_\RR\e^{-\alpha s}\d Q_t(s).
\]
\assuming{C3}{CThree}{The map 
\[
\RR\ni t\mapsto\ell_{\omega_t|\omega}\in C(\cX)
\]
is differentiable at $t=0$. 
}
The entropy production observable, or the phase space contraction rate, is defined by 
$$
\sigma=\frac{\d}{\d t}\ell_{\omega_t|\omega}\big|_{t=0}
=\frac{\d}{\d t}c^t\big|_{t=0}.
$$
\bep \label{thm:noel}
Suppose that~\COne--\CTwo{} hold. 
\ben
\item For all $t,s\in\RR$, 
\begin{align*}
\Delta_{\omega_{t+s}|\omega}
&=\Delta_{\omega_s|\omega}\circ\phi^{-t}\Delta_{\omega_t|\omega}\\[4pt]
\ell_{\omega_{t+s}|\omega}
&=\ell_{\omega_s|\omega}\circ\phi^{-t}+\ell_{\omega_t|\omega},\\[4pt]
c^{t+s}
&=c^s+c^t\circ\phi^s.
\end{align*}
\item $\Ent(\omega_t|\omega)=-\omega(c^t)$.
\een
In the remaining statements we assume that~\CThree{} also holds.
\ben
\setcounter{enumi}{2}
\item 
\[
\ell_{\omega_t|\omega}=\int_0^t\sigma_{-s}\d s,\qquad c^t=\int_0^t \sigma_s \d s,
\]
and 
$$
\Ent(\omega_t|\omega)=-\int_0^t\omega_s(\sigma)\d s.
$$
\item $\omega(\sigma)=0$.
\een
In the remaining statements we assume that, in addition, $(\cX,\phi,\omega)$ 
is \TRI{} with time-reversal $\vartheta$.
\ben
\setcounter{enumi}{4}
\item $c^t\circ\vartheta=c^{-t}$ and $\sigma\circ \vartheta=-\sigma$. 
\item For all $t\in\RR$ and $\alpha\in\CC$, 
$$
\fF_t(\alpha)=\fF_t(1-\bar\alpha).
$$
\item 
Let $\fr:\RR\to\RR$ be the reflection $\fr(s)=-s$ and $\overline{Q}_t=Q_t\circ\fr$. 
Then, for any $t\in\RR$, the measures $Q_t$ and $\overline{Q}_t$ are equivalent, and 
\beq 
\frac{\d\overline{Q}_t}{\d Q_t}(s)=\e^{-s}.
\label{str-2-quant-cl}
\eeq
\een
\eep

The NESS of $(\cX,\phi,\omega)$ are defined as the weak-limit points of the net 
\[
\left\{\frac{1}{T}\int_0^T \omega_t \d t\,|\, T>0\right\}_{T>0}.
\]
The set of NESS is non-empty and any NESS is $\phi$-invariant. Moreover, 
for any NESS $\omega_+$ one has 
\[
\omega_+(\sigma)\geq 0.
\]

The two basic questions~\QThree{} and~\QFour{} of   
Section~\ref{non-eq-genset} have the same formulation in the classical case. 
In fact, the classical case can be directly embedded in the quantum case 
by considering commutative $C^\ast$-dynamical systems, see~\cite{Benoist2024a}.

%%%%%%%%%%%%%%%%%%%%%%%%%%%%%%%%%%%%%%%%%%%%%%%%%%%
\section{GNS-representation and modular structure}
\label{sec-modular}
%%%%%%%%%%%%%%%%%%%%%%%%%%%%%%%%%%%%%%%%%%%%%%%%%%%

Some of  the material below has been already discussed; see in particular
Sections~\ref{sec:Tomita--Takesaki theory} and~\ref{sec:Entropy production}.

We  denote by $(\cH_\omega,\pi_\omega,\Omega_\omega)$ the GNS-representation 
of $\cO$ associated to $\omega$ and by $\fM_\omega=\pi_\omega(\cO)^{\prime\prime}$ 
the enveloping von~Neumann algebra of bounded  operators on $\cH_\omega$. We drop subscript $\omega$ when no confusion is possible. 
Since the state $\omega$ is  assumed to be modular, the  vector $\Omega$ is separating 
for $\fM$,\footnote{See~\cite[Corollary~5.3.9]{Bratteli1981}} and in particular 
$\|\pi(A)\|=\|A\|$ for all $A\in\cO$. Whenever the meaning is clear from the context, 
we will denote $\pi(A)$ by $A$.

$\cN$ denotes the set of all normal states on $\fM$, \ie the states described by 
density matrices on $\cH$. Obviously elements of $\cN$ also define states on $\cO$. 
Any state on $\cO$ that arises in this way is called $\omega$-normal. Again, whenever 
the meaning is clear from the context, we will denote such states by the same letter. 
In particular, the vector state $\fM\ni A\mapsto\langle\Omega,A\Omega\rangle$ is  
denoted by $\omega$. 

We will assume that the reader is familiar with the basic notions of Tomita-Takesaki 
modular theory; see Section~\ref{sec:Tomita--Takesaki theory} and any of the 
references~\cite{Bratteli1987,Bratteli1981,Derezinski2003,Haag1996,Ohya1993,
Stratila1979}. We will use the same notation and terminology as in 
Section~\ref{sec-standard-rep}. $\cH_+$ and $J$  denote the natural cone and modular 
conjugation associated to the pair $(\fM,\Omega)$. The unique vector representative of 
$\nu\in\cN$ in the natural cone is denoted by $\Omega_\nu$. The modular operator of 
$\nu\in\cN$ is denoted by $\Delta_\nu$. The relative modular operator of a pair 
$(\nu,\rho)$ of normal states is denoted by $\Delta_{\nu|\rho}$. The following 
perturbative result~\cite{Jaksic2003} is behind the entropy balance 
equation~\eqref{ep-oce}.
 
\begin{theorem} Let $\nu$ be a modular state on $\cO$ and $U\in\Dom(\delta_\nu)$ a unitary. Consider the state $\nu_U(\argdot)=\nu(U^\ast\argdot\,U)$ and 
let $P=-\i U\delta_\nu(U^\ast)$. Then 
\[
\log\Delta_{\nu_U|\nu}=\log\Delta_\nu+P.
\]
\end{theorem}
 
The relative entropy of a pair $(\nu,\rho)$ of normal faithful states is 
\[
\Ent(\nu|\rho)=\langle\Omega_\nu,\log\Delta_{\rho|\nu}\Omega_\nu\rangle.
 \]
This is the original definition of Araki~\cite{Araki1975/76}, with the sign and 
ordering convention of~\cite{Bratteli1981}; see also Section~\ref{sec:Entropy production}.

Since $\omega$ is $\varsigma_\omega$ invariant, the family 
$\{\pi\circ\varsigma_\omega^t\, |\,t \in \RR\}$ extends to a $W^\ast$-dynamics 
on $\fM$ which we again denote by $\varsigma_\omega$.\footnote{Pointwise 
$\sigma$-weakly continuous groups of $\ast$-automorphisms on $\fM$.} 
For $A\in \fM$, 
\[
\varsigma_\omega^t(A)=\Delta_\omega^{\i t}A\Delta_\omega^{-\i t}.
\]
More generally, to any faithful $\nu\in\cN$ one associates a $W^\ast$-dynamics 
$\varsigma_\nu$ by 
\[
\varsigma_\nu^t(A)=\Delta_\nu^{\i t}A\Delta_\nu^{-\i t}.
\]
$\varsigma_\nu$ is called the modular dynamics of $\nu$, and $\nu$ is a
$(\varsigma_\nu,-1)$-KMS state on $\fM$. 

We make the assumption:

\begin{quote}\textbf{(St)} The family $\{\pi\circ\tau^t\, |\,t\in\RR\}$ extends to
a  $W^\ast$-dynamics on $\fM$ which we again denote by $\tau$.
\end{quote}

This assumption is automatically satisfied for LP systems discussed  in 
Section~\ref{non-eq-genset}. It ensures that $\omega_t$ is a faithful 
$\omega$-normal state on $\fM$, and that there exists a unique selfadjoint operator 
$\cL$ on $\cH$, called the standard Liouvillean of $\tau$, such that for all 
$A\in\fM$ and $t\in\RR$, 
\[
\tau^{t}(A)=\e^{\i t\cL}A\e^{-\i t\cL}, 
\qquad 
\e^{-\i t\cL}\cH_+=\cH_+.
\]
The vector representative of $\omega_t$ in $\cH_+$ is $\e^{-\i t\cL}\Omega$. 
The standard Liouvillean is a basic example of a quantum transfer operator. 
Note that the standard Liouvillean of $\varsigma_\omega$ is $\log\Delta_\omega$. 

The basic properties of the standard Liouvillean are summarized in:
\begin{theorem}
\ben 
\item $\e^{\i t\cL}J=J\e^{\i t\cL}$.
\item The standard Liouvillean of an LP system is 
\[
\cL=\cL_\mathrm{fr}+ V-JVJ,
\]
where $\cL_\mathrm{fr}$ is the standard Liouvillean of $\tau_\mathrm{fr}$.
\item $\omega\in\cS_\tau$ iff $\cL\Omega=0$.
\item Suppose that $\omega\in\cS_\tau$. Then the quantum dynamical 
system $(\cO,\tau,\omega)$ is ergodic iff $0$ is a simple eigenvalue of $\cL$.
\een
\end{theorem}

The following well-known results identifies ergodicity with the so-called 
property of return to equilibrium.

\begin{theorem}Suppose that $\omega$ is $\tau$-invariant. Then the quantum dynamical 
system $(\cO,\tau,\omega)$ is ergodic iff, for any $\omega$-normal state $\nu$ 
and $A\in\cO$, one has
\[
\lim_{T\to\infty}\frac{1}{T}\int_0^T\nu(\tau^t(A))\d t=\omega(A).
\]
\end{theorem}

To any two faithful normal states $\nu$ and $\rho$ one associates a family of unitary operators called Connes' cocycle 
\[
[D\nu:D\rho]_{\i t}:=\Delta_{\nu|\rho}^{\i t}\Delta_{\rho}^{-\i t}, 
\qquad t\in\RR.
\]
Its basic properties are summarized in:
\begin{theorem}\label{thm:connes}
\ben
\item For all $t\in\RR$, $[D\nu:D\rho]_{\i t}$ is a unitary element of\/ $\fM$.
\item $[D\nu:D\rho]_{\i t}^\ast=[D\rho:D\nu]_{\i t}$ for all $t\in\RR$.
\item The $\varsigma_\rho$-cocycle relation
\[
[D\nu:D\rho]_{\i s}\varsigma_\rho^s([D\nu:D\rho]_{\i t})=[D\nu: D{\rho}]_{\i(s+ t)},
\]
holds for all $s,t\in\RR$.
\item The intertwining relation
\[
[D\nu:D\rho]_{\i t}\varsigma_\rho^t(A)[D\nu:D\rho]_{\i t}^\ast=\varsigma_\nu^t(A),
\]
holds for any $A\in\fM$ and $t\in\RR$.
\item If $\mu$ is another faithful normal state, then the chain rule
\[
[D\nu:D\rho]_{\i t}[D\rho:D\mu]_{\i t}=[D\nu: D\mu]_{\i t},
\]
holds for any $t\in\RR$.
\een
\end{theorem}
For the proof we refer the reader to~\cite{Araki1982}.

The family of Connes cocycles 
$\left([D\omega_t:D\omega]_\alpha\right)_{\alpha\in\i\RR}$
will be crucial for the identification of quantum phase space contraction 
in the next section. The following proposition describes its basic property,
which is the quantum counterpart of the first relation in Theorem~\ref{thm:noel}(i).

\begin{proposition}\label{start-qpsc}
For all $t,s\in\RR$ and $\alpha\in\i\RR$, 
$$
[D\omega_{t+s}:D\omega]_\alpha
=\tau^{-t}([D\omega_s:D\omega]_\alpha)[D\omega_t:D\omega]_\alpha.
$$
\end{proposition}

In the development of non-equilibrium quantum statistical mechanics, the following 
regularity assumption plays an important role:
\assuming{Reg1}{RegOne}{
For all $t\in\RR$ and $\alpha\in\i\RR$, 
\[
[D\omega_t:D\omega]_\alpha\in\pi(\cO).
\]
}
When~\RegOne{} holds, we will denote $\pi^{-1}([D\omega_t:D\omega]_{\alpha})$
by $[D\omega_t:D\omega]_{\alpha}$. For LP systems, \RegOne{} holds if 
$V\in\Dom(\delta_\omega)$.
 %%%%%%%%%%%%%%%%%%%%%%%%%%%%%%%%%%%%%%%%%%
\section{Quantum phase space contraction}
\label{sec-ph-spc-ag1}
%%%%%%%%%%%%%%%%%%%%%%%%%%%%%%%%%%%%%%%%%%

As in the classical case, our starting point is Proposition~\ref{start-qpsc}.  
We make two additional regularity  assumptions: 
\assuming{Reg2}{RegTwo}{For all $t\in\RR$ the map 
\[
\i\RR\ni\alpha\mapsto[D\omega_t:D\omega]_\alpha\in\cO
\]
is differentiable at $\alpha=0$. 
}
Let 
\[
\ell_{\omega_t|\omega}
 :=\frac{\d}{\d\alpha}[D\omega_t:D\omega]_\alpha\big|_{\alpha=0},
\]
and 
\[
c^t :=\tau^t(\ell_{\omega_t|\omega}).
\]
Note that $\ell_{\omega_t|\omega}$, and consequently $c^t$, are selfadjoint 
elements of $\cO$.  

\assuming{Reg3}{RegThree}{The map 
\[
\RR\ni t\mapsto\ell_{\omega_t|\omega}\in\cO,
\]
is differentiable at $t=0$.
}
Let
\beq
\sigma=\frac{\d}{\d t}\ell_{\omega_t|\omega}\big|_{t=0}
=\frac{\d}{\d t}c^t\big|_{t=0}.
\label{def-epo-gq}
\eeq
\begin{proposition}\label{thm-qspc-ti}
\ben
\item For all $t,s\in\RR$, one has 
\begin{align*}
\ell_{\omega_{t+s}|\omega}
&=\tau^{-t}(\ell_{\omega_s|\omega})+\ell_{\omega_t|\omega},\\[4pt]
c^{t+s}
&=c^s+\tau^s(c^t).
\end{align*}
\item For all $t\in\RR$, $\log \Delta_{\omega_t|\omega}
=\log \Delta_{\omega}+\ell_{\omega_t|\omega}$, and for $\alpha\in\i\RR$, 
\[
[D\omega_t:D\omega]_{\alpha}=\e^{\alpha(\log\Delta_\omega 
+\ell_{\omega_t|\omega})}\e^{-\alpha\log\Delta_\omega}.
\]
\item For all $t\in\RR$, $\Ent(\omega_t|\omega)=-\omega(c^t)$.
\item Setting $\sigma_t=\tau^t(\sigma)$ for $t\in\RR$, one has
\[
\ell_{\omega_t|\omega}=\int_0^t\sigma_{-s}\d s, 
\qquad 
c^t=\int_0^t\sigma_s\d s,
\]
and 
\beq
\Ent(\omega_t|\omega)=-\int_0^t\omega_s(\sigma)\d s.
\label{ent-balance-2}
\eeq
\item $\omega(\sigma)=0$.
\item If $(\cO,\tau,\omega)$ is \TRI{} with time-reversal $\Theta$, then $\Theta(c^t)=c^{-t}$ and $\Theta(\sigma)=-\sigma$. 
\een
\end{proposition}
This proposition is the direct quantum analogue of Proposition~\ref{thm:noel}.
$\sigma$ is the entropy production observable of $(\cO,\tau,\omega)$ and~\eqref{ent-balance-2} is the entropy balance equation. 

Theorem~\ref{thm-qspc-ti} sheds  light on  the introduction to non-equilibrium
quantum statistical mechanics presented in Section~\ref{non-eq-genset} in the
context of the LP systems.

\begin{proposition}
Suppose that $(\cO,\tau,\omega)$ is an LP system and that $V\in\Dom(\delta_\omega)$. 
Then~\RegOne, \RegTwo{} and \RegThree{} hold, and $\sigma$ 
in~\eqref{def-epo-gq} is equal to $\delta_\omega(V)$. 
\end{proposition}

The two basic questions of non-equilibrium quantum statistical mechanics, \QThree{} 
and \QFour{} formulated in Section~\ref{non-eq-genset}, extend in the obvious way 
to the more general setting considered here.

Motivated by the analogies between Theorem~\ref{thm:noel} and
Theorem~\ref{thm-qspc-ti}, we consider the map 
\[
\i\RR\ni\alpha\mapsto[D\omega_{-t}:D\omega]_{\alpha}\in\cO
\]
as the quantum analogue  characterization of phase space contraction for 
$(\cO,\tau,\omega)$ at time $t$.

%%%%%%%%%%%%%%%%%%%%%%%%%%%%%%%%%%%%%%%%%%%%%%%%%%%%%%%%%
\part{How should one define quantum entropy production?}
\label{sec-how-ep}
%%%%%%%%%%%%%%%%%%%%%%%%%%%%%%%%%%%%%%%%%%%%%%%%%%%%%%%%%
Entropy production quantifies irreversibility and information loss and is directly linked to the Second Law of Thermodynamics.
While entropy production is a well-defined concept in classical physics, its
interpretation becomes multifaceted and subtler in the quantum domain.
In particular, multiple non-equivalent notions of entropy production arise.

%%%%%%%%%%%%%%%%%%%%%%%%%%%%%%%%%%%%%%%%%%%%%%%%%%%%%%%%%%%%%%%%%%%%%%%%%%%%
\section{Quantum phase space contraction and entropy production observable} 
%%%%%%%%%%%%%%%%%%%%%%%%%%%%%%%%%%%%%%%%%%%%%%%%%%%%%%%%%%%%%%%%%%%%%%%%%%%%

The most direct approach to defining quantum entropy production proceeds by
quantizing the corresponding classical entropy production observable, and has
been described in Sections~\ref{non-eq-genset} and~\ref{sec-ph-spc-ag1}. 
This  route has been adopted in~\cite{Jaksic2001a,Ruelle2001} and played an
important role in the development of the subject. 

The main criticism of this approach concerns the following two points:
\begin{enumerate}%[label=\textbf{(\alph*)}]
\item The experimental status of the selfadjoint observable 
$c^t=\int_0^t\sigma_s\d s$ is questionable. 
\item The finite time fluctuation relation, the quantum analog 
of~\eqref{str-2-quant-cl}, fails for $c^t$ as an operator-valued observable. More precisely, if 
$Q_t$ is the spectral measure for $\omega$ and $c^t$, the relation 
\[ 
\frac{\d\overline{Q}_t}{\d Q_t}(s)=\e^{-s}.
\]
\textsl{does not hold} except in special cases; see~\cite[Exercise 3.3]{Jaksic2010b}.
\end{enumerate}

Point~\textbf{(a)} is connected to several perennial aspects of the so-called
standard (or orthodox) view of the conceptual foundations of quantum mechanics.
There are many excellent discussions of this view in the literature, beginning
with the classical book of von Neumann~\cite{Neumann1955}. Here we limit
ourselves to two further references: \cite[Section 7]{Haag1996}
and~\cite{Wigner1963}. Point~\textbf{(b)}, on the other hand, was emphasized 
forcefully in~\cite{Talkner2007}.

We do not think that the validity of the finite time fluctuation relation can
be raised to a fundamental principle of what is observable in quantum
statistical mechanics, and hence we do not find criticism~\textbf{(b)} tenable. 
Although~\textbf{(a)} obviously must be taken more seriously, it is reasonable 
to accept that any proposal for the average entropy production over a time 
interval $[0,t]$ of $(\cO,\tau,\omega)$ gives the same value $\omega(c^t)$, 
and this is certainly true for all the  known proposals. Thus, the questions
\QThree{} and~\QFour{} of Section~\ref{non-eq-genset} are unaffected 
by~\textbf{(a)}. The entropy production observable is also a natural
starting point in the study of the linear response theory, Onsager reciprocity
relations, and near-equilibrium fluctuation-dissipation
mechanism~\cite{Jaksic2006b,Jaksic2006c,Jaksic2006a,Jaksic2007a}; see also
Section~\ref{sec:Entropy production}. It is the analysis of far from equilibrium 
entropy production fluctuations that brings~\textbf{(a)} to focus. 

We however emphasize that some foundational issues about the choice of entropy
production observable $\sigma$ remain, even if one focuses only on the averages 
$\omega(c^t)$. One of them is our next topic.

%%%%%%%%%%%%%%%%%%%%%%%%%%%%%%%%%%%%%%%%%%%%%%%%%%%%%%%%%
\section{Ruelle's decomposition of entropy production}
%%%%%%%%%%%%%%%%%%%%%%%%%%%%%%%%%%%%%%%%%%%%%%%%%%%%%%%%%

Shortly after the introduction of NESS and entropy production observable
in~\cite{Jaksic2001a,Ruelle2001}, David Ruelle wrote an important but  little
known paper~\cite{Ruelle2002} concerning the interpretation of the  quantum entropy
production. We start with a brief summary of~\cite{Ruelle2002} adjusting certain 
points to allow for a clearer comparison  with subsequent developments.

Consider a finite open quantum system on the Hilbert space $\cK=\otimes_{j=1}^M\cK_j$,
in the directly coupled reservoir setting. With a more involved notation
the arguments easily extend to a general open quantum system. For $j=1,\ldots,
M$ we set $\cK_{\setminus j} :=\bigotimes_{i\not=j}\cK_i$. Let
\footnote{$\tr_{\cK_{\setminus j}}$ denotes the partial trace over $\cK_{\setminus j}$}
\[
\omega_{tj} :=\tr_{\cK_{\setminus j}}\omega_t
\]
be the state of the $j$-th reservoir at the time $t$ and 
\[
\omega^\dec_t :=\bigotimes_{j=1}^M\omega_{tj}.
\]
Note that $\omega^\dec_0=\omega$. Let \footnote{We recall that 
$S(\nu)=-\tr(\nu \log \nu)$ denotes the von~Neumann entropy of $\nu$.}
\[
\Delta S(t) := S(\omega_t^\dec)-S(\omega),
\]
and 
\[
\Delta\fS(t) :=-\Ent(\omega_t^\dec|\omega).
\]
Obviously, $\Delta\fS(t)\geq0$. Moreover, since $S(\omega_t^\dec)=-\tr(\omega_t^\dec\log\omega_t^\dec)=-\tr(\omega_t\log\omega_t^\dec)$
and $S(\omega)=S(\omega_t)$, one has
$$
\Delta S(t)=-\tr(\omega_t(\log\omega_t^\dec-\log\omega_t))
=-\Ent(\omega_t|\omega_t^\dec)\ge0.
$$
 The starting point 
of~\cite{Ruelle2002} is the  identity 
\begin{align}
\int_0^t\omega_s(\sigma)\d s&=-\Ent(\omega_t|\omega)
=\tr(\omega_t(\log\omega_t-\log\omega))\nonumber\\[4pt]
&=\tr(\omega_t(\log\omega_t-\log\omega_t^\dec))
+\tr(\omega_t(\log\omega_t^\dec-\log\omega))\label{ruelle-key-one}\\[4pt]
&=-\Ent(\omega_t|\omega_t^\dec)-\Ent(\omega_t^\dec|\omega)
=\Delta S(t)+\Delta\fS(t),\nonumber
\end{align}
expressing the mean entropy production over the time interval $[0,t]$ as the 
sum of  \textsl{mutual information} $\Delta S(t)$ and a relative entropy-cost  $\Delta\fS(t)$ 
associated with the reduced reservoir dynamics.  

The decomposition~\eqref{ruelle-key-one} remains valid for infinitely extended 
reservoirs and can be justified either by a thermodynamic limit argument 
or directly by an application of Corollary~5.20 and identity~(5.22) of \cite{Ohya1993}

In~\cite{Ruelle2002}, Ruelle proposes a heuristic framework in which   $\Delta S(t)$ is expected to 
dominate~\eqref{ruelle-key-one}. This perspective touches on foundational issues in non-equilibrium statistical mechanics, particularly concerning the role and structure of thermal reservoirs in open quantum systems. Ruelle’s program remains only partially understood, and much work remains to be done.

Finally, we note that the Ruelle program is not specific to the quantum setting;
the same analysis applies, and the same questions arise, in open classical
systems, where equally little is known.

%%%%%%%%%%%%%%%%%%%%%%%%%%%%%%%%%%%%%%%%%%%%%%%%%%%%%%%%%%%%
\section{Two-time measurement  entropy production (TTMEP)}
\label{sec:TTMEP}
%%%%%%%%%%%%%%%%%%%%%%%%%%%%%%%%%%%%%%%%%%%%%%%%%%%%%%%%%%%%

In the finite dimensional setting, the notion of \textsl{two-times entropy production}
goes back to~\cite{Kurchan2000,Tasaki2000}. Our presentations  follows~\cite{Jaksic2010b}.

Consider a  finite  quantum dynamical system $(\cO_\cK,\tau,\omega)$.
The observable to be measured  is described by a partition of unity
$(P_a)_{a\in\cA}$ on $\cK$\footnote{The $P_a$ are orthogonal projections such that 
$\sum_{a\in\cA}P_a=\one$.} indexed by a finite alphabet $\cA$ labeling the possible outcomes of
the measurement. The two-times measurement protocol goes as follows. At time $t=0$, when the
system was in the state $\omega$, the first  measurement is performed and the
outcome $a\in\cA$ is observed with probability 
\[
p(a)=\tr(\omega P_a).
\]
After this first measurement, the system is in the reduced state 
$$
\frac{1}{p(a)}P_a\omega P_a,
$$
which evolves  over the time interval $[0,t]$\footnote{We allow for $t<0$.} to 
\[
\frac{1}{p(a)}\e^{-\i tH}P_a\omega P_a\e^{\i tH}.
\]
The second measurement, at the time $t$, yields the outcome $a'\in\cA$ with probability 
$$
p_{t}(a'|a)=\frac{1}{p(a)}\tr\left(\e^{-\i tH}P_a\omega P_a\e^{\i tH}P_{a'}\right).
$$
Finally, the probability of observing the pair $(a,a')$ in the two-times measurement 
protocol is 
$$
p_{t}(a,a')=p_t(a'|a)p(a)=\tr\left(\e^{-\i tH}P_a\omega P_a\e^{\i tH}P_{a'}\right).
$$

Starting with the above protocol, there are two distinct routes to define
measurement based entropy production. We describe here the one which is
the main topic of this section. The second route involving repeated
measurements is reviewed in Section~\ref{sec-repeated}.

Assume that $\omega>0$  and consider the  partition of unity $(P_a)_{a\in\cA}$ that 
arises through the spectral decomposition of $\omega$,
$$
\omega=\sum_{a\in\cA}\lambda_a P_a, 
$$
where the eigenvalues $\lambda_a$ are assumed to be distinct. The entropy production random 
variable $\cE:\cA\times\cA\to\RR$ is defined by 
$$
\cE(a,a')=\log\lambda_{a'}-\log\lambda_a, 
$$
and its probability distribution with respect to $p_t$ is denoted by $Q_t$, 
\[
Q_t(s)=\sum_{\cE(a,a')=s}p_t(a,a').
\]
The statistics of TTMEP  over the time interval $[0,t]$ is described by $Q_{t}$. 
For $\alpha\in\CC$, we set 
\[
\fF_t^\ttm(\alpha):=\int_\RR\e^{-\alpha s}\d Q_t(s).
\]
We then have:
\begin{proposition}
\ben
\item 
\beq
\fF_t^\ttm(\alpha)=\omega([D\omega_{-t}:D\omega]_\alpha)=\tr(\omega_{-t}^\alpha\omega^{1-\alpha}).
\label{conc-m-1}
\eeq
\item 
\[
\int_\RR s\,\d Q_t(s)=-\Ent(\omega_t|\omega).
\]
In particular $\int_\RR s\,\d Q_t(s)\geq 0$ with equality iff $\omega=\omega_t$.
\een
In the remaining statements we assume that the system is \TRI.
\ben
\setcounter{enumi}{2}
\item 
$$
\fF_t^\ttm(\alpha)=\fF_t^\ttm(1-\bar\alpha).
$$
\item 
\beq 
Q_t(-s)=\e^{-s}Q_t(s).
\label{qes-1}
\eeq
\een
\end{proposition}

The identity~\eqref{qes-1} is sometimes called \textsl{finite time fluctuation relation.}

We now turn to a general modular quantum dynamical system $(\cO,\tau,\omega)$ satisfying~\RegOne,
\RegTwo{} and \RegThree. For $t\in\RR$ and $\alpha\in\i\RR$, we set 
$$
\fF_t^\ttm(\alpha):=\omega\left([D\omega_{-t}:D\omega]_\alpha\right).
$$
When $(\cO,\tau,\omega)$ is finite, this reduces to~\eqref{conc-m-1}.

\begin{proposition}
\ben 
\item There exists unique probability measure $Q_t$ on $\RR$ such that 
\[
\fF_t^\ttm(\alpha)=\int_\RR\e^{-\alpha s}\d Q_t(s).
\]
\item $\int_\RR s\,\d Q_t(s)=-\Ent(\omega_t|\omega)$. In particular, $\int_\RR s\,\d Q_t(s)\geq0$ with 
equality iff $\omega=\omega_t$. 
\item For $\alpha\in\i\RR$, 
\[
\fF_t^\ttm(\alpha)=\langle\Omega,\Delta_{\omega_{-t}|\omega}^\alpha\Omega\rangle.
\]
In particular, $Q_t$ is the spectral measure of the selfadjoint operator 
$-\log \Delta_{\omega_{-t}|\omega}$ for the vector $\Omega$. 
\een
In the remaining statements we assume that $({\cal O}, \tau, \omega)$ is \TRI.
\ben
\setcounter{enumi}{3}
\item The map $\i\RR\ni\alpha\mapsto\fF_t^\ttm(\alpha)$ has an analytic extension 
to the vertical strip $0<\Re z<1$ that is bounded and continuous on its closure.
\item For any $\alpha$ satisfying $0\leq\Re\alpha\leq1$, 
$$
\fF_t^\ttm(\alpha)=\fF_t^\ttm(1-\bar\alpha).
$$
\item Let $\fr:\RR\to\RR$ be the reflection $\fr(s)=-s$ and $\overline{Q}_t=Q_t\circ\fr$. 
Then the measures $Q_t$ and $\overline{Q}_t$ are equivalent and 
$$
\frac{\d\overline{Q}_t}{\d Q_t}(s)=\e^{-s}.
$$
\een
\end{proposition}

The measure $Q_t$ describes the statistics of the two-time measurement entropy production 
of $(\cO,\tau,\omega)$ over the time interval $[0,t]$.

%%%%%%%%%%%%%%%%%%%%%%%%%%%%%%%%%%%%%%%%%%%%%%%%%%%%%%%%%%%%%%%%%%%%
\section{Thermodynamic limit of TTMEP in open quantum spin systems}
%%%%%%%%%%%%%%%%%%%%%%%%%%%%%%%%%%%%%%%%%%%%%%%%%%%%%%%%%%%%%%%%%%%%

The general modular theoretic approach to $Q_t$ described in the previous
section requires justification of the thermodynamic limit for  concrete models. 
In this section we address this question in the context of open quantum lattice 
spin systems introduced in Section~\ref{sec-olqss}. We denote by $Q_{t,\Lambda}$ 
the law of the TTMEP for  $(\cO_\Lambda,\tau_\Lambda,\omega_\Lambda)$ over 
the time interval $[0,t]$ with $\omega_\Lambda$ defined by~\eqref{ru-gh} 
and~\eqref{ru-spin}. Convergence of probability measures is understood in the weak sense.

\begin{theorem}
Suppose that each $\omega_{\beta_j}$ is a thermodynamic limit point $\beta_j$-KMS 
state on $\cO_{R_j}$. Let 
\[
\Lambda_a= S\cup\left(\bigcup_{j=1}^M \Lambda_{j,a}\right)
\]
be a net in $\fGfin$ such that for each $j$, 
\[
\lim_{\Lambda_{j,a}\to\infty}\overline{\omega}_{\Lambda_{j,a}}=\omega_{\beta_j}.
\]
Then, for all $t\in\RR$,
\[
\lim_{\Lambda_a\to\infty}Q_{t,\Lambda_a}=Q_t.
\]
\end{theorem}
For the proof and additional information, see~\cite{Benoist2024b}.

%%%%%%%%%%%%%%%%%%%%%%%%%%%%%%%%%%%
\section{Ancilla state tomography}
%%%%%%%%%%%%%%%%%%%%%%%%%%%%%%%%%%%

\textsl{Ancilla state tomography} is a technique which extracts information on
the state of a quantum system from the outcome of measurements performed on an
auxiliary system, the \textsl{ancilla}, coupled to the system of interest. 
In this section, following~\cite{Benoist2024a}, we show how this technique 
allows for an indirect measurement of the function $\fF_t^\ttm$. 

We start with the finite system of Section~\ref{sec:TTMEP}.
The state of the ancilla is described by  a spin $1/2$ on $\CC^2$. We denote by $v_\pm$ the
eigenvectors of the Pauli matrix $\sigma_z$ associated to the eigenvalues
$\pm1$. The state of ancilla is described by a density matrix $\rho_a$ such 
that $\langle v_+,\rho_av_-\rangle\not=0$. The Hilbert space of the coupled 
system is $\widehat\cK=\cK\otimes\CC^2$, and its initial state is 
$\widehat\omega=\omega\otimes\rho_a$. We introduce the following family of 
Hamiltonians, indexed by $\alpha\in\i\RR$,
$$
\widehat H_\alpha=\e^{\frac{\alpha}{2}\log\omega\otimes\sigma_z}
\left(H\otimes\one\right)\e^{-\frac{\alpha}{2}\log\omega\otimes \sigma_z}.
$$
Setting $N := |v_-\rangle\langle v_+|$, a computation gives
\beq
\tr\left(\e^{-\i t\widehat H_\alpha}\widehat\omega\e^{\i t\widehat H_\alpha}(\one\otimes N)\right)
=\langle v_+,\rho_a v_-\rangle\,\fF_t^\ttm(\alpha).
\label{a-basic-ab}
\eeq
The left-hand side in~\eqref{a-basic-ab} is interpreted as \textsl{entropic ancilla state tomography} 
of the finite quantum system over the interval $[0,t]$.  

In the particular case of a finite LP system where $H=H_\mathrm{fr}+V$ with 
$[\log\omega,H_\mathrm{fr}]=0$, one has 
\[
\widehat H_\alpha=H\otimes\one+\widehat W_\alpha,
\]
where 
\beq
\widehat W_\alpha=\tfrac12W_\alpha\otimes(\one+\sigma_z)+\tfrac12W_{-\alpha}\otimes(\one-\sigma_z),
\label{tuluz-1}
\eeq
and
\beq
W_\alpha=\varsigma_\omega^{-\i\alpha/2}(V)-V.
\label{tuluz-2}
\eeq

Turning to the general modular quantum dynamical system of Section~\ref{sec:TTMEP}, 
we now consider a LP system satisfying $V\in\Dom(\delta_\omega)$.
Let $\widehat\cO :=\cO\otimes\cO_{\CC^2}$ and $\widehat\tau^t=\tau^t\otimes\Id$.
For $\alpha\in\i\RR$, $\widehat W_\alpha$ and $W_\alpha$ are given by~\eqref{tuluz-1} 
and~\eqref{tuluz-2}. Note that $W_\alpha\in\cOs$ and $\widehat W_\alpha\in\widehat{\cO}_\mathrm{sa}$. 
Let $\widehat\tau_\alpha$ be the perturbation of $\widehat\tau$ by $\widehat W_\alpha$. Let 
$N$ as above and $\widehat\omega=\omega\otimes\rho_a$. Then, again, for all $t\in \RR$, 
\[
\widehat\omega\left(\widehat\tau_\alpha^t(\one\otimes N)\right)
=\langle v_+,\rho_a v_-\rangle\,\fF_t^\ttm(\alpha).
\]
We refer the reader to~\cite{Dorner2013,Campisi_2013,Johnson2016,Roncaglia2014,Chiara2015,
Goold2014,Mazzola2013} for related theoretical studies in the physics literature
and to~\cite{An2014,Batalhao2014,Batalhao2015,Peterson2019} for experimental
implementations.

%%%%%%%%%%%%%%%%%%%%%%%%%%%%%
\section{Limitations of  TTMEP}
%%%%%%%%%%%%%%%%%%%%%%%%%%%%%

The statistics of the two-time measurement entropy production was initially
introduced in the context of finite quantum systems (or, slightly more
generally, confined quantum systems with a possibly infinite discrete energy
spectra). Experimental studies have been carried out in this same setting, often
relying on ancilla state tomography. The thermodynamic limit of these  statistics 
is stable and preserves the connection with modular theory and ancilla state tomography, 
although -- to the best of our knowledge -- the statistics
of the two-time measurement entropy production of infinitely extended system has
not been experimentally explored. Obviously, the ancilla state tomography link
is central to the physical relevance of the two-time measurement protocol for
large quantum systems. It tells us that if the protocol could have been
implemented, then the resulting statistics would coincide with those predicted 
by the ancilla-based tomography. This is certainly satisfactory. 

On the other hand, the TTMEP exhibits a surprising degree stability (or rigidity)
with respect to the instant of the first measurement -- an effect with no
classical counterpart -- which has been recently investigated in~\cite{Benoist2023a,
Benoist2024b, Benoist2024a, Benoist2024c}. Suppose  that the first measurement
of the entropy production observable is not performed at $t=0$, when the system was 
in the reference state $\omega$ from which the observable is defined, but instead 
at the later time $T$, when the system is in the state $\omega\circ\tau_T$. A 
second measurement is then performed at time $T+t$. After passing to the 
thermodynamic limit, and under a general and natural dynamical ergodicity assumption, 
the statistics of the TTMEP turn out to be independent of the instant $T$ of the
first measurement. This effect arises from the invasive decoherence induced by 
the first measurement, which dominates the resulting statistics in the thermodynamic limit. 
This stability should be further examined both theoretically and experimentally
before drawing any final conclusions regarding the foundational role  of
TTMEP in large quantum systems.

%%%%%%%%%%%%%%%%%%%%%%%%%%%%%%%%%%%%%%%%%%%%%%%%%%%
\section{Bessis-Moussa-Villani entropy production}
%%%%%%%%%%%%%%%%%%%%%%%%%%%%%%%%%%%%%%%%%%%%%%%%%%%

To motivate the introduction of this notion of entropy production, we return to
the classical setting of Section~\ref{sec-classical}. The variational formula 
for the classical relative entropy\footnote{See~\cite{Ohya1993} or Theorem~2.1 
in~\cite{Jaksic2011}.} gives that for $\alpha\in\RR$, 
$$
\log\left[\int_\cX\e^{-\alpha\int_0^t\sigma_s\d s}\d\omega\right]
=\sup_\nu\left[S(\nu|\omega)-\alpha\int_\cX\int_0^t\sigma_s\d s\d\nu\right],
$$
where the supremum is taken over all Borel probability measures $\nu$ on $\cX$. 
For finite quantum systems, one has 
\[
\log\tr\left[\e^{\log\omega-\alpha\int_0^t\sigma_s\d s}\right]
=\sup_\nu\left[S(\nu|\omega)-\alpha\int_0^t\nu(\sigma_s)\d s\right],
\]
where the supremum is taken over all density matrices $\nu$ on the system 
Hilbert space $\cK$. We set
\beq
\fF_t^\BMV(\alpha)=\tr\left[\e^{\log\omega-\alpha\int_0^t\sigma_s\d  s}\right].
\label{BMV}
\eeq
This function has been introduced and studied in~\cite{Jaksic2010b}. The acronym BMV 
in~\eqref{BMV} reflects the connection of the right-hand side with the celebrated 
Bessis--Moussa--Villani conjecture~\cite{Bessis1975}\footnote{See 
also~\cite{Lieb2004}.}, which was proved by Herbert Stahl~\cite{Stahl2013} in 2013.

\bet 
Let $A$ and $B$ be $n\times n$ Hermitian matrices, and assume that $B$ is positive 
semidefinite. Define
\[
[0,\infty)\ni \alpha \mapsto f(\alpha)=\tr\left(\e^{A-\alpha B}\right).
\]
Then there exists a non-negative Borel measure $\mu$ on $[0,\infty)$ such that
\[
f(\alpha)=\int_0^\infty\e^{-\alpha s}\,d\mu(s)
\]
for all $\alpha\ge0$.
\eet

Stahl's theorem gives that there exists a Borel probability measure $Q_t^\BMV$ on 
$\RR$ such that, for $\alpha\in\CC$, 
\beq
\fF_t^\BMV(\alpha)=\int_\RR\e^{-\alpha s}\d Q_t^\BMV(s).
\label{stahl-ss}
\eeq
We will refer to $Q_t^\BMV$ as the statistics of the BMV entropy production over 
the time interval $[0,t]$. Except in trivial cases, the measure $Q_t^\BMV$ is not 
atomic. We have 
\begin{proposition}\label{bir-d}
\ben 
\item $\int_\RR s\,\d Q_t^\BMV(s)=-\Ent(\omega_t|\omega)$.
\een
In the remaining statements we assume that the finite quantum system 
$(\cO_\cK,\tau,\omega)$ is \TRI.
\ben
\setcounter{enumi}{1}
\item For $\alpha\in\CC$, 
\[
\fF_t^\BMV(\alpha)=\fF_t^\BMV(1-\bar\alpha).
\]
\item Let $\fr:\RR\to\RR$ be the reflection $\fr(s)=-s$ and 
$\overline{Q}_t^\BMV=Q_t^\BMV\circ\fr$. Then the measures $Q_t^\BMV$ 
and $\overline{Q}_t^\BMV$ are equivalent and 
$$
\frac{\d\overline{Q}_t^\BMV}{\d Q_t^\BMV}(s)=\e^{-s}.
$$
\een
\end{proposition}

In the general setting of a modular $C^\ast$-quantum dynamical systems satisfying~\RegOne,
\RegTwo{}, and \RegThree, one has, for $\alpha\in\RR$,
\beq
\log\|\e^{\frac12(\log\Delta_\omega-\alpha\int_0^t\sigma_s\d s)}\Omega\|^2
=\sup_\nu\left[S(\nu|\omega)-\alpha\int_0^t\nu(\sigma_s)\d s\right],
\label{fri-vs}
\eeq
where the supremum is taken over all states $\nu$  on $\cO$. For finite quantum 
systems the left-hand side of~\eqref{fri-vs} is a rewriting of~\eqref{fri-vs} 
in modular terms. For $\alpha\in\RR$,  we define
\[
\fF_t^\BMV(\alpha)=\|\e^{\frac12(\log\Delta_\omega-\alpha\int_0^t\sigma_s\d s)}\Omega\|^2.
\]
It follows from Araki's perturbation theory of the KMS structure that $\fF_t^\BMV(\alpha)$ 
is finite for all $\alpha\in\RR$, and that the map 
$\RR\ni\alpha\mapsto\fF_t^\BMV(\alpha)$ extends to an entire analytic function. We also have 
\[
\frac{\partial}{\partial\alpha}\fF_t^\BMV(\alpha)\bigg|_{\alpha=0}
=\Ent(\omega_t|\omega).
\] 
In addition, if $(\cO,\tau,\omega)$ is TRI, then for all $\alpha\in\CC$,
\beq
\fF_t^\BMV(\alpha)=\fF_t^\BMV(1-\bar\alpha).
\label{bmv-ss}
\eeq
The existence of the BMV measure satisfying~\eqref{stahl-ss} is not known except for 
finite quantum systems and in the thermodynamic limit setting; see the conjecture 
at the end of this section. 

Returning to the thermodynamic limit of non-equilibrium open quantum spin systems, we have
\begin{theorem} 
Suppose that each $\omega_{\beta_j}$ is a thermodynamic limit point 
$\beta_j$-KMS state on $\cO_{R_j}$. Let 
\[
\Lambda_a=S\cup\left(\bigcup_{j=1}^M\Lambda_{j,a}\right)
\]
be a net in $\fGfin$ such that for each $j$, 
\[
\lim_{\Lambda_{j,a}\to\infty}{\omega}_{\Lambda_{j,a}},
=\omega_{\beta_j}.
\]
where ${\omega}_{\Lambda_{j,a}}$ is the same as in Theorem \ref{jj-3}. 
Then, for all $t\in\RR$, the limit 
\[
Q_t^\BMV :=\lim_{\Lambda_a\to\infty}Q_{t,\Lambda_a}^\BMV
\]
exists and for all $\alpha\in\RR$, 
\[
\fF_t^\BMV(\alpha)=\int_\RR\e^{-\alpha s}\d Q_t^\BMV.
\]
\end{theorem}
The proof of this result is implicit in \cite{Benoist2024b}.

The BMV entropy possesses  several appealing features. By
Proposition~\ref{bir-d}(i), it correctly predicts the entropy production of 
NESS, and the symmetry~\eqref{bmv-ss}, with its suitable 
generalization,  leads to the linear  response theory of NESS;
see~\cite{Jaksic2010b}. The function $\fF_t^\BMV$ is entire analytic
without invoking any additional regularity assumptions. 

Many interesting questions remain open; we mention here perhaps the two most important ones.

\noindent{\bf Question 1.} For finite quantum systems, the physical status of the statistics $Q_t^{\BMV}$ is unclear. Are these statistics experimentally accessible?

\noindent{\bf Question 2.} At present, for infinitely extended systems the existence of the probability measures $Q_t^{\BMV}$ can only be justified by thermodynamic limit arguments. These measures would exist for any modular $C^\ast$-dynamical system $(\mathcal O,\tau,\omega)$ provided that the following KMS extension of the BMV conjecture holds.
\begin{quote}{\bf Conjecture.} Let $(\fM,\tau)$ be a $W^\ast$-dynamical system 
on a Hilbert space $\cH$ and $\omega(A)=(\Omega,A\Omega)$ a faithful 
$(\tau,\beta)$-KMS vector state on $\fM$. Let $\cL$ be the standard Liouvillean 
of $\tau$, $V\in\fM$ selfadjoint, and consider the perturbed $W^\ast$-dynamics 
on $\fM$ given by 
\[
\tau_V^t(A)=\e^{\i t(\cL+V)}A\e^{-\i t(\cL+V)}.
\]
Let
\[
\omega_V(A)=\frac{(\Omega_V,A\Omega_V)}{\|\Omega_V\|^2}, 
\qquad 
\Omega_V=\e^{-\frac{\beta}{2}(\cL+V)}\Omega,
\]
be the perturbed $(\tau_V,\beta)$-KMS vector state on $\fM$. Then there exists 
a Borel probability measure $P$ on $\RR$ such that for all $\alpha\in\RR$, 
\[
\|\Omega_{\alpha V}\|^2=\int_\RR\e^{-\alpha s}\d P(s).
\]
\end{quote}

In finite dimension, the above conjecture is equivalent to the BMV conjecture
positively resolved by Stahl~\cite{Stahl2013}. The Stahl method, however, is 
restricted to finite-dimensional matrices, and novel ideas are needed to address
the above general conjecture. 

The material discussed in this section has not previously appeared  in print. 
%%%%%%%%%%%%%%%%%%%%%%%%%%%%%%%%%%%%%%%%%%%%%%%%%%%%%%%%%%%%%%
\section{Entropy production of repeated quantum measurements}
\label{sec-repeated}
%%%%%%%%%%%%%%%%%%%%%%%%%%%%%%%%%%%%%%%%%%%%%%%%%%%%%%%%%%%%%%

Besides the dynamical system approach to classical entropy production, in which
the reference state plays an important role, an altogether different random path 
approach has been developed in~\cite{Kurchan1998,Lebowitz1999, Maes1999}, that 
does not make use of the reference state and is applicable to stochastic processes. 
Its quantum formulation in the setting of repeated quantum measurement processes
goes back to~\cite{Crooks2008}, and was elaborated in~\cite{Benoist2017c,Benoist2021}.
The advent of experimental methods in cavity and circuit QED, and in particular the
experimental breakthroughs of the Haroche--Raimond and Wineland groups 
\cite{Haroche2013,Haroche2006,Wineland2013}, make this complementary approach
particularly relevant. 

Let $\cH$ be a Hilbert space and let $\cB(\cH)$ denote the algebra of bounded
operators on $\cH$. A linear completely positive  map
\[
\Phi:\cB(\cH)\to\cB(\cH)
\]
is called a \textsl{quantum channel}. The channel $\Phi$ is said to be 
\textsl{unital} if it preserves the identity operator, \ie satisfies $\Phi(\one)=\one$. 
The predual of a quantum channel $\Phi$ is the map $\Phi^\ast$ on the trace 
class operators on $\cH$ such that
$$
\tr(\rho\Phi(A))=\tr(\Phi^\ast(\rho)A)
$$
for all $A\in\cB(\cH)$. If $\Phi$ is unital, then $\Phi^\ast$ is trace preserving
and maps density matrices to density matrices.

\begin{definition}
Given a unital quantum channel $\Phi$ on a separable Hilbert space $\cH$ and a Polish 
space $\bA$, a $\boldsymbol{(\Phi,A)}${\bf-instrument} is a $\sigma$-additive map $\cJ$ 
from the Borel $\sigma$-algebra $\cA$ of $\bA$ to the set of completely positive maps
on $\BH$ satisfying 
$$
\cJ(\bA)=\Phi.
$$
\end{definition}

An instrument models a repeatable quantum measurement as follows. Let the system
be in the state (density matrix) $\rho$ at time $t=1$. A measurement is performed, 
and a random outcome $\omega_1\in\bA$ is observed with the law 
$\tr(\cJ(\d\omega_1)^\ast(\rho))$. After the measurement, the system state conditioned 
on $\omega_1\in A_1$ is
$$
\rho_{A_1}=\frac{\cJ(A_1)^\ast(\rho)}{\tr(\cJ(A_1)^\ast(\rho))}.
$$
The law of the outcome $\omega_2$ of the next measurement at time $t=2$ is
$\tr(\cJ(\d\omega_2)^\ast(\rho_{A_1}))$, and the state of the system  after the
second measurement conditioned on $(\omega_1,\omega_2)\in A_1\times A_2$ is
$$
\rho_{A_1A_2}=\frac{\cJ(A_2)^\ast(\rho_{A_1})}{\tr(\cJ(A_2)^\ast(\rho_{A_1}))}.
$$
The probability that the observed $(\omega_1,\omega_2)$ is in $A_1\times A_2$ is
\[
\tr(\cJ(A_1)^\ast(\rho))\tr (\cJ(A_2)^\ast(\rho_{A_1}))
=\tr(\cJ(A_2)^\ast\circ\cJ(A_1)^\ast(\rho))
=\tr(\rho\cJ(A_1)\circ\cJ(A_2)(\one)).
\]
Continuing in this way, one derives that, after $n$ repeated measurements,
the probability of observing a sequence of outcomes
$(\omega_1,\ldots,\omega_n)\in A_1\times\cdots\times A_n$
is given by
$$
\PP_n(A_1\times\cdots\times A_n)=\tr\left(\rho\cJ(A_1)\circ\cdots\circ\cJ(A_n)(\one)\right).
$$
This defines a probability measure $\PP_n$ on $\Omega_n=\bA^n$, and the unitality
of $\cJ(\bA)=\Phi$ implies that the family $(\PP_n)_{n\in\NN}$ is consistent.
Let $\PP$ be the unique probability measure induced on $\Omega=\bA^\NN$, 
equipped with the usual product $\sigma$-algebra $\cF$, and
the filtration $(\cF_n)_{n\in\NN}$ generated by the cylinders
$$
[A_1,\ldots,A_n]=\{\omega\in\Omega\mid(\omega_1,\ldots,\omega_n)\in A_1\times\cdots\times A_n\},
\qquad
A_1,\ldots,A_n\in\cA.
$$
We assume that the  state $\rho$ is $\Phi$-invariant. Then $\PP\circ\phi^{-1}=\PP$ 
where $\phi$ is the left shift on $\Omega$. The dynamical system $(\Omega,\phi,\PP)$ 
thus describes the outcomes of the repeated measurement process. The measure $\PP$ 
is sometimes called the $\rho$-\textsl{statistics} of the instrument $\cJ$.

A \textsl{local reversal} on $\bA$ is a measurable involution $\theta:\bA\to\bA$.
The associated $\theta$-time reversal on $\Omega_n$ is the involution
\[
\theta_n(\omega_1,\ldots,\omega_n)=(\theta(\omega_n),\ldots,\theta(\omega_1)).
\]
The family of probability measures $(\wP_n)_{n\in\NN}$ defined by
$$
\wP_n=\PP_n\circ\theta_n
$$
is also  consistent. By Kolmogorov's extension theorem, there exists   a
unique $\phi$-invariant $\wP\in\cP_\phi(\Omega)$ which describes the 
statistics of the $\theta$-time reversal of the dynamical system 
$(\Omega,\phi,\PP)$.

We shall say that the pair $(\Omega,\PP)$ is $\theta${\sl-time-reversal invariant}
if $\PP=\wP$, \ie if $\PP_n=\wP_n$ for all $n\in\NN$.

The relative entropy of two probability measures $\PP$ and $\QQ$
on $(\Omega,\cF)$, defined by
$$
\Ent(\PP|\QQ)
=\begin{cases}
\ds\int_{\Omega}\log\frac{\d\PP}{\d\QQ}(\omega)\,\d\PP(\omega)&\text{if }\PP\ll\QQ;\\[16pt]
\infty&\text{otherwise,}
\end{cases}
$$
is non-negative and vanishes iff $\PP=\QQ$. The $\theta$-{\sl entropy
production} of $(\Omega,\phi,\PP)$ in the discrete-time interval
$\llbracket1,n\rrbracket$ is defined by
\[
\Ep(\PP_n,\theta)=\Ent(\PP_n|\wP_n).
\]
The (possibly infinite) non-negative number
\[
\ep(\PP,\theta)=\limsup_{n\to\infty}\tfrac1n\Ep(\PP_n,\theta)
\]
is called the $\theta$-{\sl entropy production rate} of $(\Omega,\phi,\PP)$.

At the current level of generality, the $\theta$-entropy production rate can
exhibit pathological behavior; see ~\cite{Andrieux2016}. The following
\textsl{upper-decoupling} property excludes these pathologies, ensuring that
$\theta$-time-reversal invariance is equivalent to the vanishing of entropy
production, and in this sense is characteristic of equilibrium.

\assuming{UD}{UD}{There is a constant $C>0$ such that for any $n\in\NN$,
$A\in\cF_n$ and $B\in\cF$,
$$
\PP(A\cap\phi^{-n}(B))\leq C\,\PP(A)\PP(B).
$$
}
Note that if~\UD{} holds for $\PP$, then it also holds for $\wP$. 
\UD{} automatically holds if $\dim\cH<\infty$, $\rho>0$, and the set 
$\bA$ is finite.\footnote{These assumptions cover many cases of 
physical interest.} 

The following result is an immediate extension  of Theorem~2.1 and
Proposition~2.2 in~\cite{Benoist2017c} using the Donsker--Varadhan 
variational formula for the relative entropy. For details 
see~\cite[Proposition~1.3]{Benoist2025a}
\bep Suppose that~\UD{} holds. Then:
\ben
\item The following (possibly infinite) limit exists:
\[
\ep(\PP,\theta)=\lim_{n\to\infty}\tfrac1n\Ep(\PP_n,\theta).
\]
\item Assume in addition that $\PP$ is $\phi$-ergodic. Then $\ep(\PP,\theta)=0$
iff~ $\PP=\wP$.
\een
\eep
This result is a starting point of~\cite{Benoist2017c,Benoist2021} 
which focus on the non-equilibrium case $\ep(\PP,\theta)>0$. 
In~\cite{Benoist2025a} the vanishing of $\ep(\PP,\theta)$ for a suitable 
class of instruments is linked to the detailed balance condition 
induced by the inner product~\eqref{dbc}.

%%%%%%%%%%%%%%%%%%%%%%%%
%%%%%%%%%%%%%%%%%%%%%%%%
\section{Conclusion}
\label{sec-conclusions}
%%%%%%%%%%%%%%%%%%%%%%%%
The study of entropy production in the setting of $C^\ast$-quantum dynamical systems
has revealed a somewhat unexpected structural richness of non-equilibrium
quantum statistical mechanics. There exist several distinct notions that correctly predict the mean entropy
production of NESS and its linear response theory, while at the same time
capturing different facets of the subject.
The richness of the subject is further enhanced by the study of entropic
fluctuations in the large-deviations regime and their links to quantum
fluctuation theorems~\cite{Benoist2024a} and to hypothesis testing of the arrow
of time~\cite{Jaksic2012}.

In Parts~V--VIII, we have touched on the complexities and subtleties that arise
in the study of quantum entropy production; much theoretical and experimental
work remains to be done to achieve a deeper understanding of this notion.

%%%%%%%%%%%%%%%%%%%%%%%%%%%%%%%
\printbibliography
%\bibliography{MASTER}
%\bibliographystyle{capalpha}
%%%%%%%%%%%%%%%%%%%%%%%%%%%%%%%

\end{document}